\documentclass{article}

\usepackage{arxiv}

\usepackage{amsmath, amssymb, epsfig}
\usepackage{enumerate, amsthm}
\usepackage{bbm}
\usepackage{graphicx}
\usepackage{calrsfs}
\usepackage{nicefrac}
\usepackage{lscape}
\usepackage{pdflscape}
\usepackage{subcaption}
\usepackage[utf8]{inputenc} % allow utf-8 input
\usepackage[T1]{fontenc}    % use 8-bit T1 fonts
\usepackage{hyperref}       % hyperlinks
\usepackage{nicefrac}       % compact symbols for 1/2, etc.
\usepackage{microtype}      % microtypography
\usepackage{amsfonts}       % blackboard math symbols
\usepackage{lipsum}         % Can be removed after putting your text content
\usepackage{doi}

\usepackage{graphicx}
\usepackage{subcaption}

\usepackage{booktabs}
\usepackage{xcolor}
\usepackage{multirow}
\usepackage{bm}
\usepackage{epstopdf}
\usepackage{algpseudocode}
\usepackage[english,onelanguage,ruled]{algorithm2e}
\usepackage{epstopdf}

\usepackage{url}

\graphicspath{{../SimPlot/}}

\usepackage{natbib}
\usepackage[utf8]{inputenc}
\usepackage[english]{babel}
\usepackage[normalem]{ulem}

\newcommand{\prog}[1]{\textsf{#1}}

\newcommand{\code}[1]{\texttt{#1}}

\def\bSig\boldsymbol{\Sigma}

\newcommand{\iid}{\stackrel {{\rm iid.}}{\sim}}
\newcommand{\ind}{\stackrel {{\rm ind.}}{\sim}}

\newcommand{\sumas}{\sum^n_{i=1}}

\newcommand{\ii}{i\in\{1,\ldots,n\}}

\newtheorem{proposition}{Proposition}
\newtheorem{Theorem}{Theorem}
\newcommand{\se}{\boldsymbol{s}}

\newcommand{\bI}{\boldsymbol{I}}
\newcommand{\bmu}{\boldsymbol{\mu}}

\newcommand{\bSigma}{\boldsymbol{\Sigma}}

\newcommand{\bvarepsilon}{\mbox{\boldmath $\varepsilon$}}

\newcommand{\bbeta}{\mbox{\boldmath $\beta$}}
\newcommand{\btheta}{\mbox{\boldmath $\theta$}}

\newcommand{\bDelta}{\mbox{\boldmath $\Delta$}}

\newcommand{\N}{\textrm{N}}
\newcommand{\CN}{\textrm{CN}}
\newcommand{\TCN}{\textrm{TCN}}
\newcommand{\TN}{\textrm{TN}}

\newcommand{\x}{\boldsymbol{x}}
\newcommand{\w}{\boldsymbol{w}}

\newcommand{\A}{\boldsymbol{A}}
\newcommand{\B}{\boldsymbol{B}}
\newcommand{\D}{\boldsymbol{D}}
\newcommand{\bV}{\boldsymbol{V}}

\newcommand{\bgamma}{\mbox{\boldmath $\gamma$}}
\newcommand{\bGamma}{\mbox{\boldmath $\Gamma$}}

\newcommand{\bC}{\boldsymbol{C}}
\newcommand{\yp}{\boldsymbol{y}}
\newcommand{\xp}{\boldsymbol{x}}
\newcommand{\y}{\boldsymbol{y}}
\newcommand{\Y}{\boldsymbol{Y}}

\newcommand{\C}{\boldsymbol{C}}

\newcommand{\blambda}{\mbox{\boldmath $\lambda$}}

\newcommand{\Z}{\boldsymbol{Z}}

\newcommand{\X}{\boldsymbol{X}}
\newcommand{\bW}{\boldsymbol{W}}
\newcommand{\bw}{\boldsymbol{w}}
\newcommand{\bE}{\boldsymbol{E}}

\newcommand{\E}{\textrm{E}}
\newcommand{\ap}{\boldsymbol{a}}
\newcommand{\bp}{\boldsymbol{b}}
\newcommand{\bx}{\boldsymbol{x}}
\newcommand{\bu}{\boldsymbol{u}}
\newcommand{\tr}{\textrm{tr}}
\newcommand{\binfty}{\boldsymbol{\infty}}
\newcommand{\RR}{\mathbbm{R}}

\newcommand{\EE}{{\rm E}}
\newcommand{\W}{\boldsymbol{W}}

\usepackage{xcolor}

\title{Heckman selection - contaminated normal model}

\author{ \href{https://orcid.org/0000-0002-9573-8424}{\includegraphics[scale=0.06]{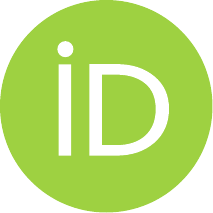}\hspace{1mm}Heeju Lim}\\ %\thanks{Use footnote for providing further information about author (webpage, alternative address)---\emph{not} for acknowledging funding agencies.} \\
	Department of Statistics\\
	University of Connecticut\\
	CT, U.S.A. \\
	\texttt{heeju.lim@uconn.edu} \\
	\And
	\href{https://orcid.org/0000-0002-5724-8918}{\includegraphics[scale=0.06]{orcid.pdf}\hspace{1mm}José Alejandro Ordoñez} \\
	Department of Statistics\\
        Pontificia Universidad Catolica de Chile\\
	Santiago, Chile\\
	\texttt{jose.ordonez@uc.cl} \\
 \And
	\href{https://orcid.org/0000-0001-7742-1821}{\includegraphics[scale=0.06]{orcid.pdf}\hspace{1mm}Antonio Punzo} \\
	Department of Economics and Business\\
        University of Catania\\
	Catania, Italy\\
	\texttt{antonio.punzo@unict.it} \\
 \AND
	\href{https://orcid.org/0000-0002-7239-2459}{\includegraphics[scale=0.06]{orcid.pdf}\hspace{1mm}V\'{\i}ctor H.~Lachos} \\
	Department of Statistics\\
	University of Connecticut\\
	CT, U.S.A. \\
	\texttt{hlachos@uconn.edu} \\
}

\renewcommand{\shorttitle}{Heckman selection - contaminated normal model}

\hypersetup{
pdftitle={EM algorithm for regularization problems in high-dimensional LMMs},
pdfsubject={stat.ME},
pdfauthor={Daniela C. R.~Oliveira, Fernanda L.~Schumacher, Victor H.~Lachos},
pdfkeywords={EM algorithm, High-dimensional data,  Mixed-effects models, R package glmnet,  Regularized variable selection methods},
}

\date{September 18, 2024}

\begin{document}
\maketitle

\begin{abstract}
The Heckman selection model is one of the most well-renounced econometric models in the analysis of data with sample selection. 
This model is designed to rectify sample selection biases based on the assumption of bivariate normal error terms. 
However, real data diverge from this assumption in the presence of heavy tails and/or atypical observations. 
Recently, this assumption has been relaxed via a more flexible Student's t-distribution, which has appealing statistical properties.
This paper introduces a novel Heckman selection model using a bivariate contaminated normal distribution for the error terms. 
We present an efficient ECM algorithm for parameter estimation with closed-form expressions at the E-step based on truncated multinormal distribution formulas. 
The identifiability of the proposed model is also discussed, and its properties have been examined. 
Through simulation studies, we compare our proposed model with the normal and Student's t counterparts and investigate the finite-sample properties and the variation in missing rate. Results obtained from two real data analyses showcase the usefulness and effectiveness of our model. The proposed algorithms are implemented in the \prog{R} package \code{HeckmanEM}.
\\
%\HJ{
%The Heckman selection model is perhaps the most popular econometric model in the analysis of data with sample selection. 
%The analyses of this model are based on the normality assumption for the error terms.
%However, in some applications, this assumption is violated, for instance, in the presence of heavy tails and/or mildly %atypical observations.  
%Recently, this assumption has been relaxed to more flexible models based on the Student's t-distribution, which has appealing statistical properties. 
%This paper proposes a novel Heckman selection model where the random errors follow a bivariate contaminated normal %distribution.  
%We develop an analytically tractable and efficient EM-type algorithm for iteratively computing maximum likelihood estimates of the parameters, with standard errors as a by-product. The algorithm has closed-form expressions at the E-step, that rely on formulas for the mean and variance of the truncated multinormal distribution. The identifiability of the proposed model is also discussed. Simulation studies are conducted to examine the performance of the Heckman selection-contaminated normal model and to compare it with the Normal and Student's t counterparts. 
%Two real datasets are analyzed to illustrate our results. 
%The proposed algorithms and methods are implemented in the \prog{R} package \code{HeckmanEM}.
%}
\end{abstract}

% keywords can be removed
\keywords{EM-type algorithms\and Heckman selection model \and  Multivariate contaminated normal \and Robustness \and \prog{R} package \texttt{HeckmanEM}}

%\bibliographystyle{unsrtnat}
%\bibliography{references}  %%% Uncomment this line and comment out the ``thebibliography'' section below to use the external .bib file (using bibtex) .

\section{Introduction}
\label{Intro}

Sample selection bias is a prevalent challenge across multiple disciplines, including economics, finance, sociology, and biostatistics. This bias occurs when a variable of interest is observable only within a specific subset of the population. In the sampling process, a relationship between observable and unobservable variables result in biases in data interpretation. To better describe this phenomenon and improve discernability between observable and unobservable variables, the classical Heckman SL model was introduced by \citet{heckman1974shadow}, where he proposed a parametric approach to parameter estimation under the assumption of bivariate normality (SLn).

However, this assumption is not always valid in describing certain phenomena, yielding residuals with a distribution with heavy tails or skewness. 
In the context of parametric models, \citet{marchenko2012heckman} introduced the Heckman selection-$t$ model (SLt) that extends the conventional SLn model to have a bivariate Student's-$t$ error distribution. 
This model provides greater flexibility for modeling heavier-tailed data than the SLn model with a single extra parameter: the degrees of freedom that controls the distribution's tails. 
More recently, \citet{lachosHeckman} propose a novel, simple, and efficient Expectation Maximization (EM)-type algorithm for iteratively computing maximum likelihood (ML) estimates of the parameters in the SLt model, where the E-step reduces to computing the first two moments of a truncated bivariate Student's-$t$ distribution.

%As well-documented in \citep{farcomeni2016robust}, real data inadequacies often occur in the form of contamination, namely anomalous values, such as inliers and outliers, that are in disagreement with the data generating mechanism.
%In real data applications, data are often contaminated by atypical points such as inliers and outliers.
%In the context of data with high kurtosis, these are often contaminated by 

In real-life applications, data are often contaminated by anomalous values (also referred to as atypical or ``bad'' points herein, \citealp{Aitk:Wils:Mixt:1980}), (i.e. or most importantly) inliers and outliers, that are in disagreement with the data-generating mechanism.
Inliers, which are comparatively less renowned than the outliers, roughly represent atypical data points that lie in the interior of a statistical distribution \citep{farcomeni2016robust}.
Identifying inliers poses a greater challenge due to their similarity to regular data.
Outliers are widely discussed in the literature and can be roughly distinguished as gross or mild \citep{ritter2014robust}. 
Gross outliers are unpredictable and incalculable such that the statistician cannot model them by a distribution but chooses a method for suppressing them. 
Mild outliers are present due to a possibly incorrect data-generating process from some populations far from the model assumed, so the statistician usually suggests choosing a more flexible model to accommodate all data points, including them.

In the context of robustness against mild outliers in SL models, many authors have considered some elliptically symmetric and fat-tailed distributions to replace the routinely used multivariate normal (N) distribution for statistical modeling, for example, the SLt model, the symmetric SL model \citep{saulo2023bivariate}, or the Birnbaum–Saunders SL model \citep{bastos2020birnbaum}.   
The multivariate contaminated normal (CN) distribution, introduced by \citet{tukey1960survey}, contains two extra parameters with respect to the parameters of the multivariate N distribution and has been widely applied in robust statistics in the case of elliptically heavy-tailed distributions.

The multivariate CN distribution can be viewed as a two-component multivariate N mixture in which one component captures a main proportion of typical observations, and the other comprises relatively few atypical points with the same mean and an inflated/deflated covariance matrix due to the occurrence of outliers/inliers. 

Some crucial properties of the multivariate CN distribution are discussed in \citet{wang2023multivariate}.

In this article, we introduce the Heckman selection-contaminated normal (SLcn) model that extends the conventional SLn model to have a bivariate CN error distribution.  
We consider the ML estimation of the SLcn model by using the expectation conditional maximization (ECM) algorithm \citep{Meng93}. 
For the developed ECM algorithm, the E-steps rely on some properties of the multivariate CN and truncated multivariate CN (TCN) distributions as well as the moments of truncated multivariate N distribution. 
The general formulas for these moments were derived efficiently by \citet{galarza2021moments}, for which we use the \texttt{MomTrunc} package in \texttt{R}. Meanwhile, the CM-steps present closed-form expressions for the updated estimators of all parameters. The likelihood function is easily computed as a byproduct of the E-step and is used for monitoring convergence and model selection. 
Furthermore, we consider a general information-based method for obtaining the asymptotic covariance matrix of the ML estimate. 
Our motivation for considering the MCN distribution is also prompted by the existence of a link between the continuous part of the selection model and the 
multivariate extended skew-contaminated (ESCN) distribution, which has been studied in recent literature \cite{arellanovalle2006uvs, arellano2010multivariate}. 
The method proposed in this paper is implemented in the \prog{R} package \texttt{HeckmanEM}, which is available for download from GitHub (\url{https://github.com/marcosop/HeckmanEM}).

The remainder of the paper is organized as follows. Section~\ref{model} briefly discusses preliminary results related to the multivariate CN distribution and its truncated version, including the ESCN distribution and key properties.  
Section~\ref{SLnModel}, for the sake of completeness,  presents the SLn model proposed by \citet{heckman1979} and some of its characterizations. 
In Section~\ref{SLtModel}, we present the SLcn model, including the ECM algorithm for ML estimation, and derive the empirical information matrix analytically to obtain the standard errors. 
In addition, the link between the SLcn and the family of ESCN distributions is discussed in detail, and following the recent work by \citet{miao2016identifiability}, we show that identifiability holds for the SLcn model. 
In Sections~\ref{sec:Simulation study} and \ref{secApp}, numerical examples, using both simulated and real data, are given to illustrate the performance of the proposed method. 
Finally, some concluding remarks are presented in Section~\ref{sec:6}. 
The relevant analytical results are given in the Appendices.

%\begin{figure}[htb]
%\centering
%\centerline{\includegraphics[width=342pt,height=9pc,draft]{empty}}
%\includegraphics[width=12cm,height=12cm]{Introd-Ap2.eps}
%\caption[FIGURE 1]{Correlation plot of covariates for riboflavin dataset, with $n=71$ and $p=101$ ($p>n$). }%Dots in black mean that the absolute value of the correlation between covariates is between 0.6 and 1, in gray, are between 0.2 and 0.6 and in white it is between 0 and 0.2.}
%\label{Ap2Introd}
%\end{figure}

\section{Background}
\label{model}

This section presents some useful results associated with the $p$-variate  CN distribution and its truncated version. We begin our exposition by defining the notation and presenting some basic concepts used throughout our methodology development. As is usual in probability theory and its applications, we denote a random variable by an upper-case letter and its realization by the corresponding
lower case and use boldface letters for vectors and matrices. Let $\mathbf{I}_p$ represent a $p\times p$ identity matrix, $\A^{\top}$ be the transpose of $\A$. 
Throughout this paper, $\N_p(\bmu, \bSigma)$ denotes the $p$-variate normal distribution with mean vector $\bmu$ and covariance matrix $\bSigma$ with $\phi_p\left(\cdot\mid\bmu,\bSigma \right)$ and $\bm{\Phi}_p(\cdot \mid \bm{\mu}, \bm{\Sigma}) $
denoting its probability density function ({pdf}) and cumulative density function ({cdf}), respectively. When $p=1$ we drop the index $p$, and in this case, if $\mu=0$ and $\sigma^2=1$, we write $\phi(\cdot)$
for the {pdf} and
$\Phi(\cdot)$ for the {cdf}. 
 For multiple integrals, we use the shorthand notation $$ \int_{\ap}^{\bp}f(\x)d\x=\int_{a_1}^{b_1}\ldots\int_{a_p}^{b_p}f(x_1,\ldots,x_p)\mathrm{d} x_p\ldots \mathrm{d} x_1,
 $$
where  $\ap=(a_1,\ldots,a_p)^\top$  and $\bp=(b_1,\ldots,b_p)^\top$. Consider the following Borel set in $\mathbb{R}^p$ of the form:
\begin{equation} \label{hyper1}
\mathbb{A} = \{(x_1,\ldots,x_p)\in \mathbb{R}^p:\,\,\, a_1\leq x_1 \leq b_1,\ldots, a_p\leq x_p \leq b_p \}=\{\xp \in\mathbb{R}^p:\ap\leq\xp\leq\bp\},
\end{equation}
a vector belonging to this set is represented as $\{\Y \in \mathbb{A}\}=\{\ap\leq \Y\leq\bp\}$.

\subsection{The multivariate contaminated normal distribution}
\label{subsec:MCN}
%orange{}
A $p$-dimensional random vector $\X$ is said to follow a CN distribution with location vector $\bmu$, positive-definite scale-covariance matrix $\bSigma$, mixing proportion $\nu_1\in (0,1)$, and degree of contamination $\nu_2\in (0,1)$, denoted by $\X \iid
\CN_p(\bmu,\bSigma,\nu_1,\nu_2)$. 
Here, $\iid$ indicates that the random vector is independent and identically distributed. 
The distribution has the probability density function (pdf) as follows:
\begin{equation}
{f}^{\text{CN}}_p(\xp\mid\bmu,\bSigma,\nu_1,\nu_2)=\nu_1\phi_p(\xp\mid\bmu,\nu_2^{-1}\bSigma)
+
(1-\nu_1)\phi_p(\xp\mid\bmu,\bSigma),\quad \xp\in {\mathbb R}^p.
\label{eq:CN distribution}
\end{equation}

% The MCN is a member of the class of elliptical distributions (see, e.g., \citealp{hawkins1982topics} and \citealp{kollo2006advanced}).
% In particular, it can be obtained as a special case of the multivariate normal scale mixture family under a convenient Bernoulli mixing random variable.
%orange{}
\subsubsection{Parametrization and interpretation}
\label{subsec:Parameterization}

As we can see from \eqref{eq:CN distribution}, the model is represented as a simple two-component normal mixture, with components sharing the same $\bmu$.
According to the parameterization we adopt, which is the one used, amongst others, by \citet{wang2023multivariate}, we have the following insightful interpretation of the model; for other parameterizations of the $p$-variate CN distribution, see, e.g., \citet{Punz:McNi:Robu:2016} and \citet{Farc:Punz:Robu:2020}.

Because of the constraint $\nu_2\in (0,1)$, the first component on the right-hand side of \eqref{eq:CN distribution}, the one with mixing proportion $\nu_1$, is a proportionally inflated (by $1/\nu_2>1$) version of the second component; as such, this is the group with the largest variability.
Now, if $\nu_1\leq 0.5$, the first component is trying to model the mild outliers, and this way to interpret the multivariate CN distribution is very common in the literature (see, e.g., \citealp{Aitk:Wils:Mixt:1980}, \citealp{Punz:McNi:Robu:2016}, and \citealp{Farc:Punz:Robu:2020}).
In this case, $\N_p\left(\bmu,\bSigma\right)$ can be considered as the reference distribution for the good points (for a discussion about the concept of reference distribution, see \citealp{Davi:Gath:Thei:1993}).
Instead, if $\nu_1>0.5$, then we flip: the first component, the larger one in terms of both mixing proportion and variability, models the typical data, while the second component is trying to capture the inliers, and this is in line with \citet{falkenhagen2019likelihood}.
% We remember that an inlier represents an atypical data point that lies in the interior of a statistical distribution.
% Identifying inliers poses a greater challenge due to their similarity to regular data.
In this case, $\N_p\left(\bmu,\nu_2^{-1}\bSigma\right)$ becomes the reference distribution for the typical points.
This falls under the robust statistics perspective that good points should be more than bad points (see, e.g., \citealp{templ2019evaluation} and \citealp{tomarchio2022mixtures}),
Finally, as for $\nu_2$, it can be considered as a degree of contamination (or degree of atypicality).
When $\nu_1\leq 0.5$, the lower $\nu_2$, the larger the atypicality of the outliers.
%In this case, $\N_p\left(\bmu,\bSigma\right)$ can be considered as the reference distribution for the good points (for a discussion about the concept of reference distribution, see \citealp{Davi:Gath:Thei:1993}).
When $\nu_1>0.5$, the lower $\nu_2$, the larger the concentration of the inliers in the interior/central part of the distribution of the typical points.
%Therefore, with respect to $\N_p\left(\bmu,\nu_2^{-1}\right)$, which can be considered as the reference distribution for the model in this case (for a discussion about the concept of reference distribution, see \citealp{Davi:Gath:Thei:1993}), the two additional parameters ($\nu_1$ and $\nu_2$) have an interpretation of practical interest.
%For other parameterizations of the $p$-variate CN distribution, see, e.g., \citet{Punz:McNi:Robu:2016} and \citet{Farc:Punz:Robu:2020}.

Hence, when contrasted with other heavy-tailed extensions of the multivariate normal distribution $\N_p\left(\bmu,\bSigma\right)$, the multivariate CN distribution's two extra parameters, $\nu_1$ and $\nu_2$, offer a practical interpretation that is advantageous.

\subsubsection{Identifiability}
\label{subsubsec:Identifiability}

A natural question that arises when looking at \eqref{eq:CN distribution} is if the model is identifiable.
Without identifiability, the parameters might not be estimated and interpreted, and, more generally, the inference might be meaningless \citep{wang2014note}.
Following the arguments in \citet{tortora2024laplace}, below we show that model~\eqref{eq:CN distribution} is identifiable.

Because the model is a finite mixture (of two components), two issues for its identifiability need to be taken into account: label-switching and overfitting \citep[see][Chapter~1.3, for details]{Fruh:Fine:2006}. 

As for the label-switching issue, it is overcome by using the constraint $\nu_2 \in (0,1)$ as explained below.
Suppose we relax this assumption on $\nu_2$ so that $\nu_2>0$.
Under this new condition, model~\eqref{eq:CN distribution} is nonidentifiable due to label-switching because, if $\tilde\bmu=\bmu$, $\tilde \bSigma=\nu_2^{-1}\bSigma$, $\tilde \nu_1=1-\nu_1$, and $\tilde \nu_2=1/\nu_2$, then $f_p^{\text{CN}}\left(\bx\mid\bmu,\bSigma,\nu_1,\nu_2\right)=f_p^{\text{CN}}\left(\bx\mid\tilde\bmu,\tilde\alpha,\tilde\bSigma,\tilde\nu_1,\tilde\nu_2\right)$. 
This tricky label-switching case, which is the only one possible, can be avoided by adding at least one of the following constraints:
\begin{enumerate}
    \item $\nu_1<0.5$ (as $\nu_1=1-\tilde\nu_1$, it follows that $\tilde\nu_1>0.5$ and we obtain a contradiction);
    \item $\nu_2\in(0,1)$ (as $\nu_2=1/\tilde\nu_2$, it follows that $\tilde\nu_2 > 1$ and we obtain a contradiction).
\end{enumerate}
Hence, either $\nu_1<0.5$ or $\nu_2\in(0,1)$ effectively addresses the label-switching issue. 
It's noteworthy that considering both constraints simultaneously, as commonly practiced in literature, imposes an additional limitation on the parameter space, consequently reducing the CN family's members/models. 
Therefore, in alignment with the practical rationale outlined in Section~\ref{subsec:Parameterization}, particularly the capability to accommodate both inliers and outliers, we opt solely for the constraint $\nu_2\in(0,1)$.

To complete the discussion on identifiability, it is important to realize that the two (trivial but important) conditions $\nu_2\neq 1$ and $\nu_1\in \left(0,1\right)$ also prevent overfitting \citep[a potential problem for identifiability first noted by][]{Craw:Anap:1994}. 
Indeed, identifiability problems may occur due to empty components (i.e., when either $\nu_1=0$ or $\nu_1=1$), where their parameters cannot be uniquely determined, and due to components with equal component parameter vectors (i.e., when $\nu_2=1$), where different values for $\nu_1$ are possible \citep[see][Chapter~1.3, for details]{Fruh:Fine:2006}.
%}
%\todo{Antonio can you check this paragraph?}

\subsubsection{Some properties and representations}
\label{subsubsec:MCN Some properties}

Let %$\Phi_p(\ap,\bp\mid\bmu,\bSigma)$ represent
$$
\Phi_p(\ap,\bp\mid\bmu,\bSigma)=\int_{\ap}^{\bp}{{\phi}_p}(\x\mid\bmu,\bSigma)\textrm{d}\bx,
$$ where  $\ap=(a_1,\ldots,a_p)^\top$  and $\bp=(b_1,\ldots,b_p)^\top$. 
Thus, we denote
$$
{F}^{\text{CN}}_p(\ap,\bp\mid \bmu,\bSigma,\nu_1,\nu_2)=\nu_1\Phi_p(\ap,\bp\mid\bmu,\nu_2^{-1}\bSigma)+(1-\nu_1)\Phi_p(\ap,\bp\mid~\bmu,\bSigma).\label{lsdefAB1}
$$
When $\ap=-\binfty$ we will  simply write $\Phi_p(\bp\mid~\bmu,\bSigma,\nu)$, and in this case the cdf of $\X$ is
$$
{F}^{\text{CN}}_p(\bp\mid \bmu,\bSigma,\nu_1,\nu_2)=\nu_1\Phi_p(\bp\mid\bmu,\nu_2^{-1}\bSigma)+(1-\nu_1)\Phi_p(\bp\mid\bmu,\bSigma).
\label{lsdefAB1}
$$

It is known that as {$\nu_1 \to 0$} or {$\nu_2 \to 1$}, $\X \iid \CN_p(\bmu,\bSigma,\nu_1,\nu_2)$ converges in distribution to $\N_p(\bmu,\bSigma)$. 
An important property of the random vector $\X$ is that it can be
written as a scale mixture of the $p$-variate N random vector
coupled with a positive random variable $U$, {i.e.},
\begin{equation}
\label{stoNI1}
\X=\bmu+U^{-1/2}\Z,
\end{equation}
where $\Z \iid \N_p(\mathbf{0},\bSigma)$, and is independent of
$U\iid \nu_1I_{\nu_2}(U)+(1-\nu_1)I_1(U)$, where $I_{\nu_2}(U)$ is an indicator variable with value 1 if $U=\nu_2$ and $0$ otherwise,  and $I_{1}(U)$ is an indicator variable with value 1 if $U=1$ and $0$ otherwise. 
From \eqref{stoNI1}, it follows that $\EE[U\mid\X]=\nu_2\mathcal{ P}_{\nu_2}+\mathcal{ P}_1$ and $\EE[U^2\mid\X]=\nu^2_2\mathcal{ P}_{\nu_2}+\mathcal{ P}_1$, where
\begin{equation}
\label{propor}
    \mathcal{ P}_{\nu_2}=P(U=\nu_2~|~\X)=\frac{\nu_1\phi_p(\xp\mid~\bmu,\nu_2^{-1}\bSigma)}{{f}^{\text{CN}}_p(\xp\mid ~\bmu,\bSigma,\nu_1,\nu_2)}\quad\text{and}\quad \mathcal{ P}_{1}=P(U=1~|~\X)=1- \mathcal{ P}_{\nu_2}.
\end{equation}

%============================================

The following properties of the  CN distribution are useful for our theoretical developments of the SLcn model, including the implementation of the ECM algorithm. We start with the marginal-conditional decomposition of a  $p$-variate CN random vector. 
The proof of the following propositions can be found in
\citet{wang2023multivariate}.

\begin{proposition}\label{prop1}
	Let  $\X\iid\CN_p(\bmu,\bSigma,\nu_1,\nu_2)$ partitioned as
	$\X^{\top}=(\X^{\top}_1,\X^{\top}_2)^{\top}$ with $\mathrm{dim}(\X_1) =
	p_1$, $\mathrm{dim}(\Y_2) = p_2$, where $p_1 + p_2 = p$. Let
	$\bmu=(\bmu^{\top}_1,\bmu^{\top}_2)^{\top}$ and
	$\bSigma=\begin{pmatrix}
			\bSigma_{11} & \bSigma_{12} \\
	\bSigma_{21} & \bSigma_{22}
		\end{pmatrix}$ be the corresponding
	partitions of $\bmu$ and $\bSigma$. 
 Then, we have\\	
	{\rm (i)} $\X_1\iid \CN_{p_1}(\bmu_1,\bSigma_{11},\nu_1,\nu_2)$; and \\

	{\rm (ii)} $\X_2 \mid (\X_1=\xp_1)\iid \CN_{p_2}\left(\bmu_{2.1},{\bSigma}_{22.1},\omega_{\nu_2},\nu_2
		\right),
		$\\
		where $\bmu_{2.1}=\bmu_2+\bSigma_{21}\bSigma^{-1}_{11}(\xp_1-\bmu_1)$, 
		$\bSigma_{22.1}=\bSigma_{22}-\bSigma_{21}\bSigma^{-1}_{11}\bSigma_{12}$, and $\omega_{\nu_2}=\displaystyle\frac{\nu_1\phi_{p_1}(\xp_1\mid~\bmu_1,\nu_2^{-1}\bSigma_{11})}{{f}^{\text{CN}}_{p_1}(\xp_1\mid ~\bmu_1,\bSigma_{11},\nu_1,\nu_2)}.$ 
  %\todo{In my opinion the subscripts of the two densities (numerator and denominator) should be $p_1$: You are correct Antonio VH}
	%\end{itemize}
\end{proposition}

\subsection{The truncated multivariate contaminated normal  distribution}
\label{sec:TMCN}

A $p$-dimensional random vector $\Y$ is said to follow a doubly truncated contaminated normal (TCN) distribution with location vector $\bmu$, scale-covariance matrix $\bSigma$, mixing proportion $\nu_1$, and inflation parameter $\nu_2$, over the truncated region $\mathbb{A}$ defined in \eqref{hyper1}, denoted by $\Y\iid
\TCN_{p}(\bmu,\bSigma,\nu_1,\nu_2;(\ap,\bp))$, if it has the pdf:
$$
f^{\text{TCN}}_p(\yp\mid\bmu,\bSigma,\nu_1,\nu_2;(\ap,\bp))=\frac{f^{\text{CN}}_p(\yp\mid\bmu,\bSigma,\nu_1,\nu_2)}{{F}^{\text{CN}}_p(\ap,\bp\mid ~\bmu,\bSigma,\nu_1,\nu_2)},\quad \yp \in \mathbb{A},
$$
where $\mathbb{A}$ is defined in \eqref{hyper1}. 

The following propositions are related to the marginal and conditional moments of the first two moments of the TCN distributions under a double truncation, which are required to develop the ECM algorithm described later. 
The proofs are similar to those given in  \citet{wang2023multivariate}. 
In what follows, we shall use the notation  $\Y^{(0)}= 1$, $\Y^{(1)}=\Y$, $\Y^{(2)}=\Y\Y^{\top}$, and $\W \iid { \TCN}_{p}(\bmu,\bSigma,\nu_1,\nu_2;(\ap,\bp))$ stands for a $p$-variate doubly truncated contaminated normal distribution on $(\ap,\bp)\subseteq\RR^p$.

\begin{proposition}
\label{prop2}
	If $\Y\iid \TCN_{p}(\bmu,\bSigma,\nu_1,\nu_2;(\ap,\bp))$, then under the notation given in \eqref{propor}, it follows that
	$$
	\EE\left[\mathcal{ P}_{\nu_2}
	\Y^{(k)}\right] =\displaystyle
	\frac{\nu_1\Phi_p(\ap,\bp\mid\bmu,\nu_2^{-1}\bSigma)}
	{{F}^{\text{CN}}_p(\ap,\bp\mid\bmu,\bSigma,\nu_1,\nu_2)}\EE[\bW^{(k)}_{\nu_2}],
 $$

	$$
	\EE\left[(\nu_2\mathcal{ P}_{\nu_2}+\mathcal{ P}_1)
	\Y^{(k)}\right] =\nu_2\EE[\Y^{(k)}]+(1-\nu_1)(1-\nu_2)\displaystyle
	\frac{\Phi_p(\ap,\bp\mid\bmu,\bSigma)}
	{{F}^{\text{CN}}_p(\ap,\bp\mid\bmu,\bSigma,\nu_1,\nu_2)}\EE[\bW^{(k)}_1],
 $$

	where $k \in \{0,1,2\}$, $\bW_{\nu_2}\iid \TN_p(\bmu,\nu_2^{-1}\bSigma;(\ap,\bp))$ and $\bW_1\iid \TN_p(\bmu,\bSigma;(\ap,\bp))$. Here ${\TN}_p(\bmu,\bSigma;(\ap,\bp))$ stands for a $p$-variate doubly truncated normal distribution on $(\ap,\bp)\subseteq\RR^p$.
\end{proposition}
\medskip

\begin{proposition}\label{prop3} Let $\Y\iid
	\TCN_{p}(\bmu,\bSigma,\nu_1,\nu_2;(\ap,\bp))$. Consider the partition $\Y^{\top} = (\Y^{\top}_1 ,
	\Y^{\top}_2 )$ with $\mathrm{dim}(\Y_1) = p_1$, $\mathrm{dim}(\Y_2) = p_2$, $p_1 + p_2
	= p$, and the corresponding partitions of  $\ap$, $\bp$, $\bmu$,
	and $\bSigma$. Then,
	$$
	\EE\left[ \mathcal{ P}_{\nu_2}\Y_2^{(k)}\mid\Y_1\right] =\frac{a_{\nu_2}}
	{a_{\nu_2}+a_1}
	\EE[\bW_1^{(k)}],
	$$

 	$$
	\EE\left[(\mathcal{ P}_{\nu_2}+\mathcal{ P}_1)\Y_2^{(k)}\mid\Y_1\right] =\frac{1}
	{a_{\nu_2}+a_1}(\nu_2a_{\nu_2}
	\EE[\bW_1^{(k)}]+a_1\EE[\bW_2^{(k)}]),
	$$
 with 
	$$ a_{\nu_2}=\omega_{\nu_2}\Phi_{p_2}(\ap^{(2)},\bp^{(2)}|\bmu_{2.1},\nu^{-1}_2\bSigma_{22.1}), \,\, a_{1}=(1-\omega_{\nu_2})\Phi_{p_2}(\ap^{(2)},\bp^{(2)}|\bmu_{2.1},\bSigma_{22.1}),
	$$
	where $\omega_{\nu_2}$, $\bmu_{2.1}$ and $\bSigma_{22.1}$ are defined in Proposition \ref{prop1}. Moreover, $\W_1 \iid \TN_{p_2}(\bmu_{2.1},\nu_2^{-1}\bSigma_{22.1};(\ap_2,\bp_2))$ and $\W_2 \iid \TN_{p_2}(\bmu_{2.1},\bSigma_{22.1};(\ap_2,\bp_2))$.
\end{proposition}

Observe that Propositions \ref{prop2} and \ref{prop3} depend on  formulas for $\mathrm{E}[\bW]$ and $\mathrm{E}[\bW\bW^{\top}]$, where $\bW \iid  \TN_p(\bmu,\bSigma;\mathbb{A})$. 
The general formulas for these moments were derived efficiently by \citet{morales2022moments}, for which we use the \texttt{MomTrunc} package in \prog{R}.

\citet{marchenko2012heckman} established a link between SL models and the families of extended skew-elliptical distributions. 
Next, we investigate the properties of the ESCN distribution since it is related to the continuous part of the SLcn model.

\subsection{The multivariate extended skew contaminated normal distribution}
\label{sec:ESCN}

A $p$-variate random vector $\Y$ is said to follow an ESCN distribution with $p$-variate location vector $\bmu$, $p\times p$ positive definite dispersion matrix $\bSigma$, mixing proportion $\nu_1$, inflation parameter $\nu_2$, $p$-variate skewness parameter vector $\blambda$, and shift parameter $\tau\in\mathbb{R}$, denoted by $\Y\iid \textrm{ESCN}_p(\bmu,\bSigma,\blambda,\nu_1,\nu_2,\tau),$ if its has pdf:
\begin{equation}
\label{denESCN}
{f}^{\text{ESCN}}_p(\yp\mid ~\bmu,\bSigma,\nu_1,\nu_2,\tau)=\frac{\nu_1\phi_p(\y\mid\bmu,\nu_2^{-1}\bSigma)
\Phi(\sqrt{\nu_2}(\tau+\bDelta))+(1-\nu_1)\phi_p(\y\mid\bmu,\bSigma)
\Phi(\tau+\bDelta)}{\nu_1\Phi(\sqrt{\nu_2}\tau/\sqrt{1+\blambda^{\top}\blambda})+(1-\nu_1)\Phi(\tau/\sqrt{1+\blambda^{\top}\blambda})},
\end{equation}
where $\bDelta=\blambda^{\top}\bSigma^{-1/2}(\y-\bmu)$.
 Note that when $\tau=0$, we retrieve the skew contaminated normal (SCN) distribution, denoted by $\Y\iid \textrm{SCN}_p(\bmu,\bSigma,\blambda,\nu_1,\nu_2),$ which has the following pdf \citep{Lachos_Ghosh_Arellano_2009}:
\begin{equation*}
%\label{denESN}
{f}^{\text{SCN}}_p(\yp\mid ~\bmu,\bSigma,\nu_1,\nu_2) = 2\nu_1\phi_p(\y\mid\bmu,\nu_2^{-1}\bSigma)
\Phi(\sqrt{\nu_2}\bDelta)+2(1-\nu_1)\phi_p(\y\mid\bmu,\bSigma)
\Phi(\bDelta).
\end{equation*}
Moreover, when either $\nu_1\rightarrow0$ or $\nu_2\rightarrow1$ in \eqref{denESCN}, we retrieve the extended skew-normal (ESN) distribution, denoted by $\Y\iid \textrm{ESN}_p(\bmu,\bSigma,\blambda,\tau),$  which has the following pdf \citep{morales2022moments}:
\begin{equation}\label{denESN}
{f}^{\text{ESN}}_p(\yp\mid ~\bmu,\bSigma,\tau) = \frac{\phi_p(\y\mid\bmu,\bSigma)
\Phi(\tau+\bDelta)}{\Phi(\tau/\sqrt{1+\blambda^{\top}\blambda})}, 
\end{equation}
with the mean vector given by
\begin{equation}\label{meanESN}
\EE[\Y] = \mu+\eta\bSigma^{1/2}\blambda, 
\end{equation}
where $\eta = {\phi(\tau\mid0,1+\blambda^{\top}\blambda)}/\Phi(\tau/\sqrt{1+\blambda^{\top}\blambda})$. 

From \citet{arellano2010multivariate}, it is straightforward to see that
$$  
{f}^{\text{ESCN}}_p(\yp\mid ~\bmu,\bSigma,\nu_1,\nu_2,\tau) {\longrightarrow}{f}^{\text{CN}}_p(\yp\mid ~\bmu,\bSigma,\nu_1,\nu_2),\quad \text{as $\tau\rightarrow +\infty$}.
$$
Also, from \eqref{denESCN} and \eqref{meanESN}, it is straightforward to show that the mean vector of $\Y\iid \textrm{ESCN}_p(\bmu,\bSigma,\blambda,\nu_1,\nu_2,\tau)$ is
\begin{equation}\label{meanESCN}
\EE[\Y] =\bmu+ \eta_1\bSigma^{1/2}\blambda,
\end{equation}
with $$\eta_1 = \frac{\displaystyle\frac{\nu_1}{\nu_2}\phi(\tau\mid0,\nu_2^{-1}(1+\blambda^{\top}\blambda))+(1-\nu_1)\phi(\tau\mid 0,(1+\blambda^{\top}\blambda))}{\nu_1\Phi(\sqrt{\nu_2}\tau/\sqrt{1+\blambda^{\top}\blambda})+(1-\nu_1)\Phi(\tau/\sqrt{1+\blambda^{\top}\blambda})}.$$

\section{Classical Heckman sample selection model}
\label{SLnModel}

The classical SL model consists of a linear
equation for the outcome, and a Probit equation for the sample selection mechanism for $i \in \{1, \ldots, n\}$. The outcome equation is
\begin{equation}\label{1HS}
Y_{1i}=\x^{\top}_i\bbeta+\epsilon_{1i},
\end{equation}
and the sample selection mechanism is characterized by the following latent linear equation,
\begin{equation}\label{2HS}
Y_{2i}=\w^{\top}_i\bgamma+\epsilon_{2i}.
\end{equation}
The vectors $\bbeta\in {\mathbb R}^p$ and 
$\bgamma \in {\mathbb R}^q$ are unknown regression parameters, while $\x^{\top}_i=(x_{i1},\ldots,x_{ip})$ and  $\w^{\top}_i=(w_{i1},\ldots,w_{iq})$ are known characteristics. 
The covariates in $\x_i$ and $\w_i$ may overlap with each other, and the exclusion restriction holds when at least one of the elements of $\w_i$ is not in $\x_i$. 
\citet{lachosHeckman} consider the approach proposed by \citet{vaida2009fast} and \citet{Matos11} to represent the SLn model belong to the structure of a 
censored linear model. 
Thus, the observed data for the
$i$th subject is given by $(\bV_i, C_i),$ where $\bV_i=(V_{1i},V_{2i})$ represents
the vector of censored readings  and $C_i=I_{\{Y_{2i}>0\}}$ is the censoring indicators. In other words,
% \begin{equation}
% Y_{1i}= V_{1i}, \mbox{ if } C_{i}=1, \; Y_{1i}=V_{2i}=NA, \mbox{ if }
% C_{i}=0,\label{CensL1}
% \end{equation}
%\HJ{
\begin{equation}
Y_{1i}=
\begin{cases}
 V_{1i} & \mbox{if $C_{i}=1$,} \\
 V_{2i} = \texttt{NA} & \mbox{if $ C_{i}=0$,}
\end{cases}
\label{CensL1}
\end{equation}\\
%}
for all $i \in \{ 1, \ldots, n\}$.  
Notice that $V_{2i}=\texttt{NA}$ is equivalent to write $-\infty<V_{2i}< \infty$. 

%The indicator for sample selection is $C_i = I(Y_{2i} > 0)$. Let $V_{1i}$ be the observed outcome, we observe the
%outcome $V_{1i}$ if and only if $C_i > 0$, i.e., $Y_{1i} = {V_{1i}}$ if $C_i = 1$, and $Y_{1i} = NA$ if $C_i = 0$, where $NA$ %indicates missing data.  Thus, the observed data for the
%$i$th subject is given by $(V_i, C_i),$ where $V_i$ represents the censored readings and $C_i$ is the censored indicator.

\citet{heckman1979} assumes a bivariate normal distribution for the error terms (SLn), as follows:
		\begin{eqnarray}\label{nerror}
		\begin{pmatrix}
           \epsilon_{1i}  \\
			\epsilon_{2i}
         \end{pmatrix} \iid {\N}_{2}\left( \bm{0},
		\bSigma
		\right),\ \text{with} \ \bSigma=\begin{pmatrix}
			\sigma^2 & \rho\sigma \\
			\rho\sigma & 1 
		\end{pmatrix} ,
	\end{eqnarray}
%\end{center}
where %and the second diagonal element of $\bSigma$ is fixed at 1 in order to achieve full identifiability.  The
the second diagonal element equals 1  because we only observe the sign of $Y_{2i}$, which is insufficient information to estimate $\btheta=(\bbeta^{\top},\bgamma^{\top},\sigma^2,\rho)^{\top}$, i.e.,  to achieve full identifiability.  
The SLn model \eqref{1HS}--\eqref{nerror} is referred to as the ``Heckman SL model''. 
The absence of the selection effect ($\rho=0$) implies that the outcomes are missing at random, and the observed outcomes are representative of the inference of the population given the observed covariates.

Using Bayes' rule, it is straightforward to show \citep{lachosHeckman} that the conditional pdf of an observed outcome $Y_{1i}={V_{1i}}\mid C_i=1$ is
\begin{equation}\label{condNormal}
f\left(Y_{1i} \mid C_i=1,\x_i,\w_i;\btheta\right)={\phi(V_{1i}\mid \x^{\top}_i\bbeta,\sigma^2)\Phi\left(\displaystyle\frac{\w^{\top}_i\bgamma+\displaystyle\frac{\rho}{\sigma}(V_{1i}-\x^{\top}_i\bbeta)}{\sqrt{1-\rho^2}}\right)} / \Phi(\w^{\top}_i\bgamma),    
\end{equation}
which is related to a density belonging to the ESN family of distributions, as discussed in Section \ref{sec:ESCN}. Thus, 
$$
Y_{1i}={V_{1i}}\mid C_i=1\iid \text{ESN}(\mu=\x^{\top}_i\bbeta,\Sigma=\sigma^2,\lambda=\rho/\sqrt{1-\rho^2},\tau=\w^{\top}_i\bgamma/\sqrt{1-\rho^2}).
$$
From \eqref{meanESN}, we have that $\eta=\sqrt{1-\rho^2}\phi(\w^{\top}_i{\bgamma})/\Phi(\w^{\top}_i{\bgamma})$ and $\Sigma^{1/2}\lambda=\sigma\rho/\sqrt{1-\rho^2}$, hence the mean equation for the observed outcomes  is
\begin{equation}\label{correH}
\EE\left[Y_{1i} \mid C_i=1,\x_i,\w_i;\btheta\right]=\x^{\top}_i\bbeta+\rho\sigma \lambda^N(\w^{\top}_i{\bgamma}),
\end{equation}
where $\lambda^N(a)={\phi(a)}/{\Phi(a)}$ is the inverse Mills ratio. 
The SLn problem can be treated as a model misspecification problem because the mean equation for the outcomes of the selected samples is a linear function $\x^{\top}_i\bbeta$
with a nonlinear correction term $\rho\sigma \lambda^N(\w^{\top}_i{\bgamma})$. 
Based on \eqref{correH}, \citet{heckman1979} proposed a 2-step procedure, which is less efficient than the ML estimation, but it is robust to the deviation of the joint normality of the error terms.  
In the two-step procedure, the standard probit model $P(C_i=c_i|\w_i;\bgamma)=\left[\Phi(\w^{\top}_i\bgamma)\right]^{c_i}\left[1-\Phi(\w^{\top}_i\bgamma)\right]^{1-c_i}$  provides the estimate $\widehat \bgamma$, then the quantity $\lambda^N(\w^{\top}_i{\widehat\bgamma})$ is taken as additional covariate in \eqref{correH}, and the least squares coefficient of $\lambda^N(\w^{\top}_i{\widehat\bgamma})$ gives an estimate of $\rho\sigma$. 
The 2-step procedure is implemented, for instance, in the \prog{R} package \texttt{sampleSelection} \citep{HenningsenCRAN}. 
On the other hand, the ML estimate of $\btheta$  can be computed by maximizing the likelihood function of $\btheta=(\bbeta^{\top},\bgamma^{\top},\sigma^2,\rho)^{\top}$,  given the observed data $(\bV,\C)$, where $\bV=(\bV_1,\ldots,\bV_n)$ and $\bC=(C_1,\ldots,C_n)$, which is expressed as:
\begin{align}
L(\btheta\mid\bV,\bC)&=\prod_{i=1}^{n}\left\{\phi(V_{1i}|\x^{\top}_i\bbeta,\sigma^2)\Phi\left(\displaystyle\frac{\w^{\top}_i\bgamma+\displaystyle\frac{\rho}{\sigma}(V_{1i}-\x^{\top}_i\bbeta)}{\sqrt{1-\rho^2}}\right)\right\}^{C_i}\left\{\Phi(-\w^{\top}_i\bgamma)\right\}^{1-C_i}\label{equ8.1}
\end{align}
or via the ECM algorithm implemented in the the \texttt{R} package \texttt{HeckmanEM} \citep{lachosHeckman}.

\section{The Heckman selection-contaminated normal model} 
\label{SLtModel}

In order to protect the reference SLn model from the occurrence of atypical points and heavier-than-normal error tails, we propose a novel Heckman selection-contaminated normal model, replacing the normal assumption of error terms in \eqref{nerror} by a bivariate CN distribution as follows:
	\begin{equation}
	  \label{modeleqt}
	\begin{pmatrix}
           \epsilon_{1i}  \\
			\epsilon_{2i}
         \end{pmatrix}
          \iid
         {\CN}_{2}\left( \bm{0},
	\bSigma,\nu_1,\nu_2
	\right), \quad i \in \{0, \ldots, n\},  
	\end{equation}
where the dispersion matrix $\bSigma$ is defined in \eqref{nerror} and, as such, depends on an unknown parameter vector $(\sigma,\rho)$.
Based on \eqref{modeleqt} and \eqref{1HS}--\eqref{CensL1}, the proposed model can be equivalently formulated as
%Suppose that we have observations on $n$ independent units
\begin{equation}
%\Y_1, \ldots , \Y_n \ind \mathcal{cN}_2(\bmu_i,\bSigma), \label{modeleq}
\Y_i \mid \x_{i}, \w_{i} \iid \CN_2(\bmu_i,\bSigma,\nu_1,\nu_2), \quad i \in \{ 1, \ldots , n\},\label{modeleqCN}
\end{equation}
where $\Y_i =(Y_{1i},Y_{2i})^{\top}$ is the vector of independent responses for sample unit $i$, and
\begin{equation}
 \bmu_i=\X_{ic}\bbeta_c, \label{eq:dynamic mean}  
\end{equation}
with
$$
\X_{ic}=
\begin{pmatrix}
\x^{\top}_i & 0 \\
0 & \w^{\top}_i \\
\end{pmatrix}\quad\text{and}\quad \bbeta_c=\begin{pmatrix}
\bbeta \\
\bgamma\\
\end{pmatrix}.
$$
%and the dispersion matrix $\bSigma$ depends on an unknown parameter vector $(\sigma,\rho)$. 
The model defined in \eqref{modeleqt}--\eqref{eq:dynamic mean} is hereafter called the SLcn model.

\subsection{The likelihood function}
\label{Likelihood_tMLC2}

To obtain the likelihood function of the SLcn model, first from Proposition \ref{prop1} note that if $C_i=1$, then $Y_{1i}\iid \CN(\x^{\top}_i\bbeta,\sigma^2,\nu_1,\nu_2)$ and $Y_{2i}|Y_{1i}=V_{1i}\iid \CN(\mu_{ti},\sigma^2_{t},\omega_{\nu_{2i}},\nu_2)$, where
\begin{equation}\label{condiy2}
\mu_{ti}=\w^{\top}_i\bgamma+\frac{\rho}{\sigma}(V_{1i}-\x^{\top}_i\bbeta),\quad \sigma^2_{t}=(1-\rho^2),\quad \text{and}\quad \omega_{\nu_{2i}}=\frac{\nu_1\phi(V_{1i}\mid\x^{\top}_i\bbeta,\nu_2^{-1}\sigma^2)}{f^{\text{CN}}(V_{1i}\mid\x^{\top}_i\bbeta,\sigma^2,\nu_1,\nu_2)}.
\end{equation}
Thus, the contribution of the $i$th unit to the likelihood function of $\btheta=(\bbeta^{\top},\bgamma^{\top},\sigma^2,\rho,\nu_1,\nu_2)^{\top}$, given the observed sample $(\bV, \C)$, is
$$f(Y_{1i}|\btheta)\Pr(Y_{2i}>0|Y_{1i}=V_i)={f}^{\text{CN}}(V_{1i}\mid\x^{\top}_i\bbeta,\sigma^2,\nu_1,\nu_2){F}^{\text{CN}}(-\infty,0\mid-\mu_{ti},\sigma^2_{t},\omega_{\nu_{2i}},\nu_2).$$

If $C_i=0$, then the contribution in the likelihood function is
$$\Pr(Y_{2i}\leq 0)={F}^{\text{CN}}(-\infty,0\mid\w^{\top}_i\bgamma,1,\nu_1,\nu_2).$$

Therefore, the likelihood function of $\btheta$, given the observed sample $(\bV, \C)$, is
\begin{equation}
L(\btheta\mid\bV,\bC) = \prod_{i=1}^{n}\left[{f}^{\text{CN}}(V_{1i}\mid\x^{\top}_i\bbeta,\sigma^2,\nu_1,\nu_2){F}^{\text{CN}}(-\infty,0\mid -\mu_{ti},\sigma^2_{t},\omega_{\nu_{2i}},\nu_2)\right]^{C_i}\left[{F}^{\text{CN}}(-\infty,0\mid\w^{\top}_i\bgamma,1,\nu_1,\nu_2)\right]^{1-C_i}.\label{equ8.2}
\end{equation}
%where $\bV=(\bV_1,\ldots,\bV_n)$ and $\bC=(C_1,\ldots,C_n)$.  
The log-likelihood function for the observed data is given by $\ell(\btheta)=\ell(\btheta\mid\bV,\bC)=\ln L(\btheta\mid\bV,\bC)$, that is,
\begin{align}
    \ell(\btheta)&=\sum_{i=1}^{n}\left[C_i\ln {f}^{\text{CN}}(V_{1i}\mid\x^{\top}_i\bbeta,\sigma^2,\nu_1,\nu_2)+C_i\ln {F}^{\text{CN}}(-\infty,0\mid -\mu_{ti},\sigma^2_{t},\omega_{\nu_{2i}},\nu_2)\right]\nonumber\\
    &\quad+\sum_{i=1}^{n}(1-C_i)\ln\left[ {F}^{\text{CN}}(-\infty,0\mid\w^{\top}_i\bgamma,1,\nu_1,\nu_2)\right]. \label{equ8.1}
\end{align}

\begin{figure}[!thb]
    \centering
    \includegraphics[scale = 0.7]{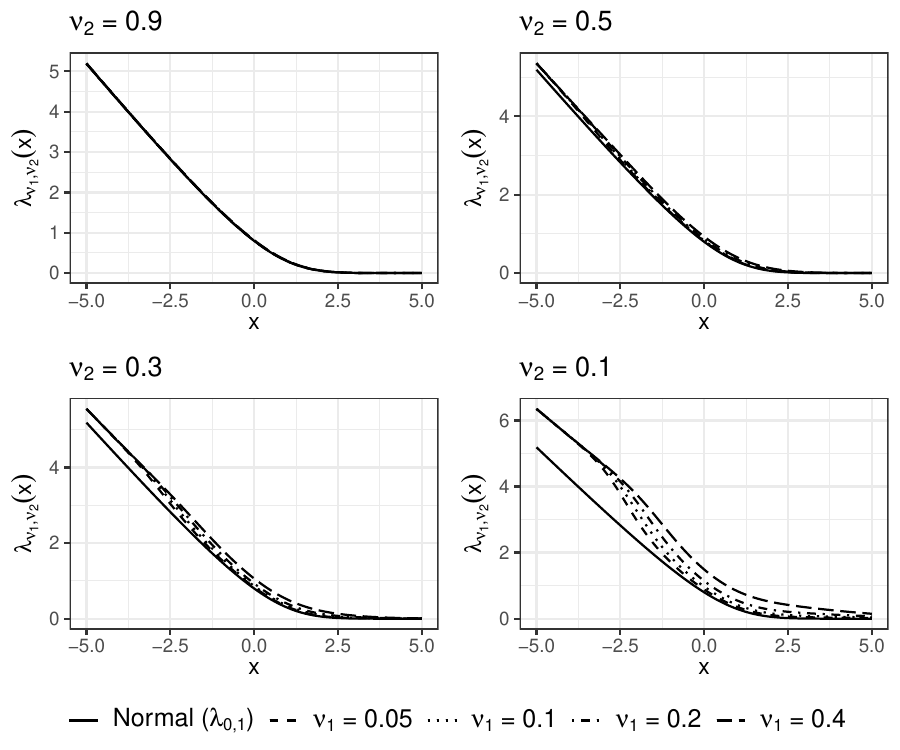}
    \caption{Plot of $\lambda_{\nu_1,\nu_2}(\cdot)$ for different values of $\nu_1$ and $\nu_2$ with $\lambda_{0,1} (\cdot)$ corresponding to the normal case.}
    \label{lambdacn}
\end{figure}

\begin{figure}[!htb]
    \centering
    \includegraphics[scale = 0.7]{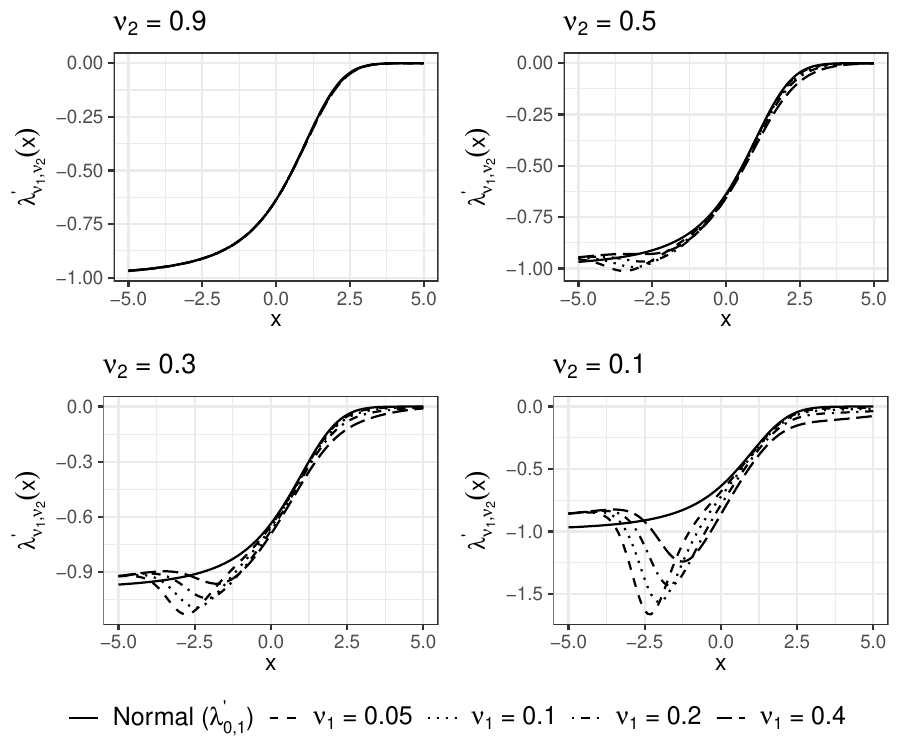}
    \caption{Plot of $\lambda^{\prime}_{\nu_1,\nu_2}(\cdot)$ for different values of $\nu_1$ and $\nu_2$ with $\lambda^{\prime}_{0,1} (\cdot)$ corresponding to the normal case.}
    \label{derlambdacn}
\end{figure}

\subsection{Characterizations of the SLcn model}\label{Car_tMLC2}

In this section, we provide some characterizations of the SLcn model. 
The following proposition shows a link between the continuous part of the SLcn model and the ESCN distribution.

\begin{proposition}
The conditional pdf of an observed outcome $Y_{1i}={V_{1i}}\mid C_i=1$ under the SLcn model is
\begin{equation*}
\label{condCNormal}
f\left(Y_{1i} \mid C_i=1,x_i,\w_i;\btheta\right)
=\displaystyle\frac{\nu_{1}\phi(V_{1i}\mid\x^{\top}_i\bbeta,\nu_2^{-1}\sigma^2)\Phi(\sqrt{\nu_2}\Delta_i)+(1-\nu_{1})\phi(V_{1i}\mid\x^{\top}_i\bbeta,\sigma^2)\Phi(\Delta_i)}{\nu_{1}\Phi(\sqrt{\nu_2}\w^{\top}_i\bgamma)+(1-\nu_{1})\Phi(\w^{\top}_i\bgamma)}.
\end{equation*}
where $\Delta_i=\displaystyle\frac{\w^{\top}_i\bgamma+\displaystyle\frac{\rho}{\sigma}(V_{1i}-\x^{\top}_i\bbeta)}{\sqrt{1-\rho^2}}$. 
That is, 
$$
Y_{1i}={V_{1i}}\mid C_i=1\iid \text{ESCN}(\mu=\x^{\top}_i\bbeta,\Sigma=\sigma^2,\lambda=\rho/\sqrt{1-\rho^2},\nu_1,\nu_2,\tau=\w^{\top}_i\bgamma/\sqrt{1-\rho^2}).
$$
\end{proposition}
\begin{proof}
   See \appendixname~A.
\end{proof}

The conditional expectation of the observed data and the marginal effects of the predictors on $Y_{1i}$ in the observed sample
are often of interest in practice. These expressions are derived in the next proposition.

\begin{proposition}
The mean equation for the observed outcomes under the SLcn model is
\begin{equation}\label{correH1}
\EE\left[Y_{1i}\mid C_i=1,\x_i,\w_i;\btheta\right] = \x_i^{\top}\bbeta + \rho\sigma\lambda^{\text{CN}}_{\nu_1,\nu_2}(\w_i^{\top}\bgamma),
\end{equation}
where 
$$
\lambda^{\text{CN}}_{\nu_1,\nu_2}(x) =\displaystyle\frac{\displaystyle\frac{\nu_1}{\sqrt{\nu_2}}\phi\left(\sqrt{\nu_2} x\right) + (1-\nu_1)\phi\left(x\right)}{{F}^{\text{CN}}(0,+\infty \mid-x,1,\nu_1,\nu_2)}. 
$$ 

Moreover, the marginal effects of the predictors on $Y_{1i}$ in the observed sample are given by 
\begin{equation*}
\frac{\partial}{\partial x_{k}} \EE\left[Y_{1i}|C_i=1,\x_i,\w_i;\btheta\right] =    \beta_k + \rho\sigma\lambda^{\prime \text{CN}}_{\nu_1,\nu_2}(\w_i^{\top}\bgamma),
\end{equation*}
where 
$$
\lambda^{\prime CN}_{\nu_1,\nu_2}(x) = -\displaystyle\frac{f^{\text{CN}}( x |0,1, \nu_1, \nu_2)}{F^{\text{CN}}(x |0,1, \nu_1,\nu_2)}\left[x + \lambda^{\text{CN}}_{\nu_1,\nu_2}(x)\right].
$$
% \todo{
% Note that the symbols are different: $\lambda_{CN}^{\prime}$ and $\lambda^{\prime CN}_{\nu_1,\nu_2}$.
% }
\end{proposition}

\begin{proof}
   The proof of \eqref{correH1} follows from \eqref{meanESCN}. 
   See \appendixname~A for an alternative proof.
\end{proof}

\figurename~\ref{lambdacn} and \ref{derlambdacn} show, respectively, the expected outcome and the marginal effects of the predictors on the response of the selected samples for different values of $\nu_1$ and $\nu_2$. 
We can see from these plots that the conditional expectation will be, in general, underestimated when using the SLn model instead of the SLcn model, this is especially notorious for negative values of the predictor $\w^{\top}\bgamma$ and for small values of $\nu_2$. 
In contrast, the marginal effect of the predictor $x_k$ on the response will be, in general, overestimated by the SLn model. 
This is also especially notorious for negative values of the predictor $\w^{\top}\bgamma$ and for small values of $\nu_2$.

Next, we establish the identifiability of the SLcn model.
\begin{Theorem} The three identifiability conditions given in \citet{wang2016iden}, i.e.,
\begin{description}
\item[\textit{Condition 1.}] 
For any $\delta>0$, $\lim_{z \rightarrow-\infty} F(z) / e^{\delta z}=0 $ or $+\infty$.
\item[\textit{Condition 2.}]
For any $\theta_0 \in(-\infty,+\infty), \theta_1>0, \delta_1, \delta_2 \geq 0$, and $\delta_1+\delta_2 \neq 0$, the limit $\lim _{z \rightarrow+\infty}\left\{F\left(\theta_0+\theta_1 z\right)-\right.$ $F(z)\} / e^{-\delta_1 z^2-\delta_2 z}=0$ or $+\infty$.
\item[\textit{Condition 3.}]
For any $\theta_0 \in(-\infty, \infty), \theta_1>0$, and $M>0$, at most one of $\lim _{z \rightarrow+\infty} z^M\left\{F\left(\theta_0+\theta_1 z\right)-F(z)\right\}$ and $\lim _{z \rightarrow-\infty} z^M\left\{1-F\left(\theta_0+\theta_1 z\right) / F(z)\right\}$ is finite and positive.
\end{description}
hold for the SLcn model defined in \eqref{1HS}--\eqref{CensL1} and \eqref{modeleqt}. 
Hence, the SLcn model is identifiable. 
\end{Theorem}

\begin{proof}
See Appendix A.
\end{proof}

%\varepsilon_{1i}\left(\right\omega_1\phi(\varepsilon_{1i}|\rho\sigma\varepsilon_{2i})
%

\subsection{Parameter estimation via the ECM algorithm}
\label{EM-tlcm}

The ML estimate of the vector of unknown parameters $\btheta=(\bbeta^{\top},\bgamma^{\top},\sigma^2,\rho,\nu_1,\nu_2)^{\top}$ can be calculated by maximizing the log-likelihood given in \eqref{equ8.1}. 
Many optimization procedures are available in standard programs, such as the \texttt{optim} routine in \prog{R}, which need only the original estimating function. 
A disadvantage of direct maximization of the log-likelihood function is that it may not converge unless	good starting values are used. 
Thus, we propose the ECM algorithm for parameter estimation that accounts for censored and missing values together with all hidden variables as different sources of incomplete information. 
An interesting property of the EM algorithm is that it can be used to automatically discover the expected values of the censored and missing values to proceed with further inferences. 
Moreover, the EM estimates are quite insensitive to the starting values, as discussed by \citet{lachosHeckman} regarding the SLn and SLt models.

Although we could directly consider the binary random variable $U$ in \eqref{stoNI1} as a source of incompleteness for the implementation of the ECM algorithm, in the fashion of \citet{punzo2021modeling} and \citet{punzo2020high}, and for the sake of interpretation of the results, it is convenient to consider the Bernoulli random variable 
$$
\varepsilon=\frac{U-1}{\nu_2-1} = 
\begin{cases}
      1 & \text{with probability $\nu_1$}\\
      0 & \text{with probability $1-\nu_1$,}\\
 	\end{cases}
$$
which is a linear transformation of $U$.

So, advantageously, we have an indicator of membership to the inflated normal component ($\varepsilon_i=1$) or not ($\varepsilon_i=0$) for each unit $i$, $i \in \left\{1,\ldots,n\right\}$.

This gives rise to the alternative following hierarchical representation
\begin{align}
\varepsilon_i & \iid \mbox{Bernoulli}\left(\nu_1\right), \nonumber \\
\Y_i \mid \x_i,\w_i,\varepsilon_i & \ind \begin{cases} 
	       {\N}_{2}(\bmu_i,\nu_2^{-1}\bSigma)  &   \text{if $\varepsilon_i=1$}\\
				 {\N}_{2}(\bmu_i,\bSigma) &   \text{if $\varepsilon_i=0$,}
				 \end{cases}	
\label{eqn tCMcomplete}
\end{align}
where $\bmu_i$ is defined as in \eqref{eq:dynamic mean}.

Let $\yp=(\yp^{\top}_1,\ldots,\yp^{\top}_n)^{\top}$,
$\bV=(\bV_1,\ldots,\bV_n)$, $\bC=(C_1,\ldots,C_n)$, $\bvarepsilon=(\varepsilon_1,\ldots, \varepsilon_n)$, and assume that we observe $(\bV_i,C_i)$ for the $i$th subject.  
In the estimation procedure, $\y$ and $\bu$ are
treated as hypothetical missing data, and augmented with the
observed data set we have
$\yp_c=(\bC,\bV,\yp,\bvarepsilon)^{\top}$. 
Hence, the ECM algorithm is applied to the complete data log-likelihood function given by
$$
\ell_c(\btheta|\yp_c)=\sumas \ell_{ic}(\btheta),
$$
where
\begin{equation*} 
\ell_{ic}(\btheta)=-\frac{1}{2} \ln|\bSigma|+\bigl\{\varepsilon_i\ln (\nu_1)+(1-\varepsilon_i)\ln (1-\nu_1)\bigr\} +
\ln (\nu_2) \varepsilon_i-\frac{1}{2}[\nu_2\varepsilon_i+(1-\varepsilon_i)](\yp_i-\bmu_i)^{\top}\bSigma^{-1}(\yp_i-\bmu_i)+c,
\end{equation*}
where $c$ is a constant that does not depend on $\btheta$. 
The $(k+1)${th} iteration of the ECM algorithm for the SLcn model can be summarized in the following two steps.
%\bigskip

\paragraph{E-step:}

Given the current estimate  $\btheta=\widehat{\btheta}^{(k)}$,
%at the $k$th step of the algorithm, 
the E-step provides the conditional expectation of the complete data
log-likelihood function
\begin{equation*}
Q(\btheta\mid\widehat{\btheta}^{(k)})=\mathrm{E} \left[ \ell_c(\btheta|\yp_c)\mid\bV,\bC,\widehat{\btheta}^{(k)} \right]
=\sumas{Q_i(\btheta\mid\widehat{\btheta}^{(k)})},\label{eq:Em:Q}
\end{equation*}
where
\begin{align*}
Q_i(\btheta\mid\widehat{\btheta}^{(k)})=
Q_{i}(\bbeta^{\top},\bgamma^{\top},\sigma^2,\rho,\nu_1,\nu_2\mid\widehat{\btheta}^{(k)}) = &
-\frac{1}{2}\ln{|\bSigma|}+\bigl\{\widehat{\varepsilon}^{(k)}_i\ln (\nu_1)+(1-\widehat{\varepsilon}^{(k)}_i)\ln (1-\nu_1)\bigr\}+\ln (\nu_2) \widehat{\varepsilon}^{(k)}_i\\
&-\frac{1}{2}\tr\left[\widehat{\bE}^{(k)}_{1i}(\bmu)\bSigma^{-1}\right]-\frac{(\nu_2-1)}{2}\tr\left[\widehat{\bE}^{(k)}_{2i}(\bmu)\bSigma^{-1}\right], 
\end{align*}
where 
\begin{equation*}
   \widehat{\bE}^{(k)}_{1i}(\bmu)=\widehat{\yp_i^{2}}^{(k)}-\widehat{\yp}^{(k)}_i{\bmu_i}^{\top} -{\bmu_i} (\widehat{\yp}^{(k)}_i)^{\top}+{\bmu_i}{\bmu_i}^{\top}\\
\end{equation*}
and
\begin{equation*}
\widehat{\bE}^{(k)}_{2i}(\bmu)=\widehat{\varepsilon\yp_i^{2}}^{(k)}-\widehat{\varepsilon\yp}^{(k)}_i{\bmu_i}^{\top} -{\bmu_i} (\widehat{\varepsilon\yp}^{(k)}_i)^{\top}+\widehat{\varepsilon}^{(k)}_i{\bmu_i}{\bmu_i}^{\top},
\end{equation*}
with 
\begin{align*}
   \widehat{\yp}^{(k)}_i&=\mathrm{E}[\displaystyle
\Y_i\mid\bV_i,\bC_i,\widehat{\btheta}^{(k)}] \\
\widehat{\yp_i^{2}}^{(k)}&=\mathrm{E}[\displaystyle \Y_i\Y_i^{\top}\mid\bV_i,\bC_i,\widehat{\btheta}^{(k)}] \\
\widehat{\varepsilon\yp}^{(k)}_i&=\mathrm{E}[\displaystyle
\varepsilon_i\Y_i\mid\bV_i,\bC_i,\widehat{\btheta}^{(k)}] \\
\widehat{\varepsilon\yp_i^{2}}^{(k)}&=\mathrm{E}[\displaystyle \varepsilon_i\Y_i\Y_i^{\top}\mid\bV_i,\bC_i,\widehat{\btheta}^{(k)}]
\end{align*}
and 
\begin{equation}
  \widehat{\varepsilon}^{(k)}_i=\mathrm{E}[\displaystyle \varepsilon_i\mid\bV_i,\bC_i,\widehat{\btheta}^{(k)}]=P(\varepsilon_i=1\mid\bV_i,\bC_i,\widehat{\btheta}^{(k)}),
  \label{eq:posterior to be bad}
\end{equation}

with $\widehat{\varepsilon}^{(k)}_i$ being the posterior probability for the $i$th unit to belong to the inflated normal component of the multivariate CN distribution, $i \in \left\{1,\ldots,n\right\}$; see Section~\ref{sec:Automatic mild outlier detection} for details.

As in \citet{lachosHeckman}, we use the parameter transformations $\psi = \sigma^2(1 -\rho^2)$ and $\rho^* = \rho\sigma$ to get closed form expression in the CM-Step.

%\bigskip
\paragraph{CM-step:}

In this step, $Q(\btheta\mid\widehat{\btheta}^{(k)})$ is conditionally maximized with respect to $\bbeta_c,\sigma^2,\rho,\nu_1,\nu_2$ and new estimates
$\widehat{\bbeta}^{(k+1)}_c,\widehat{\sigma}^{2(k+1)},\widehat{\rho}^{(k+1)},\widehat{\nu}^{(k+1)}_1,\widehat{\nu}^{(k+1)}_2$ are obtained. Specifically, we have that
\begin{align}
\widehat{\bbeta}^{(k+1)}_c &= \left\{\sumas \left[ 1 + \left(\widehat\nu^{(k)}_2 - 1 \right) \widehat{\varepsilon}^{(k)}_i\right]\X^{\top}_{ic}\widehat\bSigma^{(k)-1}\X_{ic}\right\}^{-1}\sumas
\X^{\top}_{ic}\widehat\bSigma^{(k)-1}\left[\widehat{\yp}^{(k)}_i+\left(\widehat\nu^{(k)}_2-1\right)\widehat{\varepsilon\yp}^{(k)}_i\right],\label{eq:beta_tMLC0}\\
\widehat{\bmu}^{(k+1)}_i &= \X_{ic}\widehat{\bbeta}^{(k+1)}_c,\\
\widehat{\psi}^{(k+1)} &= \frac{1}{n}\sumas\left\{\widehat\Gamma^{(k)}_{11i}-\widehat \rho^{*(k)}(\widehat \Gamma^{(k)}_{12i}+\widehat\Gamma^{(k)}_{21i})+\widehat \rho^{*2(k)} \widehat\Gamma^{(k)}_{22i}\right\},\\
\widehat{\rho^*}^{(k+1)} &= \frac{\sumas \left( \widehat \Gamma^{(k)}_{12i}+\widehat\Gamma^{(k)}_{21i}\right)}{2\sumas \widehat\Gamma^{(k)}_{22i}}, \label{eq:beta_tMLCn2} \\
\widehat{\nu}^{(k+1)}_1 &= \frac{1}{n}\sumas \widehat{\varepsilon}^{(k)}_i,\label{eq:estimated proportion of bad points}\\
\widehat{\nu}^{(k+1)}_2&=\min\left\{\frac{2\sumas\widehat{\varepsilon}^{(k)}_i}{\sumas \tr(\widehat\bSigma^{(k)-1}\widehat{\bE}^{(k)}_{2i}(\bmu) )},1\right\},\\
\widehat{\sigma}^{2(k+1)}&=\widehat{\psi}^{(k+1)}+\widehat{\rho^*}^{2(k+1)},\\
%\; %\quad %\qquad\text{and}\qquad\qquad
	\widehat\rho^{(k+1)} &= \displaystyle\frac{\widehat{\rho}^{*(k+1)}}{\widehat{\sigma}^{(k+1)}},\label{eq:rho_tMLC}
\end{align}
where $\widehat\Gamma^{(k)}_{kli}$ is the element of position $(k,l)$ in the matrix  $\widehat\bGamma^{(k)}_{i}=\widehat{\bE}^{(k)}_{1i}(\bmu)+(\widehat{\nu}^{(k)}_2-1)\widehat{\bE}^{(k)}_{2i}(\bmu).$ 

The algorithm is terminated when the relative distance between two successive evaluations of the log-likelihood defined in \eqref{equ8.1} is less than a tolerance, i.e., $|\ell(\widehat{\btheta}^{(k+1)}\mid\bV,\bC)/\ell(\widehat{\btheta}^{(k)}\mid\bV,\bC)-1|<\epsilon$, for example, $\epsilon=10^ {-6}$.  
The initial values of the parameters $\widehat\btheta^{(0)}$ for the EM algorithm are obtained from the 2-step procedure, which is implemented through the \prog{R} package \texttt{sampleSelection} \citep{HenningsenCRAN}, setting $\widehat\nu^{(0)}_1=\widehat\nu^{(0)}_2=0.5$ or in a grid of possible pairs $(\widehat\nu^{(0)}_1,\widehat\nu^{(0)}_2)$ and then pick up the model in correspondence to the maximum value of the log-likelihood.

It is important to stress that, from Eqs.~\eqref{eq:beta_tMLC0}--\eqref{eq:rho_tMLC}, the E-step reduces to the computation of $\widehat{\yp}^{(k)}_i$, $\widehat{\yp_i^{2}}^{(k)}$,  $\widehat{\varepsilon\yp}^{(k)}_i$, $\widehat{\varepsilon\yp_i^{2}}^{(k)}$ and $\widehat{\varepsilon}^{(k)}_i$. 
To compute these expected values, first observe that they can be written in terms of $\mathrm{E}[\varepsilon_i \mid \Y_i]$, where $\Y_i\iid \CN_{p}(\bmu_i,\bSigma,\nu_1,\nu_2)$. 
For example, we have that $\widehat{\varepsilon}_i= \mathrm{E}\left[ \mathrm{E}[\varepsilon_i\mid \Y_i]\mid \bV_i,\bC_i,\widehat{\btheta}^{(k)}\right] $. 
It is straightforward to prove that $\mathrm{E}[\varepsilon_i \mid \Y_i=\yp_i]=P(\varepsilon_i =1\mid \Y_i=\yp_i)=P(U_i =\nu_2\mid \Y_i=\yp_i)=\displaystyle\frac{\nu_1\phi_2(\yp_i~|~\bmu_i,\nu_2^{-1}\bSigma)}{f^{\text{CN}}_2(\yp_i~|~\bmu_i,\bSigma, \nu_1,\nu_2)}={\cal P}_{\nu_2}$. 
Then, we can use Propositions \ref{prop2} and \ref{prop3} to obtain closed-form expressions as follows: \\ 

%\begin{itemize}
\noindent{\rm 1.} If $C_i=0$, then from Proposition~\ref{prop2} we have
	\begin{align*}
 \widehat{\yp}^{(k)}_i &= \mathrm{E}[\displaystyle \Y_i\mid\bV_i,C_i,\widehat{\btheta}^{(k)}]=\widehat{\bw}_i^{c(k)},\\
 \widehat{\yp_i^{2}}^{(k)} &= \mathrm{E}[\displaystyle \Y_i\Y^{\top}_i\mid\bV_i,C_i,\widehat{\btheta}^{(k)}]=\widehat{\bw}_i^{2 ^{c(k)}},\\
 \widehat{\varepsilon}^{(k)}_i &= \mathrm{E}[\displaystyle		\varepsilon_i\mid\bV_i,C_i,\widehat{\btheta}^{(k)}]=\widehat{\varphi}^{(k)}(\bV_i),\\
		\widehat{\varepsilon\yp}^{(k)}_i 
  &=
  \mathrm{E}[\displaystyle \varepsilon_i\Y_i\mid\bV_i,C_i,\widehat{\btheta}^{(k)}]=\widehat{\varphi}^{(k)}(\bV_i)\widehat{\bw}_{1i}^{c(k)},\\
  \widehat{\varepsilon\yp_i^{2}}^{(k)} 
  &= \mathrm{E}[\displaystyle \varepsilon_i\Y_i\Y_i^{\top}\mid\bV_i,C_i,\widehat{\btheta}^{(k)}]=\widehat{\varphi}^{(k)}(\bV_i)\widehat{\bw}_{1i}^{2 ^{c(k)}},\;  
	\end{align*}
	where 
	\begin{align*}
	\widehat{\varphi}^{(k)}(\bV_i)&=\widehat{\nu}^{(k)}_1\frac{\Phi_{2}((-\infty,-\infty),(\infty,0)~|~\widehat{\bmu}_i^{(k)},\widehat{\bSigma}^{*(k)})}{{F}^{\text{CN}}_2((-\infty,-\infty),(\infty,0)\mid ~\widehat{\bmu}_i^{(k)},\widehat{\bSigma}^{(k)},\widehat{\nu}^{(k)}_1,\widehat{\nu}^{(k)}_2)},\quad\\
	 \widehat{\bw}_i^{c(k)}&=\mathrm{E}[\bW_i\mid\widehat{\btheta}^{(k)}], \quad \widehat{\bw}_i^{2 ^{c(k)}}=\mathrm{E}[\bW_i \bW_i^\top\mid\widehat{\btheta}^{(k)}],\\
 \quad \widehat{\bw}_{1i}^{c(k)}&=\mathrm{E}[\bW_{1i}\mid\widehat{\btheta}^{(k)}], \quad \widehat{\bw}_{1i}^{2 ^{c(k)}}=\mathrm{E}[\bW_{1i} \bW_{1i}^\top\mid\widehat{\btheta}^{(k)}],
\end{align*}
 with	
$$	
\bW_i\iid
	\TCN_{2}(\widehat{\bmu}_i^{(k)},\widehat{\bSigma}^{(k)},\widehat\nu^{(k)}_1,\widehat\nu^{(k)}_2;\mathbb{A}),\quad \bW_{1i}\iid
	\TN_{2}(\widehat{\bmu}_i^{(k)},\widehat{\bSigma}^{*(k)};\mathbb{A}),\quad
	\widehat{\bSigma}^{*(k)}=\displaystyle\frac{1}{\widehat\nu^{(k)}_2}\widehat{\bSigma}^{(k)}
	$$
and	
\begin{equation} \label{hyper2}
\mathbb{A} = \left\{(x_1, x_2)\in \mathbb{R}^2: -\infty \leq x_1 \leq \infty, -\infty \leq x_2 \leq 0\right\}.
\end{equation}
% Here $\TN_p(\bmu,\bSigma, \mathbb{ A})$ denotes the $p$-variate truncated normal distribution with location $\bmu$, scale matrix $\bSigma$ over the truncation region $\mathbb{A}$.  
To  compute  $\mathrm{E}[\bW_i]$ and
	$\mathrm{E}[\bW_i\bW_i^{\top}]$ we use  the \prog{R} package \texttt{MomTrunc}  \citep{GalarzaCran}. \\
	
\noindent {\rm 2.} If $C_i=1$,
	then from Proposition~\ref{prop3} we have that
	\begin{align*}
	\widehat{\yp}^{(k)}_i&=\mathrm{E}[\displaystyle \Y_i\mid Y_{1i}, \bV_i,C_i,\widehat{\btheta}^{(k)}]=(V_{1i},\widehat{{w}}^{c(k)}_i)^{\top},\\
 \widehat{\yp_i^{2}}^{(k)}&=\mathrm{E}[\displaystyle \Y_i\Y^{\top}_i \mid Y_{1i},\bV_i,C_i,\widehat{\btheta}^{(k)}]=\left(\begin{array}{cc}
	V^2_{1i}  & \widehat{{w}}^{c(k)\top}_iV_{1i} \\
	\widehat{{w}}^{c(k)}_iV_{1i}  &\widehat{w}_i^{2 ^{c(k)}}\\
	\end{array}\right ),\\
 \widehat{\varepsilon}^{(k)}_i &= \mathrm{E}[\displaystyle
		\varepsilon_i \mid Y_{1i},\bV_i,C_i,\widehat{\btheta}^{(k)}]=\widehat{\omega}_{\nu_{2i}}^{(k)}\frac{\Phi(0,\infty~|~\widehat{\mu}^{(k)}_{ti},\widehat \nu^{(k)-1}_2 \widehat \sigma^{2(k)}_{t})}{F^{\text{CN}}(0,\infty~|~\widehat{\mu}^{(k)}_{ti},\sigma^{2(k)}_{t},\widehat{\omega}_{\nu_{2i}}^{(k)},\nu^{(k)}_2)},\\
	\widehat{\varepsilon\yp_i^{2}}^{(k)}&=\mathrm{E}[\displaystyle
	\varepsilon_i\Y_i\Y_i^{\top}\mid Y_{1i},\bV_i,C_i,\widehat{\btheta}^{(k)}]=\left(\begin{array}{cc}
	V^2_{1i} \widehat{\varepsilon}^{(k)}_i & \widehat{\varepsilon}^{(k)}_i \widehat{{w}}^{c(k)\top}_{1i}V_{1i} \\
	\widehat{\varepsilon}^{(k)}_i \widehat{{w}}^{c(k)}_iV_{1i}  &\widehat{\varepsilon}^{(k)}_i  \widehat{w}_{1i}^{2 ^{c(k)}}\\
	\end{array}\right ),\\
	\widehat{\varepsilon\yp}^{(k)}_i&=\mathrm{E}[\displaystyle \varepsilon_i\Y_i\mid Y_{1i},\bV_i,\bC_i,\widehat{\btheta}^{(k)}]=(V_{1i}\widehat{\varepsilon}^{(k)}_i ,\widehat{\varepsilon}^{(k)}_i \widehat{{w}}^{c(k)}_{1i})^{\top} ,
	\end{align*}
	where 
	$$
 {\widehat \mu}^{(k)}_{ti}=\w^{\top}_i{{\widehat \bgamma}^{(k)}}+\frac{{\widehat \rho}^{(k)}}{{\widehat \sigma}^{(k)}}(V_{1i}-\x^{\top}_i{\widehat\bbeta}^{(k)}),
 \quad \widehat \sigma^{2(k)}_{t}=(1-\widehat\rho^{2(k)}),
 \quad \widehat{\omega}_{\nu_{2i}}^{(k)}=\frac{\nu_1\phi(V_{1i}|\x^{\top}_i{\widehat\bbeta}^{(k)},\widehat \nu^{(k)-1}_2 \widehat \sigma^{2(k)}))}{f^{\text{CN}}(V_{1i}|\x^{\top}_i{\widehat\bbeta}^{(k)},\widehat \sigma^{2(k)},\widehat\nu_1^{(k)},\widehat\nu_2^{(k)})},
 $$ 
 $$
 \widehat{{w}}^{c(k)}_i=\mathrm{E}[{W}_i\mid\widehat{\btheta}^{(k)}],\quad\widehat{w}_i^{2 ^{c(k)}}=\mathrm{E}[{W}^2_i\mid\widehat{\btheta}^{(k)}],\quad \widehat{{w}}^{c(k)}_{1i}=\mathrm{E}[{W}_{1i}\mid\widehat{\btheta}^{(k)}],\quad\text{and}\quad
	\widehat{w}_{1i}^{2 ^{c(k)}}=\mathrm{E}[{W}^2_{1i}\mid\widehat{\btheta}^{(k)}],
 $$
	 with  
	${W}_i\iid
	\TCN(\widehat{\mu}^{(k)}_{ti},\widehat \sigma^{2(k)}_{t},\widehat{\omega}_{\nu_{2i}}^{(k)},\widehat\nu^{(k)}_2;[0,\infty))$ and
 ${W}_{1i}\iid
	\TN(\widehat{\mu}^{(k)}_{ti},\widehat \nu^{(k)-1}_2 \widehat \sigma^{2(k)}_{t});[0,\infty))$.
%\end{itemize}

\subsection{Standard errors}  
\label{sec SE1}

In this section, we describe how to obtain the standard errors of the ML estimates for the SLcn model. 
We follow the information-based method exploited by \citet{basford1997standard} to compute the asymptotic covariance of the ML estimates. 
The empirical information matrix, according to \citet{meilijson1989fast}'s formula, is defined as
\begin{eqnarray}\label{eq:IM}
	\bI_e(\btheta|\yp) = \sum_{i = 1}^{n} s(\yp_i|\btheta)s^\top (\yp_i|\btheta) - \frac{1}{n} S(\yp|\btheta)S^\top (\yp|\btheta),
\end{eqnarray}
where {\small$S(\yp|\btheta) = \sum_{i = 1}^{N} s(\yp_i|\btheta)$} and {\small$s(\yp_i|\btheta)$} is the empirical score function for the {\small$i$}th subject. It is noted from the result of \citet{louis1982finding} that the individual score can be determined as 
\begin{eqnarray}
	s(\yp_i|\btheta) = \E\left[\left. \frac{\partial\ell_{ic}(\btheta)}{\partial \btheta}\right| \bV_i, C_i,\btheta\right]\nonumber.
\end{eqnarray}
Using the ML estimates {$\widehat{\btheta}$ in $s(\yp_i|\btheta)$, leads to $S(\yp|\widehat{\btheta}) = 0$, so from \eqref{eq:IM} we have that
\begin{eqnarray}\label{eq:oim}
	\bI_e(\widehat{\btheta}|\yp) = \sum_{i = 1}^{n} \widehat{\se}_i \widehat{\se}^\top_i,\nonumber
	\end{eqnarray}
where $\widehat{\se}_i$ is an individual score vector given by $\widehat{\se}_i = (\widehat{s}_{i,\beta_{c}}, \widehat{s}_{i,{\sigma}},\widehat{s}_{i,{\rho}},\widehat{s}_{i,{\nu_1}},\widehat{s}_{i,{\nu_2}})$}. So, the expressions for the elements of {$\widehat{\se}_i$} are given by:
\begin{align*}
	\widehat{s}_{i,\beta_c} =& \frac{1}{2}\X^{\top}_{ic}\bSigma^{-1}{\widehat{\yp_i}}+\frac{1}{2}{\widehat{\yp_i}}^{\top}\bSigma^{-1}\X_{ic}-\X^{\top}_{ic}\bSigma^{-1}\X_{ic}\bbeta_c\\
 &+(\nu_2-1)\left[\frac{1}{2}\X^{\top}_{ic}\bSigma^{-1}{\widehat{\varepsilon\yp_i}}+\frac{1}{2}{\widehat{\varepsilon\yp_i}}^{\top}\bSigma^{-1}\X_{ic}-\widehat{\varepsilon}_i\X^{\top}_{ic}\bSigma^{-1}\X_{ic}\bbeta_c\right],\\
	\widehat{s}_{i,\sigma} =&  -\frac{1}{2}\tr\left(\bSigma^{-1}\B\right)+\frac{1}{2}\tr\left(\bGamma_i\bSigma^{-1}\B \bSigma^{-1}\right),
		\,\, \\
		\widehat{s}_{i,\rho} =&  -\frac{1}{2}\tr\left(\bSigma^{-1}\D\right)+\frac{1}{2}\tr\left(\bGamma_i\bSigma^{-1}\D \bSigma^{-1}\right),  \\
       \widehat{s}_{i,\nu_1} =& \frac{\widehat{\varepsilon}_i}{\nu_1}+ \frac{1-\widehat{\varepsilon}_i}{1-\nu_1},\\
       \widehat{s}_{i,\nu_2} =&\frac{\widehat{\varepsilon}_i}{\nu_2}-\frac{1}{2}\tr\left[\widehat{\bE}_{2i}(\bmu)\bSigma^{-1}\right] , 
\end{align*}
where $\bGamma_i$ is defined as in \eqref{eq:beta_tMLC0}--\eqref{eq:rho_tMLC},
$\B=\begin{pmatrix}
2\sigma & \rho \\
\rho & 0 \\
\end{pmatrix}$, and $\D=\begin{pmatrix}
0 & \sigma \\
\sigma & 0 \\
\end{pmatrix}$.

\subsection{Residual analysis}
\label{sec:Residual analysis}

In order to check departures from the CN error assumption, we use the normalized quantile residuals introduced by \citet{dunn1996randomized}, which is defined by
$$r_{NQ_i}=\Phi^{-1}\left\{F_{\widehat{\btheta}}(y_i)\right\},$$
for $i \in \{ 1, \ldots, n\}$, where $\Phi^{-1}(\cdot)$ is the inverse of the cdf of the standard normal distribution and $F_{\widehat \btheta}(y_i)$ is  the cdf of the response variable evaluated in the ML estimates $\widehat\btheta$, which is defined by
\begin{align*}
F_{\btheta}(y_i)&=I(Y_{2i} > 0)F_{Y_{1i}|Y_{2i}>0}(y_i)\Pr(Y_{2i}>0)+\left[1-I(Y_{2i} > 0)\right]\Pr(Y_{2i}\leq 0),\\
&=C_iF_{Y_{1i}|Y_{2i}>0}(y_i)\Pr(Y_{2i}>0)+(1-C_i)\Pr(Y_{2i}\leq 0),\quad i \in \left\{1,\ldots,n\right\},
\end{align*}
where $F_{Y_{1i}|Y_{2i}>0}(y)$ is the conditional cumulative distribution of $Y_{1i}$ given $Y_{2i}>0$ and $\Y_i=(Y_{i1},Y_{i2})^{\top}\iid \CN_2(\bmu_i,\bSigma,\nu_1,\nu_2)$ or $ \N_2(\bmu_i,\bSigma)$ for the SLcn or SLn model, respectively. 
A recent application of the normalized quantile residuals in the context of Heckman SL models can be found in \citet{bastos2020birnbaum} and \citet{lachosHeckman}. 

\subsection{Automatic inlier/outlier detection}
%\subsection{\orange{Automatic inlier/outlier detection}}
\label{sec:Automatic mild outlier detection}

Let $\widehat{\varepsilon}_i$ denote the value of $\widehat{\varepsilon}_i^{(k)}$ in \eqref{eq:posterior to be bad} at convergence of the ECM algorithm, $i \in \left\{1,\ldots,n\right\}$.
If $\widehat{\nu}_1 \leq 0.5$, 
%As already mentioned in Section~\ref{EM-tlcm}, and usually under the additional constraint that $\nu_1>0.5$ (refer to Section~\ref{subsubsec:Identifiability}), 
then $\widehat{\varepsilon}_i$ denotes the posterior probability for unit $i$ of being a mild outlier; otherwise, $1-\widehat{\varepsilon}_i$ is the probability to be an inlier.
%Instead, if 
Interestingly, $\widehat{\varepsilon}_i$ can be computed for all the units in the sample, no matter if $Y_{1i}$ is observed ($C_i=1$) or not ($C_i=0$).
Although such a probability is informative on its own, the user could be simply interested in classifying the observation as good or bad; in other more common words in the robust literature, the user may be interested in detecting inliers/outliers (see also \citealp{falkenhagen2019likelihood}, \citealp{mazza2020mixtures} and \citealp{Punz:McNi:Robu:2017}).
In such a case, it is natural to classify the $i$th unit as an outlier (inlier) if $\widehat{\nu}_1 \leq 0.5$ and $\widehat{\varepsilon}_i > 0.5$ ($\widehat{\nu}_1 > 0.5$ and $\widehat{\varepsilon}_i < 0.5$), $i \in \left\{1,\ldots,n\right\}$.
Thus, the approach reveals richer information about the role of that observation. 
Note also that, the resulting classification can be used to possibly eliminate the inliers/outliers if such an outcome is desired \citep{Berk:Bent:Esti:1988}.

 This gives further value to the SLcn model compared to other elliptical error extensions of the SLn model, such as the SLt, where no automatic/objective inlier/outlier detection rule can be defined.

\section{Simulation study} 
\label{sec:Simulation study}

In this section, we conduct various simulation studies involving three bivariate distributions: normal, contaminated normal, and slash distribution. 
Firstly, we explore the finite sample properties of the EM estimates at three levels of sample size: 250, 500, and 1000, and compare the robustness of parameter estimation for the SLcn model to that of the SLn model when data are generated from the normal distribution. 
The other studies focus on the vulnerability of the SLn model when data are generated from the contaminated normal and slash distributions, i.e., considering heavy-tailed data, and we include the SLt model for reference. 
Secondly, we consider missing variations at three levels of censoring: 10\%, 25\%, and 50\%, and compare the variability of parameter estimates for the SL models. 
For each scenario, 1000 Monte Carlo (MC) samples are generated. 
In order to enhance the reliability of the model selection, we used well-known criteria: the Akaike information criterion (AIC; \citealp{Akai:Anew:1974}) and the Bayesian information criterion (BIC; \citealp{Schw:Esti:1978}). 
We investigate whether the lowest AIC or BIC among models could identify the most suitable model for the simulated data.

\begin{table}[!ht]
\caption{Simulation study 1. Mean estimates (EM), mean standard errors (SE) and Monte Carlo standard error (MC SE) of the $1000$ Monte Carlo replicates for the data generated from the normal distribution. The mean (and SE) AIC and BIC are reported for the SLn and SLcn models.}
	\label{tab:sim_normal}
\centering
\begin{tabular}{cccccccc}
  \toprule
 & & \multicolumn{3}{c}{SLn} & \multicolumn{3}{c}{SLcn} \\
  %\cmidrule{3-5} \cmidrule{6-8}    
 Sample Size & TRUE & EM & SE & MC SE & EM & SE & MC SE \\ 
 \midrule
  %\midrule
250 & $\beta_0=1.000$ & 1.011 & 0.134 & 0.138 & 1.016 & 0.131 & 0.138 \\ 
  & $\beta_1=0.500$ & 0.506 & 0.130 & 0.125 & 0.507 & 0.129 & 0.124 \\ 
  & $\gamma_0=0.674$ & 0.682 & 0.094 & 0.094 & 0.778 & 0.159 & 0.219 \\ 
  & $\gamma_1=0.300$ & 0.305 & 0.147 & 0.148 & 0.345 & 0.182 & 0.186 \\ 
  & $\gamma_2=-0.500$ & -0.507 & 0.099 & 0.097 & -0.585 & 0.150 & 0.185 \\ 
  & $\sigma^2=1.000$ & 1.002 & 0.082 & 0.082 & 0.836 & 0.125 & 0.273 \\ 
  & $\rho=0.600$ & 0.563 & 0.222 & 0.237 & 0.558 & 0.225 & 0.234 \\
  & $\nu\,\, (\nu_1)$ &  &  &  & 0.564 & 0.032 & 0.149   \\
  & $\nu_2$ &  &  &  & 0.757 & 0.316 & 0.289 \\
  & AIC & \multicolumn{3}{c}{737.058 (22.910)} & \multicolumn{3}{c}{736.433 (22.936)}\\ 
  & BIC & \multicolumn{3}{c}{740.579 (22.910)} & \multicolumn{3}{c}{739.955 (22.936)}\\[1.5mm]
500 & $\beta_0=1.000$ & 1.004 & 0.091 & 0.090 & 1.007 & 0.091 & 0.089 \\  
  & $\beta_1=0.500$ & 0.504 & 0.091 & 0.088 & 0.503 & 0.091 & 0.087 \\  
  & $\gamma_0=0.674$ & 0.678 & 0.065 & 0.065 & 0.740 & 0.106 & 0.126 \\ 
  & $\gamma_1=0.300$ & 0.303 & 0.107 & 0.107 & 0.329 & 0.126 & 0.128 \\  
  & $\gamma_2=-0.500$ & -0.505 & 0.070 & 0.069 & -0.556 & 0.101 & 0.113 \\ 
  & $\sigma^2=1.000$ & 1.001 & 0.057 & 0.058 & 0.864 & 0.101 & 0.206 \\ 
  & $\rho=0.600$ & 0.588 & 0.153 & 0.154 & 0.584 & 0.157 & 0.149 \\ 
  & $\nu\,\, (\nu_1)$ &  &  &  & 0.564 & 0.022 & 0.104 \\  
  & $\nu_2$ &  &  &  & 0.795 & 0.321 & 0.232 \\  
    & AIC & \multicolumn{3}{c}{1486.283 (32.583)} & \multicolumn{3}{c}{1485.792 (32.610)}\\ 
  & BIC & \multicolumn{3}{c}{1490.498 (32.583)} & \multicolumn{3}{c}{1490.007 (32.610)}\\[1.5mm]
1000   & $\beta_0=1.000$ & 1.001 & 0.063 & 0.064 & 1.002 & 0.063 & 0.062 \\ 
  & $\beta_1=0.500$ & 0.501 & 0.066 & 0.065 & 0.500 & 0.066 & 0.063 \\ 
  & $\gamma_0=0.674$ & 0.678 & 0.046 & 0.044 & 0.727 & 0.084 & 0.079 \\  
  & $\gamma_1=0.300$ & 0.301 & 0.078 & 0.077 & 0.322 & 0.092 & 0.088 \\  
  & $\gamma_2=-0.500$ & -0.502 & 0.048 & 0.049 & -0.540 & 0.075 & 0.070 \\ 
  & $\sigma^2=1.000$ & 1.001 & 0.041 & 0.042 & 0.881 & 0.092 & 0.155 \\ 
  & $\rho=0.600$& 0.595 & 0.106 & 0.110 & 0.596 & 0.108 & 0.107 \\ 
  & $\nu\,\, (\nu_1)$ &  &  &  & 0.562 & 0.016 & 0.072 \\   
  & $\nu_2$ &  &  &  & 0.814 & 0.339 & 0.184 \\  
    & AIC & \multicolumn{3}{c}{2977.061 (45.110)} & \multicolumn{3}{c}{2976.600 (45.143)}\\ 
  & BIC & \multicolumn{3}{c}{2981.969 (45.110)} & \multicolumn{3}{c}{2981.508 (45.143)}\\
   \bottomrule
\end{tabular}
\end{table}

\subsection{Finite sample properties}
\label{sec:s1}

The primary objective of this simulation is to illustrate how well the proposed ECM algorithm can successfully retrieve the parameters in SL models. 
Additionally, we examine the robustness of the SL models generating data not only from the contaminated normal distribution but also from the slash distribution \citep{rogers1972understanding}, often used as a challenging distribution for a statistical procedure. 
Also, the configuration of this simulation is revisited by \citet{lachosHeckman}. 
\begin{table}[!ht]
\caption{Simulation study 1. 
Mean estimates (EM), mean standard errors (SE), and Monte Carlo standard error (MC SE) of the $1000$ Monte Carlo replicates for the data generated from the contaminated normal with $\nu_1=0.1$ and $\nu_2=0.1$. 
The mean (and SE) AIC and BIC are reported for the SL models.}
	\label{tab:sim_cn}
\centering
\begin{tabular}{ccccccccccc}
  \toprule
 & & \multicolumn{3}{c}{SLn} & \multicolumn{3}{c}{SLt} & \multicolumn{3}{c}{SLcn} \\
  %\cmidrule{3-5} \cmidrule{6-8}    
 Sample Size & TRUE & EM & SE & MC SE & EM & SE & MC SE & EM & SE & MC SE \\ 
  \midrule
250 & $\beta_0=1.000$ & 0.961 & 0.164 & 0.210 & 1.017 & 0.117 & 0.135 & 1.016 & 0.125 & 0.132 \\ 
  & $\beta_1=0.500$ & 0.517 & 0.177 & 0.175 & 0.493 & 0.137 & 0.141 & 0.494 & 0.137 & 0.139 \\ 
  & $\gamma_0=0.786$ & 0.735 & 0.096 & 0.101 & 0.862 & 0.120 & 0.128 & 0.870 & 0.124 & 0.169 \\ 
  & $\gamma_1=0.300$ & 0.260 & 0.148 & 0.151 & 0.314 & 0.174 & 0.181 & 0.314 & 0.175 & 0.182 \\ 
  & $\gamma_2=-0.500$ & -0.423 & 0.093 & 0.109 & -0.537 & 0.119 & 0.127 & -0.539 & 0.123 & 0.153 \\ 
  & $\sigma^2=1.000$ & 1.359 & 0.068 & 0.183 & 0.973 & 0.083 & 0.098 & 0.954 & 0.091 & 0.222 \\ 
  & $\rho=0.600$ & 0.611 & 0.161 & 0.342 & 0.562 & 0.196 & 0.235 & 0.563 & 0.214 & 0.229 \\ 
  & $\nu\,\, (\nu_1=0.1)$ &  &  &  & 6.3 &  & & 0.150 & 0.030 & 0.126 \\   
  & $\nu_2=0.1$ &  &  &  &  &  & & 0.131 & 0.072 & 0.106 \\ 
  & AIC & \multicolumn{3}{c}{845.484 (40.215)} & \multicolumn{3}{c}{818.429 (30.961)}& \multicolumn{3}{c}{815.303 (30.555)}\\ 
  & BIC & \multicolumn{3}{c}{849.006 (40.215)} & \multicolumn{3}{c}{821.950 (30.961)}& \multicolumn{3}{c}{818.825 (30.555)}\\[1.5mm]
500 & $\beta_0=1.000$ & 0.927 & 0.105 & 0.129 & 1.007 & 0.080 & 0.088 & 1.006 & 0.086 & 0.086 \\ 
  & $\beta_1=0.500$ & 0.537 & 0.122 & 0.120 & 0.498 & 0.096 & 0.095 & 0.500 & 0.097 & 0.094 \\ 
  & $\gamma_0=0.786$ & 0.727 & 0.066 & 0.069 & 0.860 & 0.084 & 0.091 & 0.840 & 0.081 & 0.093 \\ 
  & $\gamma_1=0.300$ & 0.259 & 0.107 & 0.111 & 0.322 & 0.130 & 0.135 & 0.309 & 0.125 & 0.131 \\ 
  & $\gamma_2=-0.500$ & -0.409 & 0.064 & 0.075 & -0.537 & 0.085 & 0.086 & -0.517 & 0.082 & 0.085 \\  
  & $\sigma^2=1.000$ & 1.372 & 0.043 & 0.129 & 0.964 & 0.058 & 0.062 & 0.979 & 0.061 & 0.139 \\ 
  & $\rho=0.600$ & 0.685 & 0.082 & 0.217 & 0.584 & 0.136 & 0.152 & 0.586 & 0.147 & 0.150 \\
  & $\nu\,\, (\nu_1=0.1)$ &  &  &  & 4.759 &  & & 0.118 & 0.021 & 0.062 \\
  & $\nu_2=0.1$ &  &  &  &  &  & & 0.111 & 0.038 & 0.051 \\ 
  & AIC & \multicolumn{3}{c}{1711.999 (53.884)} & \multicolumn{3}{c}{1652.616 (40.394)} & \multicolumn{3}{c}{1647.220 (39.854)}\\ 
  & BIC & \multicolumn{3}{c}{1716.213 (53.884)} & \multicolumn{3}{c}{1656.831 (40.394)} & \multicolumn{3}{c}{1651.435 (39.854)}\\[1.5mm]
1000   & $\beta_0=1.000$ & 0.917 & 0.070 & 0.080 & 1.007 & 0.056 & 0.063 & 1.004 & 0.060 & 0.062 \\ 
  & $\beta_1=0.500$ & 0.531 & 0.087 & 0.089 & 0.497 & 0.070 & 0.073 & 0.499 & 0.070 & 0.071 \\ 
  & $\gamma_0=0.786$ & 0.728 & 0.047 & 0.045 & 0.858 & 0.059 & 0.061 & 0.830 & 0.055 & 0.059 \\
  & $\gamma_1=0.300$ & 0.254 & 0.078 & 0.078 & 0.317 & 0.094 & 0.096 & 0.302 & 0.090 & 0.091 \\ 
  & $\gamma_2=-0.500$ & -0.407 & 0.044 & 0.052 & -0.532 & 0.058 & 0.058 & -0.507 & 0.055 & 0.056 \\  
  & $\sigma^2=1.000$ & 1.370 & 0.029 & 0.096 & 0.962 & 0.041 & 0.047 & 0.993 & 0.043 & 0.099 \\ 
  & $\rho=0.600$ & 0.710 & 0.045 & 0.135 & 0.587 & 0.096 & 0.111 & 0.592 & 0.102 & 0.108 \\
  & $\nu\,\, (\nu_1=0.1)$ &  &  &  &4.110 &  & &0.106 & 0.015 & 0.031 \\ 
  & $\nu_2=0.1$ &  &  &  & &  & &0.105 & 0.025 & 0.029 \\ 
  & AIC & \multicolumn{3}{c}{3428.475 (79.020)} & \multicolumn{3}{c}{3308.133 (57.611)} & \multicolumn{3}{c}{3298.501 (57.173)}\\ 
  & BIC & \multicolumn{3}{c}{3433.383 (79.020)} & \multicolumn{3}{c}{3313.041 (57.611)} & \multicolumn{3}{c}{3303.409 (57.173)}\\
   \bottomrule
\end{tabular}
\end{table}

\begin{figure}[!ht]
	\begin{center}
		\includegraphics[scale=0.3]{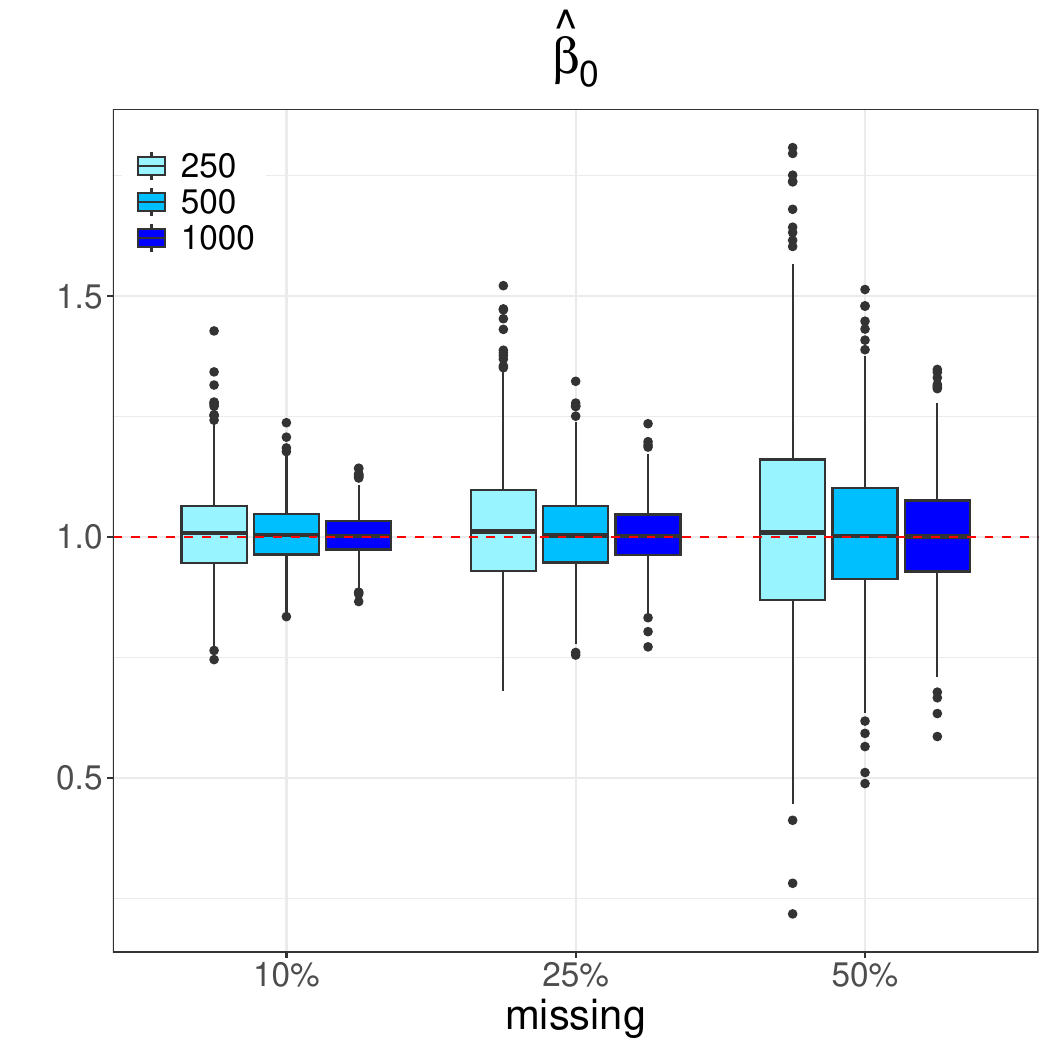}
  	\includegraphics[scale=0.3]{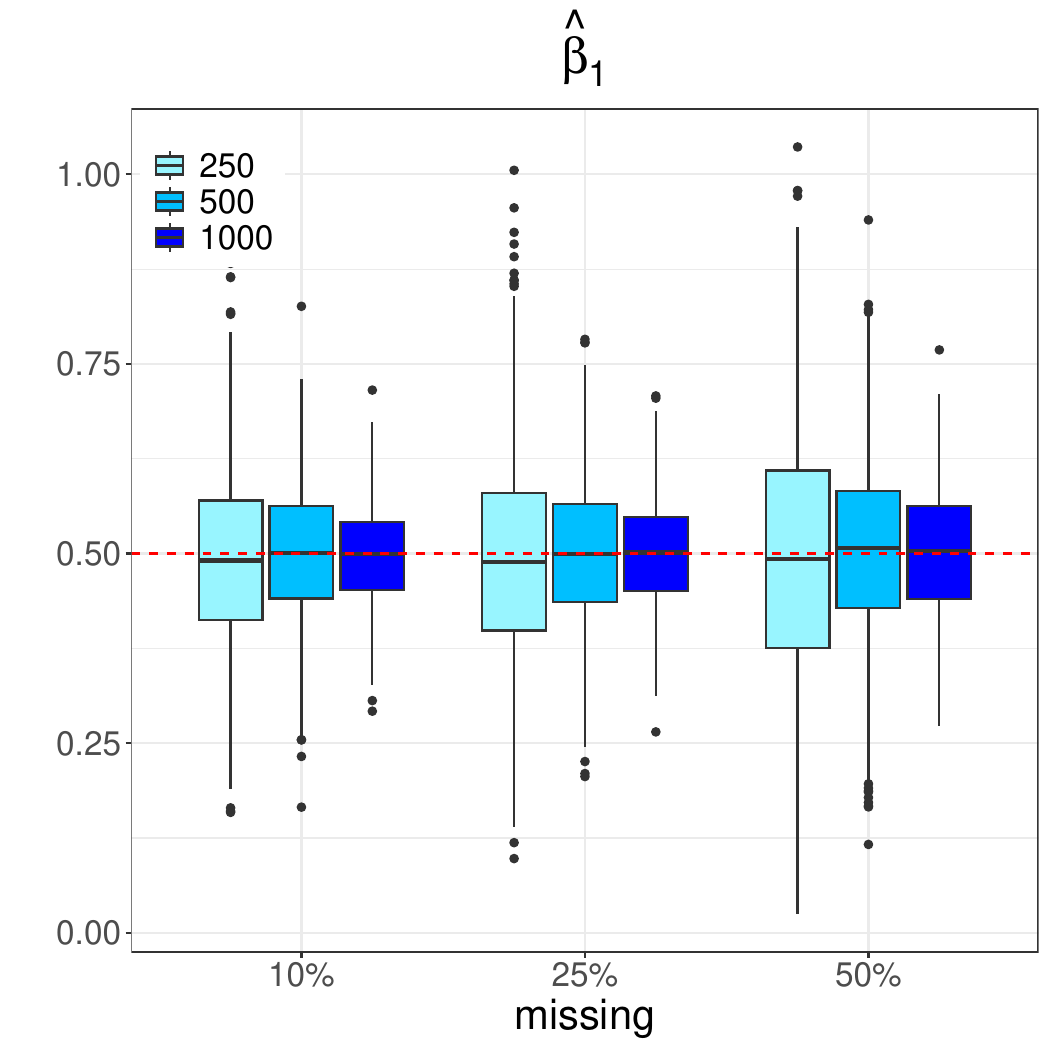}
   	\includegraphics[scale=0.3]{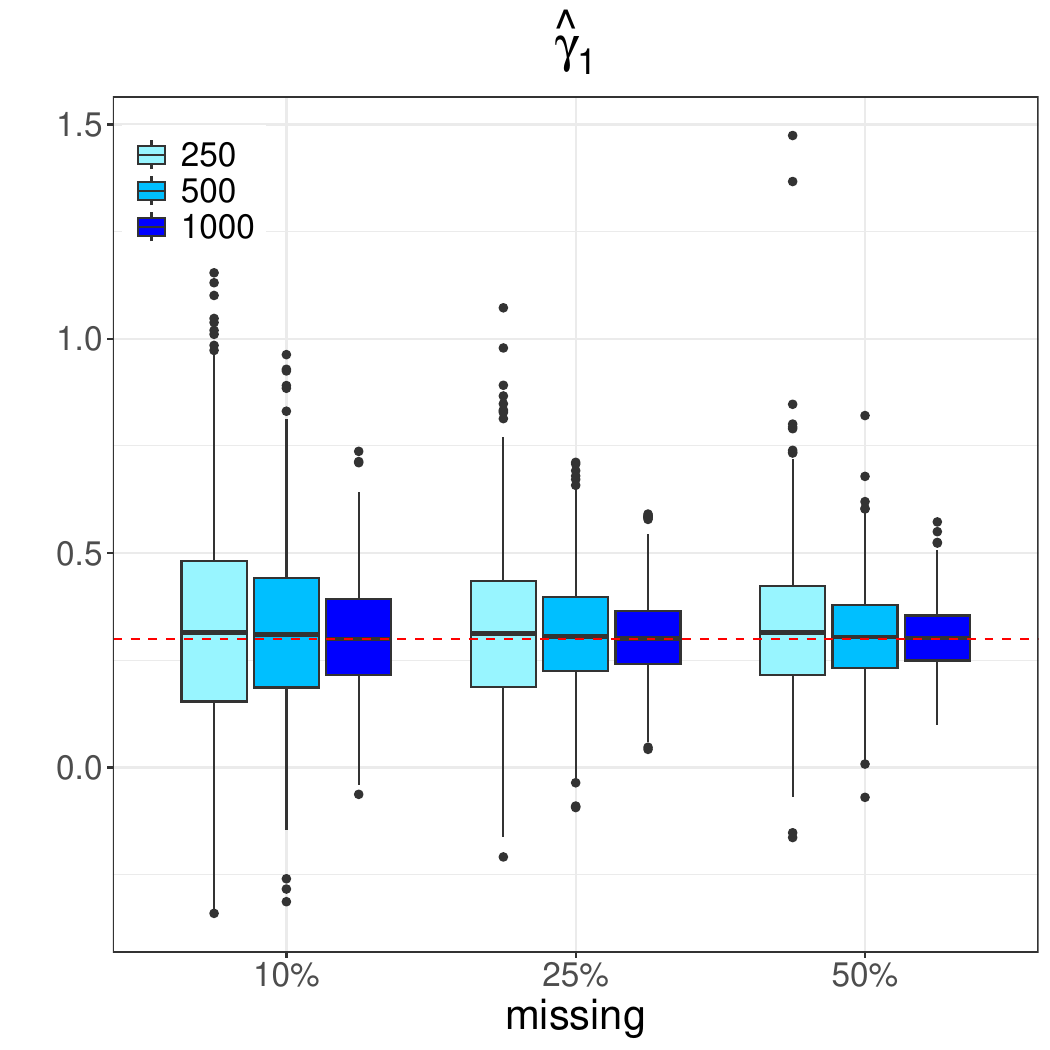}
		\includegraphics[scale=0.3]{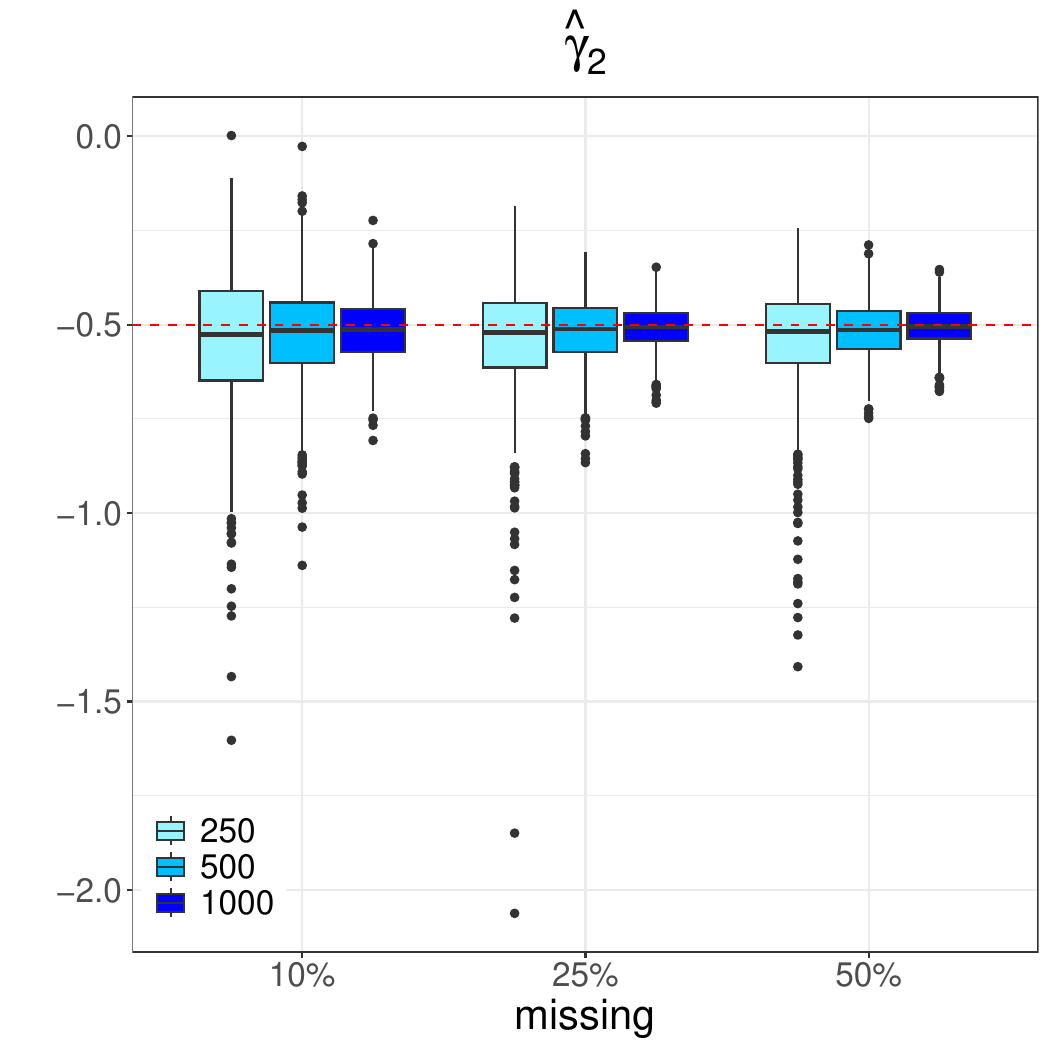}
  	\includegraphics[scale=0.3]{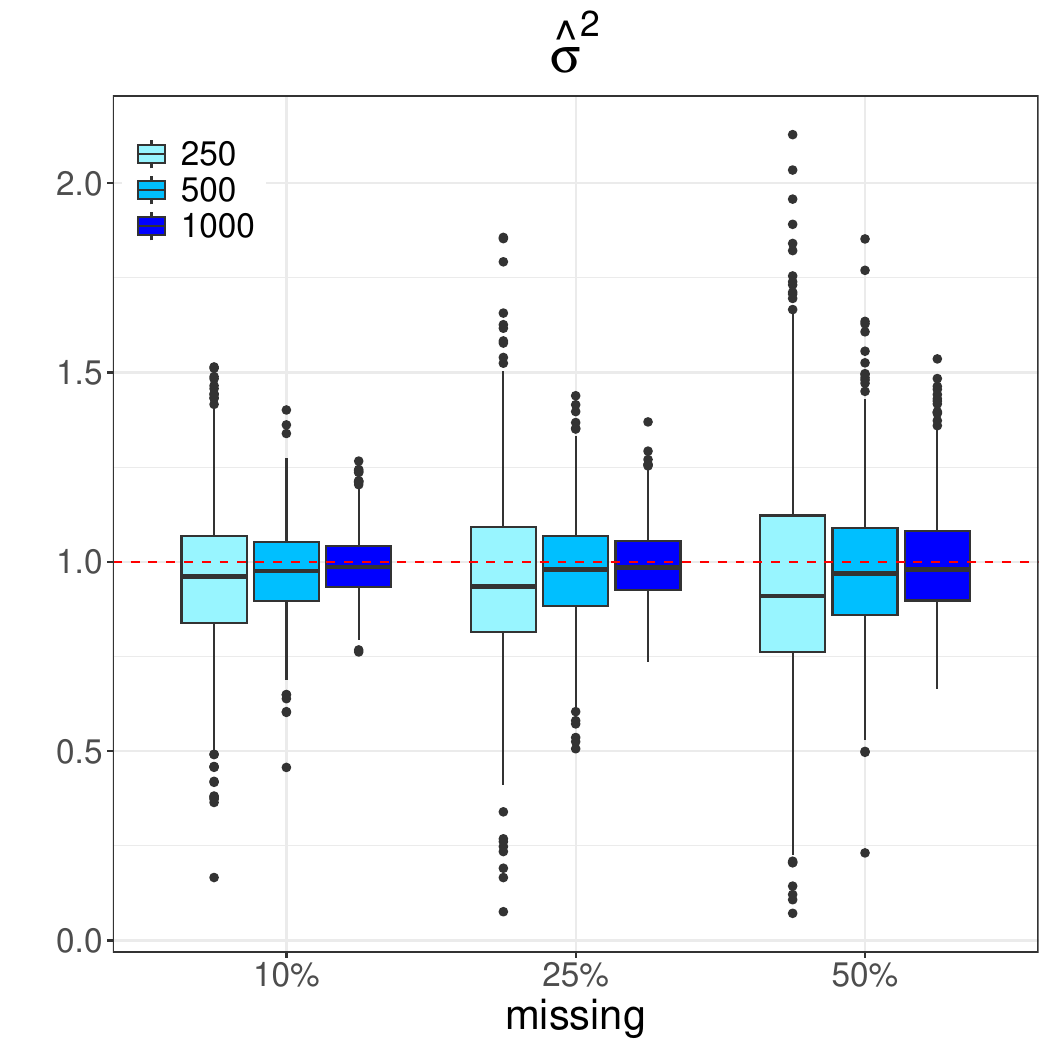}
        \includegraphics[scale=0.3]{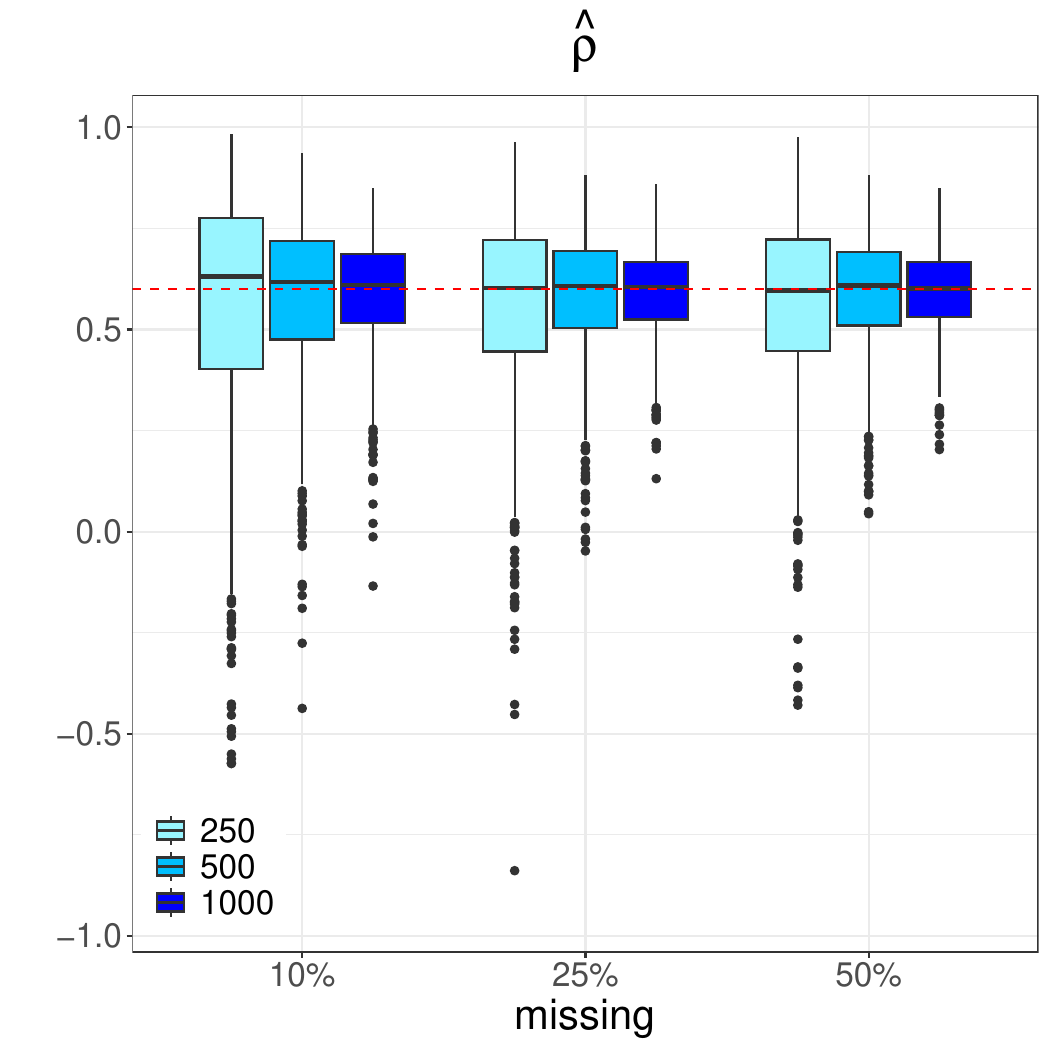}
        \caption{Boxplot of the 1000 Monte Carlo estimates of $\beta_0$, $\beta_1$, $\gamma_1$, $\gamma_2$, $\sigma$ and $\rho$ for the SLcn model, when data are generated from the contaminated normal distribution, with $n=250$, $500$, and $1000$ as sample sizes, when varying the missing proportion as $10\%$, $25\%$ and $50\%$.}
		\label{Comp1}
	\end{center}
\end{figure}

In this simulation study, data are generated from the normal, the contaminated normal, and slash distributions, respectively, as the sample sizes progressively increase: $n=250$, $500$, and $1000$ on a fixed missing rate of 25\%. 
The elements of $\w_i^\top = (1, w_{i1}, w_{i2})$ are produced from a Uniform$(-1,1)$ and $N(0,1)$, respectively. 
To consider the exclusion restriction we let $\x_i^\top = (1, w_{i1})$, $\ii$. 
Across all scenarios, we set $\bgamma =(\gamma_{0d},0.3,-0.5)$ and $\bbeta = (1,0.5)$. The value of $\gamma_{0d}$ is determined as the $-25\%$ quantile times $\sigma^2$ of the generating distribution to ensure an average missing rate of $25\%$.
Finally, the scale parameters are specified as $\sigma^2 = 1$ and $\rho = 0.6$.
We calculate the mean values (EM) and standard deviations (SE) across the estimates from the 1000 MC samples. MC SE is the average value of the approximate standard error obtained through the information-based method defined in \eqref{eq:IM}.

\begin{table}[!htb]
\caption{Simulation study 1. Mean estimates (EM), mean standard errors (SE) and Monte Carlo standard error (MC SE) of the $1000$ Monte Carlo replicates for the data generated from the slash with $1.43$ degrees of freedom. The mean (and SE) AIC and BIC are reported for the SL models.}
	\label{tab:sim_sl}
\centering
\begin{tabular}{ccccccccccc}
  \toprule
 & & \multicolumn{3}{c}{SLn} & \multicolumn{3}{c}{SLt} & \multicolumn{3}{c}{SLcn} \\
 % \cmidrule{3-5} \cmidrule{6-8}    
 Sample Size & TRUE & EM & SE & MC SE & EM & SE & MC SE& EM & SE & MC SE \\ 
  \midrule
250 & $\beta_0=1.000$ & 0.931 & 0.249 & 0.332 & 1.017 & 0.173 & 0.189 & 1.008 & 0.185 & 0.190 \\ 
  & $\beta_1=0.500$ & 0.532 & 0.239 & 0.223 & 0.506 & 0.179 & 0.181 & 0.510 & 0.181 & 0.182 \\ 
  & $\gamma_0=0.884$ & 0.619 & 0.091 & 0.096 & 0.712 & 0.106 & 0.113 & 0.794 & 0.141 & 0.260 \\ 
  & $\gamma_1=0.300$ & 0.208 & 0.143 & 0.146 & 0.248 & 0.163 & 0.168 & 0.274 & 0.185 & 0.206 \\ 
  & $\gamma_2=-0.500$ & -0.328 & 0.087 & 0.100 & -0.409 & 0.106 & 0.112 & -0.455 & 0.130 & 0.192 \\ 
  & $\sigma^2=1.000$ & 1.768 & 0.096 & 0.364 & 1.263 & 0.114 & 0.147 & 1.499 & 0.137 & 0.570 \\ 
  & $\rho=0.600$ & 0.601 & 0.195 & 0.369 & 0.562 & 0.209 & 0.243 & 0.572 & 0.223 & 0.249 \\  
  & $\nu= 1.43,\, (\nu_1)$ &  &  &  & 9.240 &  & & 0.237 & 0.029 & 0.236 \\ 
  & $\nu_2$ &  &  &  &  &  &  &0.151 & 0.080 & 0.123 \\ 
  & AIC & \multicolumn{3}{c}{930.044 (53.450)} & \multicolumn{3}{c}{901.802 (31.419)} & \multicolumn{3}{c}{899.870 (31.419)}\\ 
  & BIC & \multicolumn{3}{c}{933.566 (53.450)} & \multicolumn{3}{c}{905.324 (31.358)} & \multicolumn{3}{c}{903.392 (31.358)}\\[1.5mm]
500 & $\beta_0=1.000$ & 0.877 & 0.153 & 0.245 & 1.013 & 0.117 & 0.128 & 0.995 & 0.130 & 0.135 \\  
  & $\beta_1=0.500$ & 0.553 & 0.166 & 0.166 & 0.502 & 0.125 & 0.125 & 0.509 & 0.129 & 0.126 \\ 
  & $\gamma_0=0.884$ & 0.617 & 0.063 & 0.071 & 0.712 & 0.074 & 0.079 & 0.717 & 0.076 & 0.104 \\ 
  & $\gamma_1=0.300$ & 0.207 & 0.104 & 0.104 & 0.246 & 0.120 & 0.123 & 0.244 & 0.120 & 0.125 \\ 
  & $\gamma_2=-0.500$ & -0.316 & 0.060 & 0.073 & -0.407 & 0.075 & 0.078 & -0.400 & 0.076 & 0.089 \\  
  & $\sigma^2=1.000$ & 1.780 & 0.058 & 0.335 & 1.245 & 0.080 & 0.096 & 1.658 & 0.091 & 0.408 \\ 
  & $\rho=0.600$ & 0.680 & 0.099 & 0.269 & 0.581 & 0.148 & 0.164 & 0.602 & 0.154 & 0.173 \\ 
  & $\nu= 1.43,\, (\nu_1)$ &  &  &  & 5.023 &  & & 0.148 & 0.020 & 0.139 \\  
  & $\nu_2$ &  &  &  &  &  &  &0.132 & 0.044 & 0.089 \\ 
  & AIC & \multicolumn{3}{c}{1882.901 (89.272)} & \multicolumn{3}{c}{1820.695 (43.149)} & \multicolumn{3}{c}{1819.332 (43.520)}\\ 
  & BIC & \multicolumn{3}{c}{1887.116 (89.272)} & \multicolumn{3}{c}{1824.910 (43.149)} & \multicolumn{3}{c}{1823.546 (43.520)}\\[1.5mm]
1000   & $\beta_0=1.000$ & 0.844 & 0.098 & 0.177 & 1.012 & 0.081 & 0.087 & 0.988 & 0.090 & 0.094 \\  
  & $\beta_1=0.500$ & 0.546 & 0.119 & 0.125 & 0.498 & 0.090 & 0.091 & 0.507 & 0.093 & 0.092 \\ 
  & $\gamma_0=0.884$ & 0.614 & 0.045 & 0.049 & 0.714 & 0.052 & 0.053 & 0.696 & 0.050 & 0.060 \\ 
  & $\gamma_1=0.300$ & 0.201 & 0.075 & 0.075 & 0.244 & 0.088 & 0.089 & 0.232 & 0.084 & 0.084 \\ 
  & $\gamma_2=-0.500$ & -0.311 & 0.041 & 0.055 & -0.409 & 0.052 & 0.054 & -0.387 & 0.049 & 0.056 \\ 
  & $\sigma^2=1.000$ & 1.815 & 0.035 & 0.244 & 1.235 & 0.057 & 0.067 & 1.736 & 0.062 & 0.293 \\ 
  & $\rho=0.600$ & 0.721 & 0.047 & 0.194 & 0.584 & 0.103 & 0.110 & 0.614 & 0.104 & 0.121 \\  
  & $\nu= 1.43,\, (\nu_1)$ &  &  &  & 4.407 &  & & 0.102 & 0.014 & 0.071 \\ 
  & $\nu_2$ &  &  &  &  &  &  & 0.107 & 0.022 & 0.059 \\ 
  & AIC & \multicolumn{3}{c}{3790.217 (134.756)} & \multicolumn{3}{c}{3649.566 (59.903)} & \multicolumn{3}{c}{3650.540 (61.262)}\\ 
  & BIC & \multicolumn{3}{c}{3795.125 (134.756)} & \multicolumn{3}{c}{3654.474 (59.903)} & \multicolumn{3}{c}{3655.448 (61.262)}\\
   \bottomrule
\end{tabular}
\end{table}

\tablename~\ref{tab:sim_normal} demonstrates how both SLn and SLcn models consistently recover the original parameter values across various sample sizes, showcasing enhanced precision with larger sample sizes. 
Furthermore, it is evident that the estimated standard errors for the parameters closely approximate the MC standard deviations (MC SE). 
The lack of differentiation in the model selection criterion comes as no surprise, given that the SLn model is a specific case of the SLcn model when $\nu_2$ approaches $1$ or $\nu_1$ approaches $0$.

In the second simulation, we generate data from the contaminated normal distribution with $\nu_1=0.1$ and $\nu_2=0.1$ under the same configuration of the parameter values. 
\tablename~\ref{tab:sim_cn} displays results where the SLn model exhibits poor performance in comparison to the SLt and SLcn models. 
While the SLt and SLcn fit appear to be adequate, the SLn model exhibits biased parameter estimation that becomes more pronounced as sample sizes increase. With a larger sample size, more data points are observed in the tail or the atypical shape due to heavy tails of the contaminated normal distribution, consequently impacting the estimation of the SLn model. 
Additionally, it is evident that the model selection criteria (AIC and BIC) favor the SLcn model, as expected, given that the SLcn model is the true generating model. 
\tablename~\ref{tab:sim_select} illustrates the percentage of times the SLn, SLt, and SLcn models were selected across various sample sizes and model generations. 
In the case of the contaminated normal distribution, our proposed model, the SLcn, demonstrates superior fit (97\%) compared to the other models, even surpassing the SLt (3\%) when the sample size is 1000. 
This trend persists across all sample sizes, with favorable results of 90\% or higher.

Finally, we generate data from a slash distribution with heavy tails (degrees of freedom $1.43$) in the same setting. \tablename~\ref{tab:sim_sl} illustrates that both the SLt and the SLcn models provide a more appropriate fit under model misspecification. 
Based on the \tablename~\ref{tab:sim_select} of the model selection criterion, when the sample size is 250, the SLcn model was selected 81.5\%, indicating its superiority. 
Also, as the sample size increases, both the SLt and SLcn are chosen almost equally. 
The estimates of $\bbeta$ outperform those of $\bgamma$ across all models. However, the bias is more pronounced in the SLn model compared to the SLt and the SLcn models, particularly with larger sample sizes. 
Additionally, the SLt and the SLcn model estimates exhibit greater stability across different sample sizes. 
For instance, as sample sizes increase, the estimated bias for $\rho$ increases significantly under the SLn model, while the SLcn and SLt models show minimal variation. Finally, the standard error estimates for the SLcn and the SLt model remain consistent across all scenarios.

\begin{figure}[!ht]
	\begin{center}
		\includegraphics[scale=0.3]{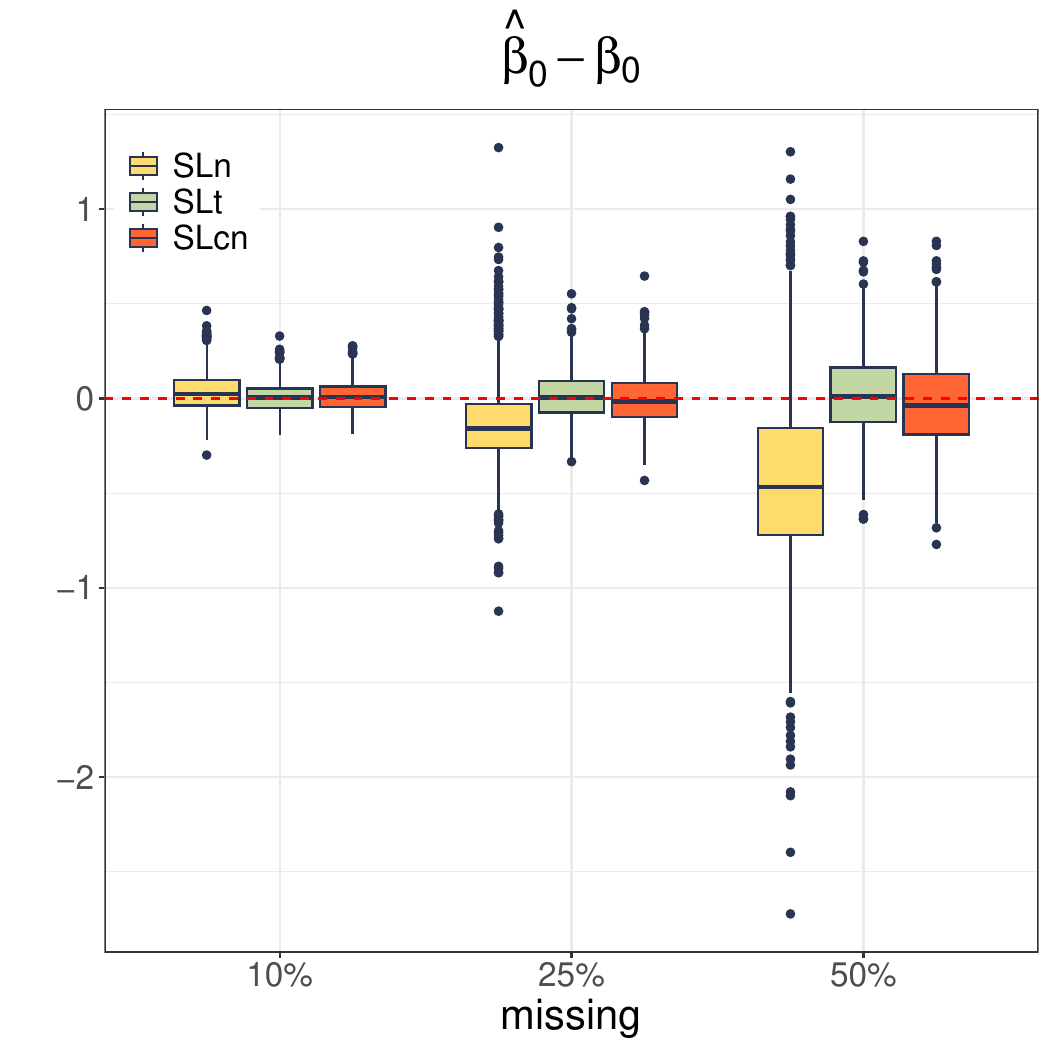}
		\includegraphics[scale=0.3]{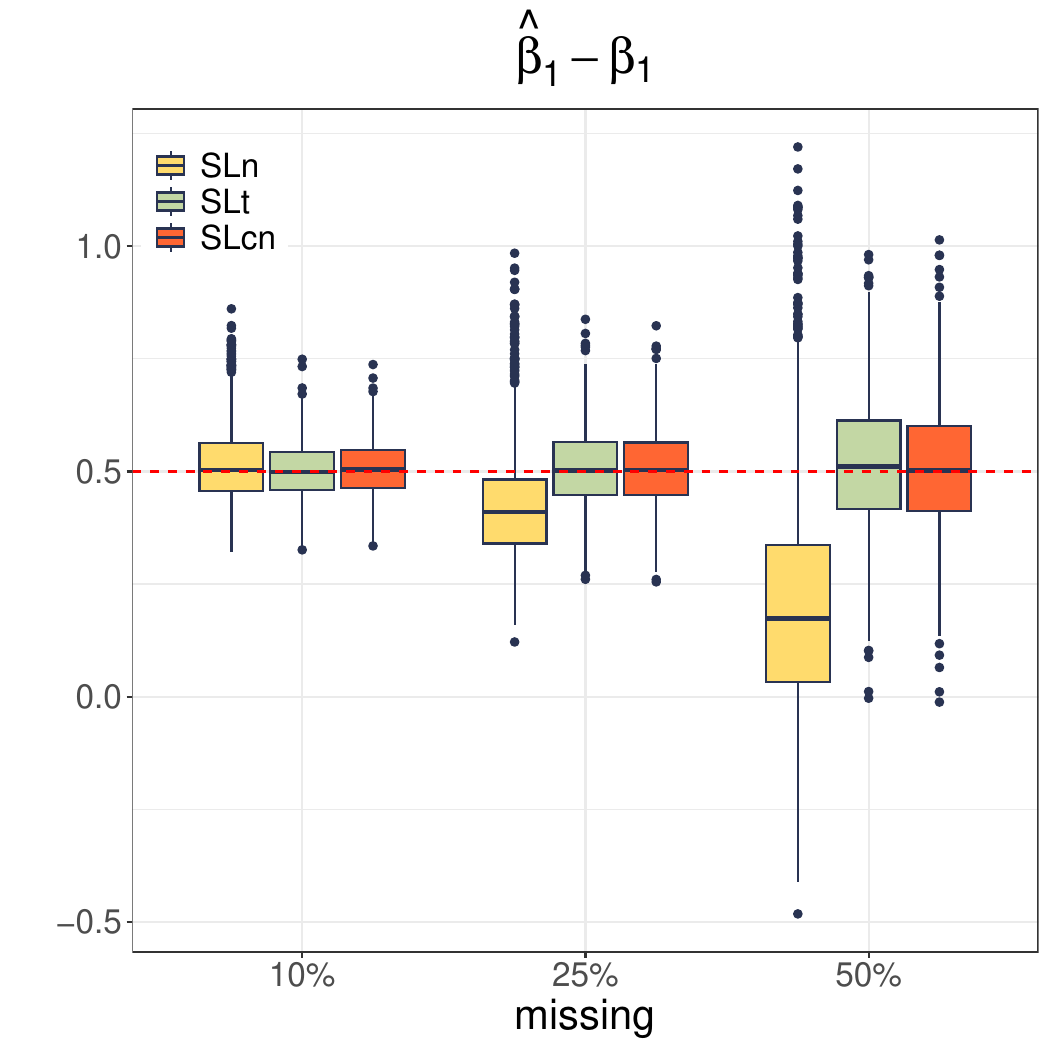}
		\includegraphics[scale=0.3]{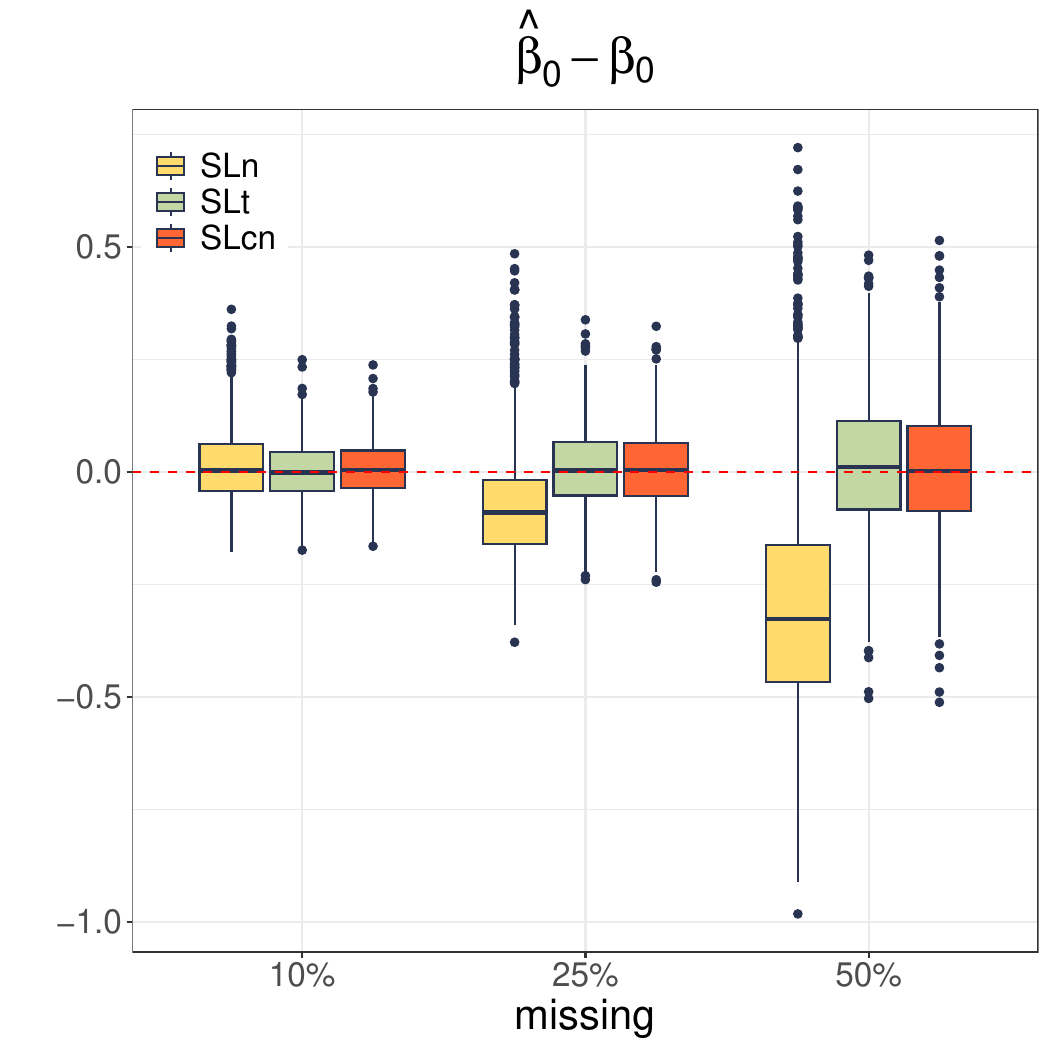}
  		\includegraphics[scale=0.3]{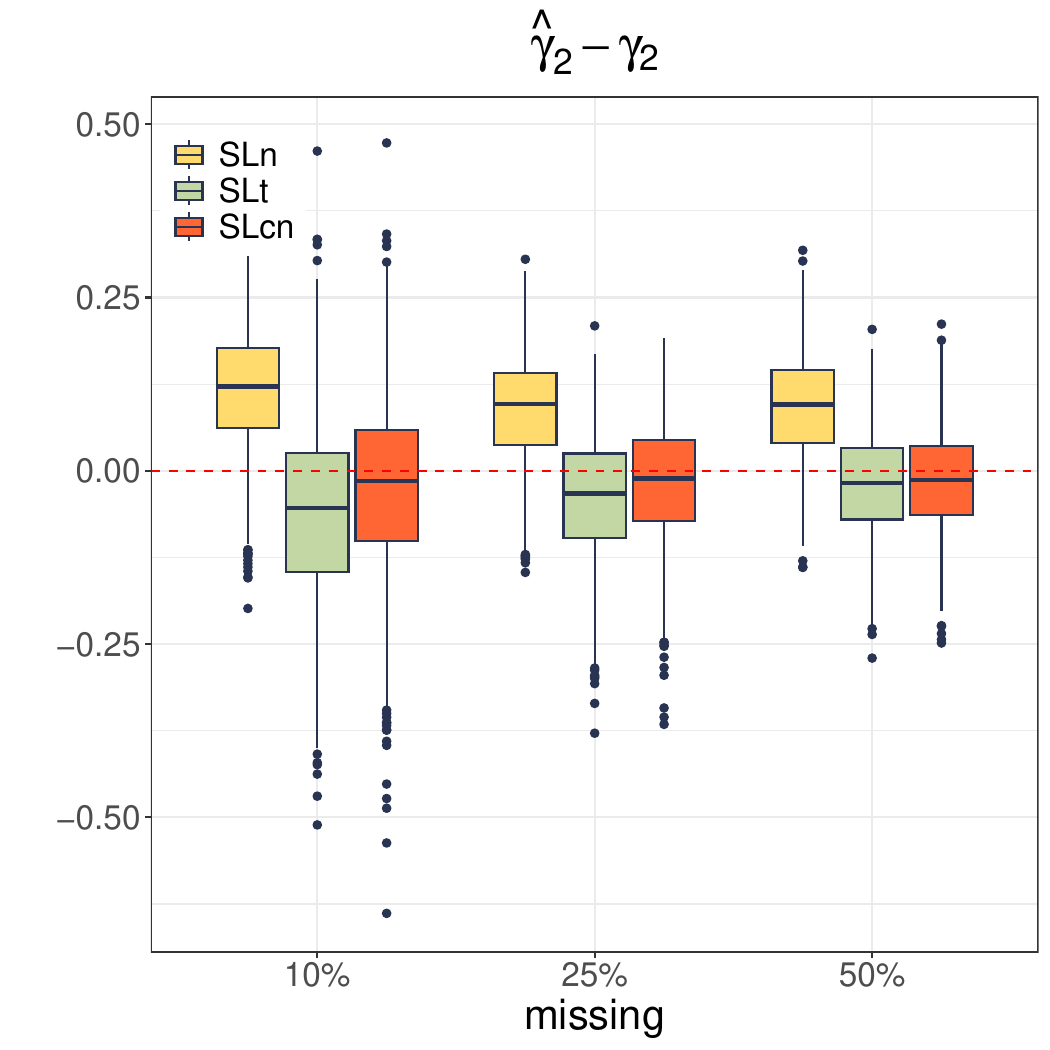}
    	\includegraphics[scale=0.3]{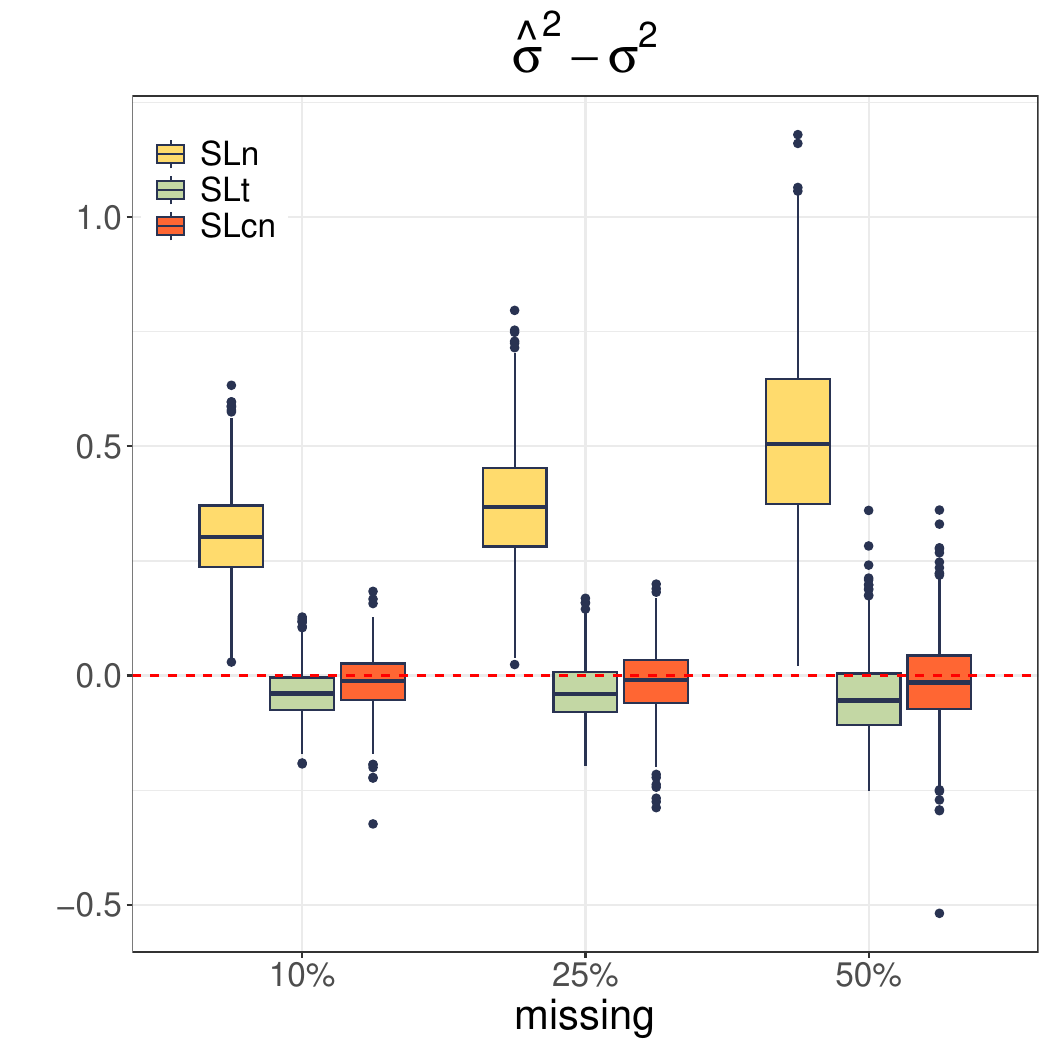}
		\includegraphics[scale=0.3]{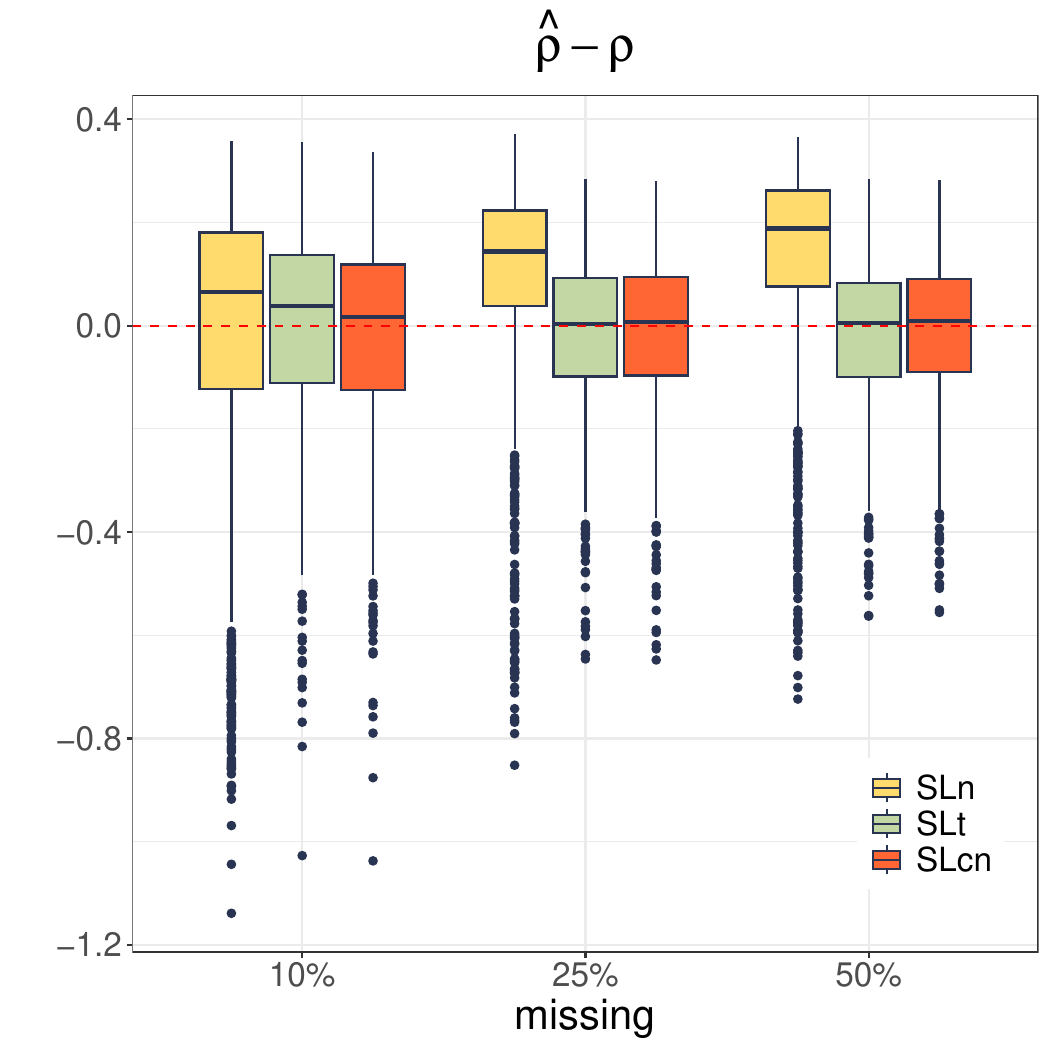}
		\caption{Boxplot of the SL models comparison of the 1000 Monte Carlo estimates for the data generated from the contaminated normal distribution. When varying the missing proportion as 10\%, 25\% and 50\%, the estimated parameters are $\beta_0$, $\beta_1$, $\gamma_1$, $\gamma_2$, $\sigma$ and $\rho$.}
		\label{Comp2}
	\end{center}
\end{figure}

\begin{table}[!hb]
\caption{Simulation study 1. Percentage of times each of the three models was selected out of $1000$ Monte Carlo replicates from the different sample sizes and generation distributions.}
	\label{tab:sim_select}
\centering
\begin{tabular}{crrrrrrr}
  \toprule
 & & \multicolumn{3}{c}{AIC} & \multicolumn{3}{c} {BIC} \\
 % \cmidrule{3-4} \cmidrule{5-6}    
 Distribution & Sample Size & SLn & SLt & SLcn & SLn & SLt & SLcn\\ 
 \midrule
  %\midrule
          & 250 & 52.9\% & 2.3\% & 44.8\% 
          & 52.9\% & 2.3\% & 44.8\% \\
Normal    & 500 & 56.4\% & 5.7\%  & 37.9\% 
          & 56.4\% & 5.7\%  & 37.9\% \\ 
          & 1000 & 58.2\% & 9.7\%  & 32.1\% 
          & 58.2\% & 9.7\%  & 32.1\%  \\[1.7mm] 
          %\hline
          & 250 & 0.8\% & 6.9\% & 92.3\% & 0.8\% & 6.9\% & 92.3\% \\ 
C-Normal & 500 & 0.1\% & 5.9\% & 94.0\% & 0.1\% & 5.9\% & 94.0\%\\ 
          & 1000& 0\% & 3\% & 97\% & 0\% & 3\% & 97\% \\[1.7mm] 
          & 250 & 1.3\% & 17.2\% & 81.5\% & 1.3\% & 17.2\% & 81.5\% \\ 
Slash     & 500 & 0\% & 49.4\% & 50.3\% & 0\% & 49.4\% & 50.3\% \\ 
          & 1000& 0\% & 47.7\% & 52.3\% & 0\% & 47.7\% & 52.3\% \\ 
   \bottomrule
\end{tabular}
\end{table}

\subsection{Missing variation}
\label{sec:s2}

In order to explore the impact of missing data on the SL models, we specify the mean percentage of missing values across different levels: 10\%, 25\%, and 50\%. 
We retain the remaining setup of the simulation as described in the preceding section. 
First, when the missing rate increases, we examine how the parameter estimates, such as $\bbeta$, $\bgamma$, $\sigma$, and $\rho$, change as the sample size increases to 250, 500, and 1000 for our proposed SLcn model. 
\figurename~\ref{Comp1} shows the boxplots of the centered estimates of the parameters for the SLcn model when the data is generated from the contaminated normal distribution in this setting. 
\figurename~\ref{app1} and \figurename~\ref{app2} in Appendix B contain boxplots for the results of the SLcn model when the data is generated from the normal distribution and the slash distribution, respectively. 
\figurename~\ref{Comp1} indicates that increasing the sample size results in a corresponding decrease in bias and variability of the parameter estimates. 
These results suggest that the finite sample properties hold. 
Furthermore, for a given sample size, an increase in the missing rate leads to an increase in the variability of the parameter estimates of $\bbeta$, while for $\bgamma$, the bias decreases. 
This aspect is intriguing; presumably, as our model estimates $\bgamma$ are contained in the data-generating part of the model, it enhances the accuracy of estimating parameters in covariates while the missing rate increases.

Secondly, when the missing rate increases, we explore any differences among the parameters across the three different SL models, specifically focusing on a fixed sample size of 500. 
This simulation aims to shed light on any disparities between the SL models from data generated from the contaminated normal distribution depicted in \figurename~\ref{Comp2}. 
\figurename~\ref{app3} and  \figurename~\ref{app4} in Appendix B contain boxplots from the normal distribution and the slash distribution, respectively. 
As previously observed, the SLn model exhibits greater bias across all parameters, with the bias increasing even as the missing rate increases, while the SLcn model adequately fits the heavy-tailed data.

In general, we confirmed that the SLcn model demonstrates greater resilience compared to the other models across various variations in characteristics regarding model misspecification (Section~\ref{sec:s1}) and missing data (Section~\ref{sec:s2}). According to \tablename~\ref{tab:sim_select}, the SLcn model exhibits remarkable performance. The SLcn model demonstrates superior performance and is selected overwhelmingly (over 92\%) in the contaminated normal distribution. 
Also, it demonstrates results comparable to those of the SLn in the normal distribution. 
Even with the slash distribution, it is chosen comparably to the SLt model. 
Consequently, we can conclude that our proposed SLcn model is a reliable method for the missing data.

\newpage

\section{Real data applications}
\label{secApp}

In this section, we illustrate two real data applications of our SLcn model.
In analogy with Section~\ref{sec:Simulation study}, we consider the SLn and SLt models for comparison.
We adopt the ML approach to estimate the parameters for all the considered models.
The analysis is entirely conducted in \prog{R} using the \texttt{HeckmanEM} package, available on \textsf{GitHub} (\url{https://github.com/marcosop/HeckmanEM}); the package allows for fitting all the SL models via EM-based algorithms.

Since the competing models have a differing number of parameters, we compare their goodness-of-fit, as usual (see also Section~\ref{sec:Simulation study}), via AIC and BIC.
%, which we remind need to be minimized.
%\textcolor{orange}{AGGIUSTARE IN ACCORDO CON APPENDICE A.3.
Moreover, we use likelihood-ratio (LR) tests to compare the SLn (null model) with the SLt and SLcn (alternative models).
The obtained $p$-value in the sequel will always be compared to the classical 5\% significance level.

%We illustrate the proposed model with the analysis of two real data sets. The analysis was performed using the \prog{R} package \texttt{HeckmanEM}  available on \textsf{GitHub} (\url{https://github.com/marcosop/HeckmanEM}). 

\subsection{RAND Health Insurance data}
\label{sec:RAND Health Insurance data}

The first application concerns a study from the RAND Health Insurance Experiment (RAND HIE), which is a comprehensive study of healthcare cost, utilization, and outcome in the United States. 
This data set is used by \citet{cameron2009microeconometrics} to analyze how the patient's use of health services is affected by types of randomly assigned health insurance. 
More recently, the data were also revisited by \citet{lachosHeckman} considering SLn and SLt models. 
Here, we revisit the RAND Health Insurance data intending to provide additional inferences and insights by using the SLcn model.

In our analysis, we choose the same set of variables as \citet{lachosHeckman}.  
We adopt the $\ln$ of the individual's medical expenses (\code{meddol}) as the outcome variable. 
The covariates in the outcome equation are:
\begin{align*}
\xp = & (\code{1, logc, idp, lpi, fmde, physlm, disea, hlthg, hlthf, hlthp, linc, lfam, educdec, xage,}
\\
& \quad\code{female, child, fchild, black}),
\end{align*}
including the $\ln$ of coinsurance rate plus 1 $(\code{logc} = \ln (\code{coins+1}))$, the dummy
for the individual deductible plan (\code{idp}), the $\ln$ of participation incentive payment (\code{lpi}), an artificial variable \code{fmde} that is 0 if $\code{idp} = 1$ and $\ln(\max(1,\code{mde}/(0.01\cdot\code{coins})))$ otherwise (where \code{mde} is the maximum expenditure offer), physical limitations (\code{physlm}), the number of chronic diseases (\code{disea}), dummy variables for good (\code{hlthg}), fair (\code{hlthf}), and poor (\code{hlthp}) self-rated health (where the baseline is excellent self-rated health), the log of family income (\code{linc}), the log of family size (\code{lfam}), education of household head in years (\code{educdec}), age of individual in years (\code{xage}), a dummy variable for female individuals (\code{female}), a dummy variable for individuals younger than 18 years (\code{child}), a dummy variable for female individuals younger than 18 years (\code{fchild}), and a dummy variable for black household heads (\code{black}). 
The selection variable is \code{binexp}, which indicates whether the medical expenses are positive and without exclusion.
We consider that $\bx=\w$.

For our analysis, a subsample was selected so that study year is 2 and \code{educdec} is not ‘‘\texttt{NA}’’. 
Out of $n=5574$ observations, 1293 of \code{meddol} (medical expenses) are 0 which means that the outcome variable $\ln$ of \code{meddol} is unobserved, and 4281 of \code{meddol} are positive (means that the outcome variable $\ln$ of \code{meddol} is available). 
%\orange{}
The data is available in the \prog{R} package \texttt{sampleSelection}.  
The results for the SLn, SLt, and SLcn models are presented in \tablename~\ref{tab:example2}. 
The ML estimates for the SLn and SLt models are the same as those reported in \citet{lachosHeckman}. 
The $95\%$ Wald confidence interval for $\rho$ is $[0.663,0.806]$ for the SLn model, $[0.575,0.760]$ for the SLt model, and $[0.548,0.693]$ for the SLcn model. 
Thus, regardless of the considered model, the interval does not contain the value $\rho=0$, suggesting the SL effect exists.
The bottom part of \tablename~\ref{tab:example2} presents some model selection criteria.
When we conducted the likelihood ratio (LR) tests for both the SLt and SLcn (alternative) models versus the SLn (null) model, the $p$-values were less than 0.001 in both cases. 
These testing results indicate the need for heavier-than-normal tails for the bivariate error distribution of the SL model.
However, the lowest AIC and BIC values among the three models support the superiority of the SLcn model, indicating a better fit than the SLn and SLt models. 

% SLcn converge iteration :1067
\begin{table}[!ht]
	\caption{RAND HIE data. 
 Parameter estimates and standard errors for SLn, SLt, and SLcn models.
 AIC and BIC values, along with LR tests comparing the SLn (null model) with the SLt and SLcn (alternative models), are provided as evaluation measures. 
 Smaller values of AIC and BIC indicate a better fit (in bold).}
	\label{tab:example2}
	\begin{center}
		%\scalebox{0.7}{
		%\resizebox{\textwidth}{!}{%
		\begin{tabular}{lrrrrrr}
			\toprule
			Parameter & \multicolumn{2}{l}{SLn} & \multicolumn{2}{l}{SLt} & \multicolumn{2}{l}{SLcn} \\
			\cmidrule(lr){2-3} \cmidrule(lr){4-5}\cmidrule(lr){6-7}
			  & EM    & SE   & EM   & SE& EM   & SE \\
			\midrule
			Outcome model ($\ln$-\code{meddol})&\multicolumn{6}{c}{} \\[1.5mm]
			%\midrule
			\code{Intercept} & 2.155& 0.253 &2.358  &  0.242&2.334 & 0.233 \\ 
			\code{logc} & -0.073& 0.033 & -0.067 & 0.032 &-0.076 & 0.031 \\ 
			\code{idp} &-0.146& 0.063 & -0.152 & 0.061 &-0.137 & 0.060 \\ 
			\code{lpi} & 0.014 & 0.011 &0.014  &0.010  &0.015 & 0.010 \\ 
			\code{fmde} & -0.024&  0.019&-0.028 &0.018  &-0.022 & 0.018 \\ 
			\code{physlm} &0.350 &0.073  &0.339  & 0.070 & 0.329 & 0.069 \\
			\code{disea} & 0.028&  0.004 & 0.028 &  0.0036 & 0.027 & 0.004 \\
			\code{hlthg} &0.156 & 0.052 &  0.145 & 0.050  &0.139 & 0.048 \\ 
			\code{hlthf} &0.442 &  0.092& 0.462 &  0.089 & 0.473 & 0.087 \\
			\code{hlthp} &  0.989&  0.167 & 0.881 & 0.165 & 0.812 & 0.166 \\
			\code{linc} &0.120& 0.024 & 0.110 &  0.023&0.106 & 0.023 \\
			\code{lfam} & -0.157 &  0.048 &  -0.180& 0.047&-0.169 & 0.045 \\
			\code{educdec} & 0.017 &  0.009&  0.016 &  0.009& 0.019 & 0.008 \\ 
			\code{xage} &0.006 &0.002 & 0.005 & 0.002 & 0.006 & 0.002 \\ 
			\code{female} & 0.540& 0.061 & 0.503 & 0.060&  0.433 & 0.058 \\
			\code{child} & -0.202& 0.098 & -0.192 &  0.093& -0.179 & 0.090 \\
			\code{fchild} &-0.554 & 0.100 & -0.526 & 0.095& -0.458 & 0.091 \\ 
			\code{black}&-0.518 &  0.070 &   -0.502& 0.072  &  -0.474 & 0.070 \\[1.5mm] 
			%\midrule
			Selection model&\multicolumn{6}{c}{} \\[1.5mm]
			%\midrule
			\code{Intercept} &-0.220 &0.191  &  -0.228&0.207  &  -0.356 & 0.325 \\
			\code{logc} &-0.108 & 0.025 &  -0.129& 0.028  & -0.221 & 0.046 \\
			\code{idp} & -0.110 & 0.048 & -0.105 &0.053  & -0.128 & 0.085 \\ 
			\code{lpi} & 0.030& 0.009 & 0.033 & 0.010  & 0.053 & 0.015 \\ 
			\code{fmde} & 0.002& 0.016 &  0.005 &  0.018 & 0.016 & 0.028 \\ 
			\code{physlm} &0.285 & 0.073 &0.335  & 0.084 & 0.540 & 0.139 \\ 
			\code{disea} & 0.021& 0.004 & 0.024 & 0.004  & 0.040 & 0.007 \\
			\code{hlthg} & 0.056&  0.043&  0.055 & 0.047  & 0.079 & 0.075 \\
			\code{hlthf} &  0.223& 0.082 & 0.247&   0.091& 0.377 & 0.145 \\
			\code{hlthp} & 0.796& 0.187 & 0.904  &   0.230 & 1.338 & 0.375 \\ 
			\code{linc} & 0.055&0.017  & 0.054 & 0.018  & 0.077 & 0.028 \\ 
			\code{lfam} & -0.032 &0.040  & -0.041 & 0.044 &-0.065 & 0.070 \\
			\code{educdec} & 0.032 & 0.008 &0.037 &  0.008  &0.059 & 0.013 \\ 
			\code{xage} &-0.001 & 0.002 & -0.001 &  0.002 &-0.001 & 0.004 \\ 
			\code{female} & 0.413& 0.053 &  0.463 &  0.059 &0.706 & 0.100 \\ 
			\code{child} &0.059 &0.079  &0.082  &  0.087 &  0.130 & 0.137 \\ 
			\code{fchild} &-0.401&0.079  & -0.456 & 0.087 & -0.697 & 0.141 \\
			\code{black}& -0.587& 0.051 & -0.646 & 0.055  & -0.996 & 0.093 \\[1.5mm] 
			%\midrule
			$\sigma$ &1.570 & 0.027&  1.374 &0.027 & 0.863 & 0.052 \\ 
			$\rho$ & 0.736 & 0.037& 0.667 & 0.047&  0.621 & 0.037 \\ 
			$\nu\,\, (\nu_1)$ & & & 8.809 & - &   0.674 & 0.007 \\ 
			$\nu_2$ & & & &  & 0.236 & 0.022 \\[1.5mm] 
   			%\midrule
%			log-likelihood	& \multicolumn{2}{c}{-10170.37} & \multicolumn{2}{c}{ -10141.06} \\
Evaluation criteria & \multicolumn{6}{c}{} \\[1.5mm]
				AIC	& \multicolumn{2}{c}{20342.22} & \multicolumn{2}{c}{20284.12}& \multicolumn{2}{c}{\bf 20252.55}\\
			BIC	& \multicolumn{2}{c}{20348.85} & \multicolumn{2}{c}{ 20290.75}& \multicolumn{2}{c}{\bf 20259.18}\\
    $p$-value (LR test of bivariate normality) & \multicolumn{2}{c}{} & \multicolumn{2}{c}{$< 0.001$} & \multicolumn{2}{c}{$< 0.001$}\\
			\bottomrule
		\end{tabular}
			\end{center}
\end{table}

\begin{figure}[!ht]
\centering
    \begin{subfigure}[b]{0.32\textwidth}            
            \includegraphics[width=\textwidth]{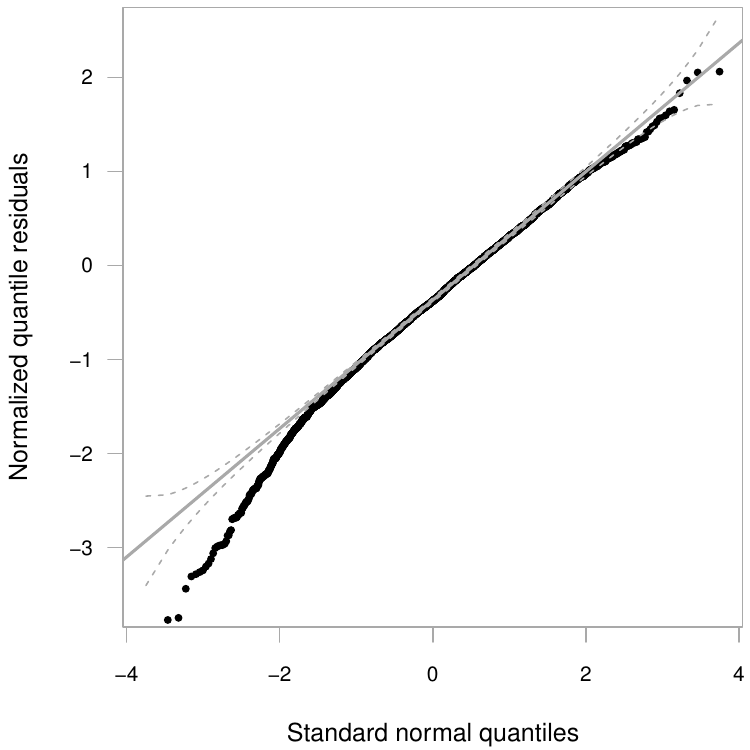}
            \caption{SLn}
            \label{fig:A1_SLn}
    \end{subfigure}%
     %add desired spacing between images, e. g. ~, \quad, \qquad etc.
      %(or a blank line to force the subfigure onto a new line)
    \begin{subfigure}[b]{0.32\textwidth}
            \centering
            \includegraphics[width=\textwidth]{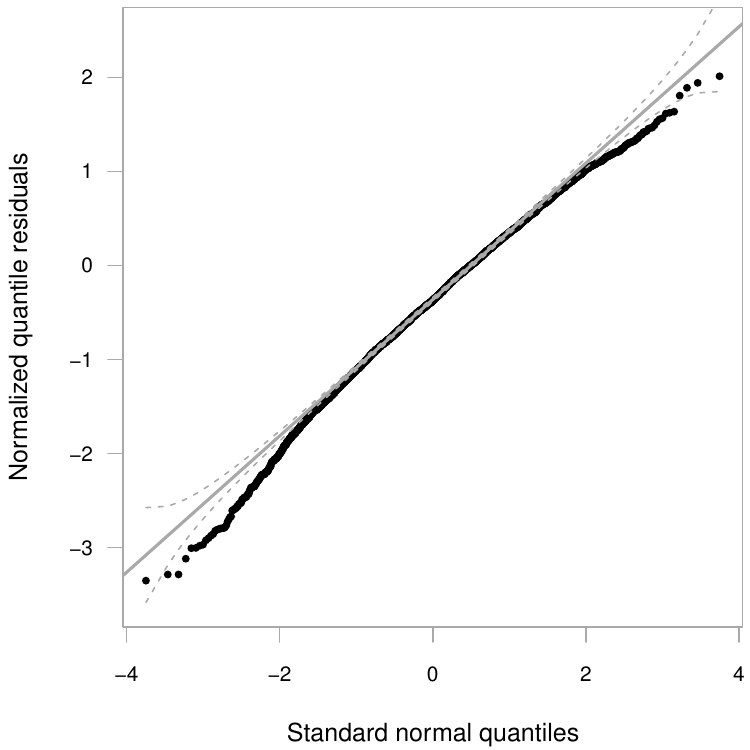}
            \caption{SLt}
            \label{fig:A1_SLt}
    \end{subfigure}
     %add desired spacing between images, e. g. ~, \quad, \qquad etc.
      %(or a blank line to force the subfigure onto a new line)
    \begin{subfigure}[b]{0.32\textwidth}
            \centering
            \includegraphics[width=\textwidth]{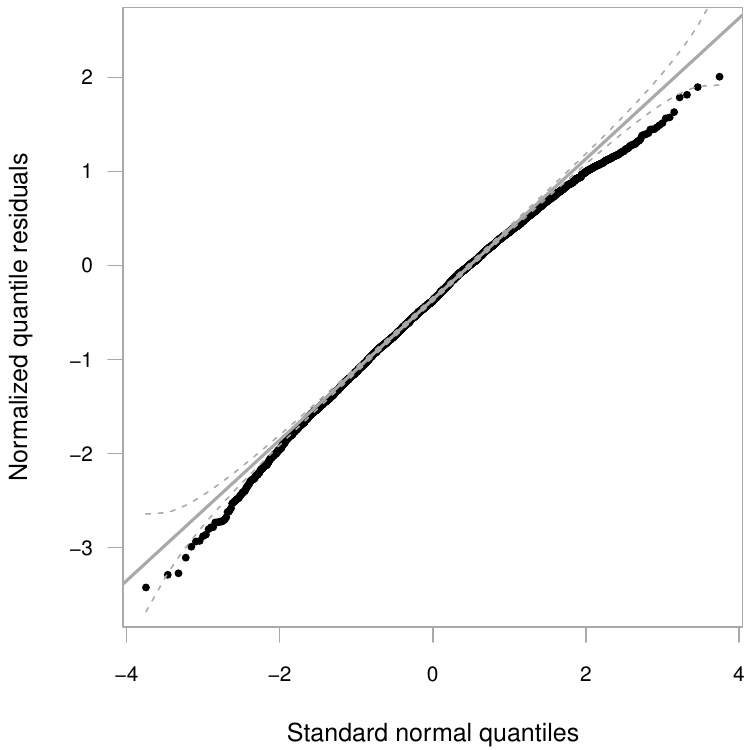}
            \caption{SLcn}
            \label{fig:A1_SLcn}
    \end{subfigure}
    \caption{RAND HIE data.  
Normal probability plot for the normalized quantile residuals, along with 95\% envelopes, for the SLn (left), SLt (middle), and SLcn (right) models.}
\label{fig:envelopes1}
\end{figure}

The above results are further corroborated by the normal probability plot of the $r_{NQ_{i}}$ residuals in \figurename~\ref{fig:envelopes1}, plotted along with the generated envelopes obtained through the \texttt{HeckmanEM} package in \prog{R}. 
Based on our observations, we can conclude again that the SLcn model better fits the data than the SLt and SLn models. 
This is evident as just fewer observations fall outside the envelopes in the left tail in the case of the SLcn model (right panel).

\begin{figure}[!ht]
\centering
    \begin{subfigure}[b]{0.49\textwidth}            \includegraphics[width=\textwidth]{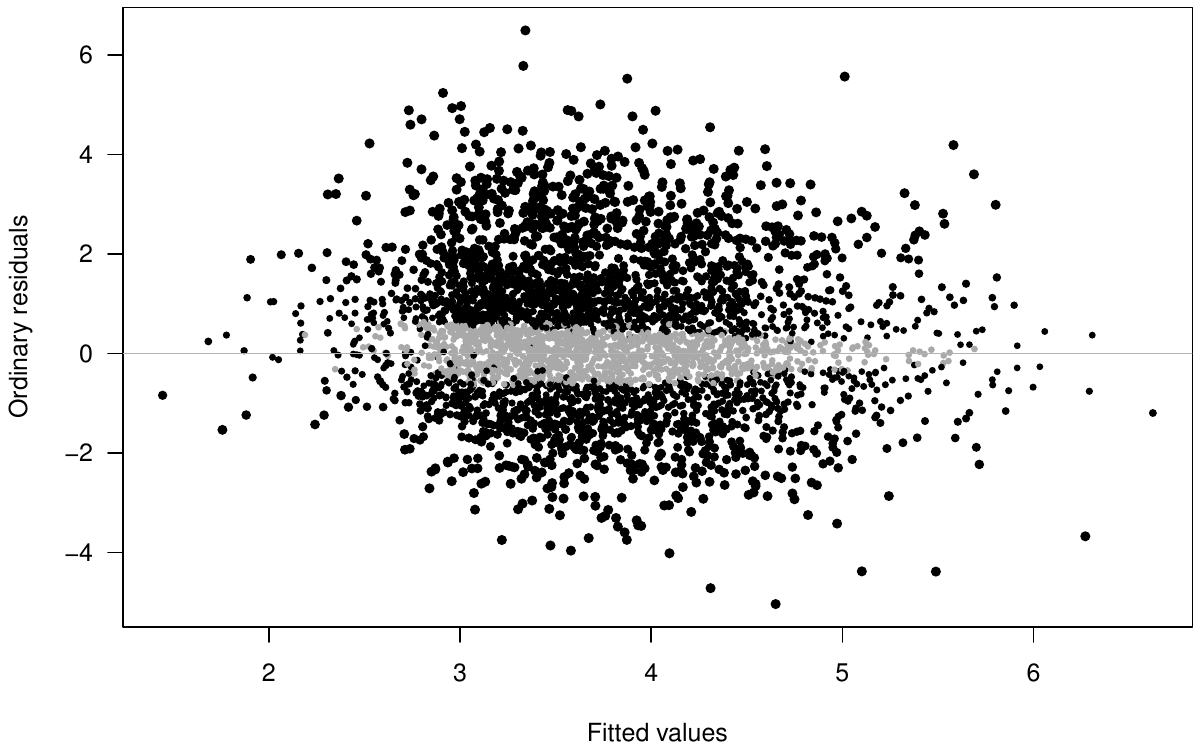}
            \caption{Ordinary residuals vs. fitted values}
            \label{fig:A1residualplot}
    \end{subfigure}%
     %add desired spacing between images, e. g. ~, \quad, \qquad etc.
      %(or a blank line to force the subfigure onto a new line)
    \begin{subfigure}[b]{0.49\textwidth}
            \centering
            \includegraphics[width=\textwidth]{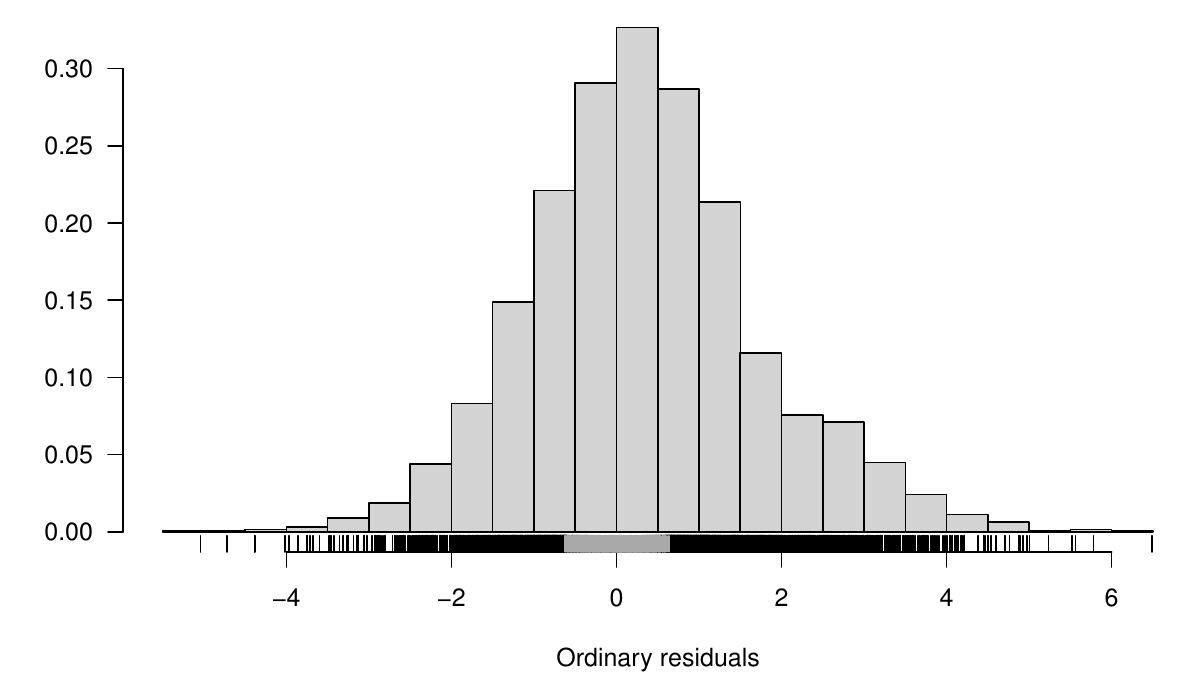}
            \caption{Histogram of ordinary residuals}
            \label{fig:A1residualhist}
    \end{subfigure}
     %add desired spacing between images, e. g. ~, \quad, \qquad etc.
      %(or a blank line to force the subfigure onto a new line)
    % \begin{subfigure}[b]{0.32\textwidth}
    %         \centering
    %         \includegraphics[width=\textwidth]{../A2QRCN.pdf}
    %         \caption{SLcn}
    %         \label{fig:A2_SLcn}
    % \end{subfigure}
    \caption{RAND HIE data.
  Classical residuals versus fitted values (on the left) and histogram of the classical residuals (on the right).
  Black and gray indicate detected good points and inliers, respectively.
  The size of the bullets in the left plot is proportional to the probability of being good in \eqref{eq:posterior to be bad}.}.
  %orange
\label{fig:RAND HIE data_residual_plot2}
\end{figure}

Finally, to obtain further insights from the estimated SLcn model, \figurename~\ref{fig:RAND HIE data_residual_plot2} shows the plot of the (classical) residuals versus the fitted values, as well as the histogram of these residuals.
First of all, it is important to note that in this real data application, we have $\widehat{\nu}_1=0.674$ (refer to \tablename~\ref{tab:example2}); because the mixing proportion is larger than 0.5, we are in a situation with inliers.
In \figurename~\ref{fig:A1residualplot}, the size of each bullet is proportional to the posterior probability of an observation being classified as good; refer to \eqref{eq:posterior to be bad}. 
Thus, larger bullets indicate points more likely to be regular points. 
Additionally, the color of the bullets is determined by the automatic inlier detection rule discussed in Section~\ref{sec:Automatic mild outlier detection}: black indicates good points (probability of being a regular point larger than 0.5), and gray denotes inliers (probability of being a regular point lower than 0.5). 
The residual plot shows what we expect from an appropriate model: a random scatter of points forming an approximately constant width band around the benchmark horizontal null-residual line, which is superimposed on the scatter.
Out of 4281 observed data points, 3001 were identified as good observations and 1280 as inliers.
Therefore, the proportion of inliers is 0.299, a value close enough to $1-\widehat{\nu}_1=1-0.674=0.326$ (refer to \tablename~\ref{tab:example2}).
The discrepancy between the two values arises because 0.299 is a ‘‘hard'' version, computed on 0/1 values from the inlier detection rule in Section~\ref{sec:Automatic mild outlier detection}, of the ‘‘soft'' proportion (0.326) computed according to \eqref{eq:estimated proportion of bad points} and, as such, based on the probabilities of being good points. 
Moreover, the small estimated degree of contamination (0.236) indicates a large variability of the good points compared to the inliers.

%\sout{This figure demonstrates that the residuals are more closely distributed in a contaminated normal distribution with bad observations rather than in a normal distribution.}

% Heeju version
% \begin{figure}[!ht]
% 	\begin{center}
% 		\centering \hspace {1cm}\centering \\
% 		\includegraphics[width=7cm, height=6cm]{../SimPlot/Rplot01.pdf}
% 		\includegraphics[width=7cm, height=6cm]{../SimPlot/Histogram1.pdf}
% 		\caption{
%   RAND HIE data.
%   Classical residuals versus fitted values (on the left) and histogram of the classical residuals (on the right).
%   Orange and black indicate detected good and bad observations, respectively.
%   The size of the bullets in the left plot is proportional to the probability of being a bad point.
%   } 
%   \label{fig:RAND HIE data_residual_plot2}
% 	\end{center}
% \end{figure}

%\newpage

\subsection{Mroz: Labor Supply Data}
\label{application_Mroz}

The second application focuses on missing econometric data, where we re-analyzed the dataset originally presented in \citet{mroz1987sensitivity} to estimate the wage offer function for married women.
The dataset -- which is very famous, and benchmark, in the sample selection literature and, more in general, in the econometric world (see, e.g., \citealp{baltagi2012econometrics}, \citealp{greene2017econometric}, \citealp{vinod2008hands}, \citealp{wooldridge2010econometric}, and \citealp{wooldridge2019introductory}) -- consists of observations on $n=753$ married white women for 21 variables and can be found in the \prog{R} package \code{AER} \citep{AERpackage}.
The outcome of interest is the logarithm of \code{wage}, which is missing for 325 individuals (43.2\% of the data) and observed for 428 individuals (corresponding to 56.8\%).  
Covariates in the outcome equation are education status (\code{educ}) and city (\code{city}), that is, $\x = (1, \code{educ}, \code{city})$. 
The selection equation uses the husband's wage (\code{hwage}), the number of children 5 years old or younger (\code{youngkids}), the marginal tax rate of the wife (\code{tax}), and the wife's father's educational attainment (\code{feduc}) as well as education status and city, that is, $\w=(\x, \code{hwage}, \code{youngkids}, \code{tax}, \code{feduc})$.

A summary of the fit -- in terms of parameter estimates, standard errors, and evaluation criteria -- for SLn, SLt, and SLcn models is given in \tablename~\ref{tab:mroz}.
For the SLt model, the estimated degrees of freedom are 3.001; this is the initial indication that the data require heavier-than-normal tails.
As for the SLcn model, we have $\widehat{\nu}_1=0.218 \leq 0.5$, and this means that we have mild outliers in the data.
Because the parameter $\nu_2$ controls the effect of contaminated points in the data, the small estimate $\widehat{\nu}_2 = 0.181$ reveals a large increase of variability due to the mild outliers. 
The 95\% confidence interval for $\rho$, spanning [-0.858, -0.702], [-0.853, -0.613], and [-0.857, -0.614] across the SLn, SLt, and SLcn models, respectively, does not contain 0, indicating an SL effect, meaning that the data are not missing at random.
The LR tests, considering the SLt and SLcn as alternative models versus the SLn as the null model, suggest the need for heavier error tails; indeed, both tests yield a $p$-value lower than $0.001$.
The superior overall performance of the SLcn model is supported by its attainment of the lowest values among the considered information criteria (AIC and BIC).

\begin{table}[!htb]
	\caption{Mroz data. 
 Parameter estimates and standard errors for SLn, SLt, and SLcn models.
 AIC and BIC values, along with LR tests comparing the SLn (null model) with the SLt and SLcn (alternative models), are provided as evaluation measures. 
 Smaller values of AIC and BIC indicate a better fit (in bold).
 }
	\label{tab:mroz}
	\begin{center}
		%\scalebox{0.7}{
		%\resizebox{\textwidth}{!}{%
		\begin{tabular}{lrrrrrr}
			\toprule
			& \multicolumn{2}{l}{SLn} & \multicolumn{2}{l}{SLt}&\multicolumn{2}{l}{SLcn} \\
			\cmidrule(lr){2-3} \cmidrule(lr){4-5}\cmidrule(lr){6-7}
			Parameter   & EM    & SE   & EM   & SE & EM   & SE \\
			\midrule
			Outcome model ($\ln$-\code{wage})&\multicolumn{6}{c}{} \\[1.5mm]
			%\midrule
			\code{Intercept} & 0.669& 0.239& 0.332 & 0.170& 0.351 & 0.170 \\
			\code{educ} & 0.066&0.018 & 0.087 &0.013 & 0.085 & 0.013 \\
			\code{city} &0.107 &0.082 & 0.094 &0.059 & 0.082 & 0.059 \\[1.5mm] 
			%\midrule
			Selection model&\multicolumn{6}{c}{} \\[1.5mm]
			%\midrule
			\code{Intercept} &3.802 & 0.764& 5.934 &0.953 & 6.171 & 0.986 \\
			\code{hwage} & -0.104& 0.015 & -0.153&  0.021& -0.158 & 0.021 \\ 
			\code{youngkids} &-0.415 &0.078 & -0.585 & 0.108& -0.621 & 0.112 \\ 
			\code{tax} &-5.782 &0.847 & -8.448 & 1.089&-8.781 & 1.124 \\ 
			\code{feduc} &-0.020 & 0.013& -0.012 &0.016 &-0.011 & 0.017 \\ 
			\code{educ} &0.112 & 0.024& 0.118 & 0.029&  0.122 & 0.029 \\ 
			\code{city} &-0.040 & 0.107& -0.097 & 0.123& -0.095 & 0.129 \\[1.5mm] 
			%\midrule
			$\sigma$ & 0.800 & 0.028& 0.501 &0.030 & 0.487 & 0.030 \\
			$\rho$ & -0.780&  0.040&-0.733 & 0.061&-0.736 & 0.062 \\ 
			$\nu\,\, (\nu_1)$ & & & 3.001 & -&0.218 & 0.018 \\ 
                $\nu_2$ & & &  & & 0.118 & 0.023 \\[1.5mm] 
			%	\midrule
%			log-likelihood	& \multicolumn{2}{c}{-5836.219} & \multicolumn{2}{c}{-5822.076} \\
Evaluation criteria & \multicolumn{6}{c}{} \\[1.5mm]			
					AIC	& \multicolumn{2}{c}{1765.604} & \multicolumn{2}{c}{1678.728}& \multicolumn{2}{c}{\bf 1675.516}\\
				BIC	& \multicolumn{2}{c}{1770.228} & \multicolumn{2}{c}{1683.352}& \multicolumn{2}{c}{\bf 1680.140}\\
    $p$-value (LR test of bivariate normality) & \multicolumn{2}{c}{} & \multicolumn{2}{c}{$< 0.001$} & \multicolumn{2}{c}{$< 0.001$}\\
			\bottomrule
		\end{tabular}
		%}
	\end{center}
\end{table}

\begin{figure}[!ht]
\centering
    \begin{subfigure}[b]{0.32\textwidth}            
            \includegraphics[width=\textwidth]{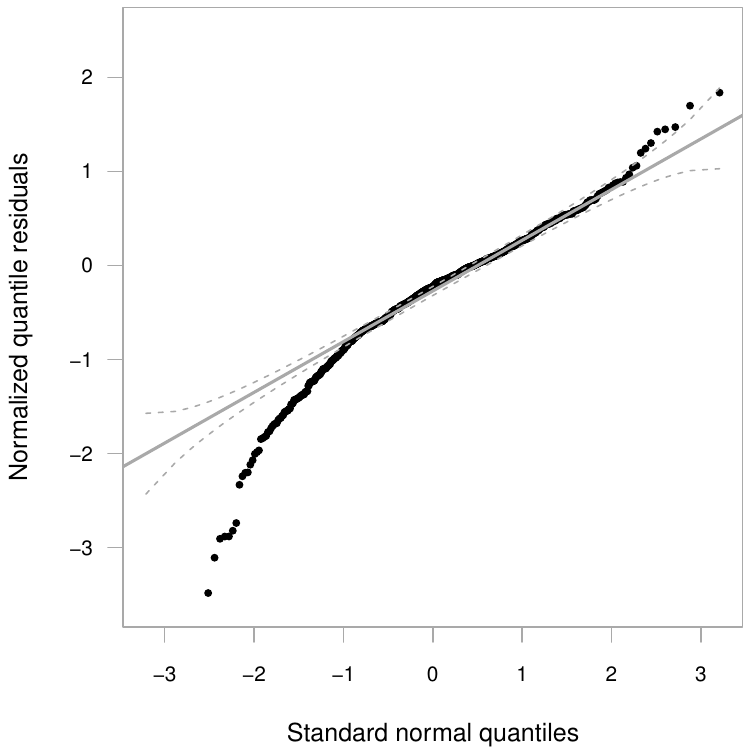}
            \caption{SLn}
            \label{fig:A2_SLn}
    \end{subfigure}%
     %add desired spacing between images, e. g. ~, \quad, \qquad etc.
      %(or a blank line to force the subfigure onto a new line)
    \begin{subfigure}[b]{0.32\textwidth}
            \centering
            \includegraphics[width=\textwidth]{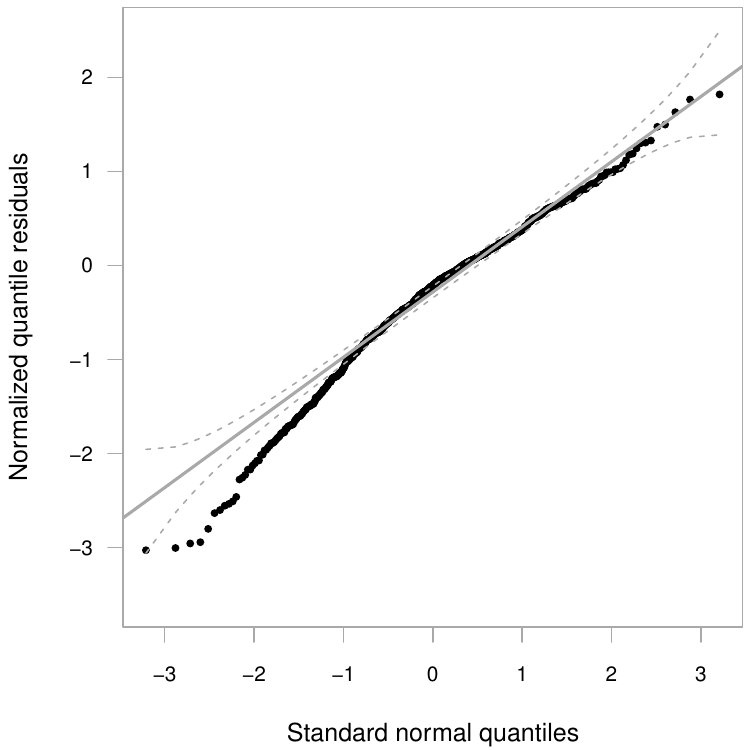}
            \caption{SLt}
            \label{fig:A2_SLt}
    \end{subfigure}
     %add desired spacing between images, e. g. ~, \quad, \qquad etc.
      %(or a blank line to force the subfigure onto a new line)
    \begin{subfigure}[b]{0.32\textwidth}
            \centering
            \includegraphics[width=\textwidth]{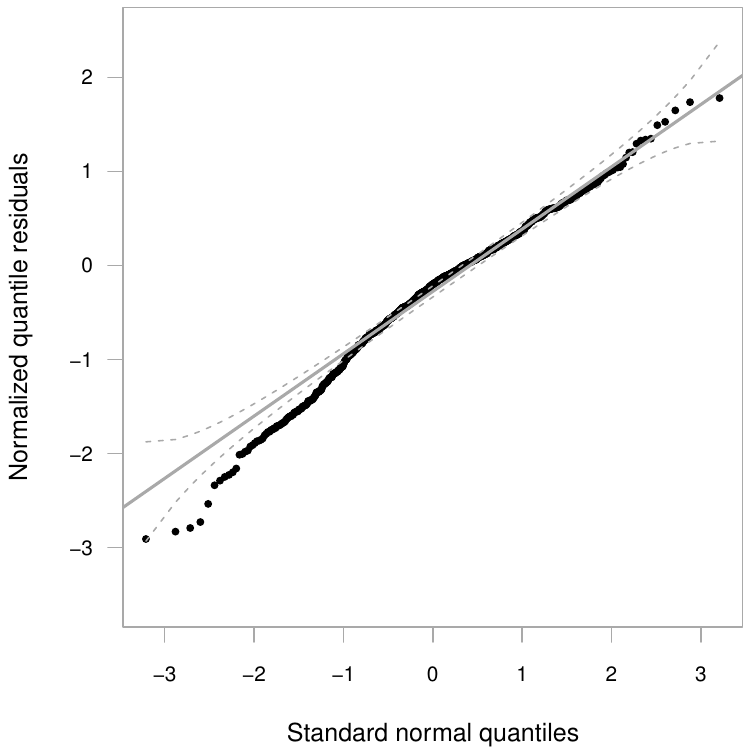}
            \caption{SLcn}
            \label{fig:A2_SLcn}
    \end{subfigure}
    \caption{Mroz data. 
  Normal probability plot for the normalized quantile residuals, along with 95\% envelopes, for the SLn (left), SLt (middle), and SLcn (right) models.}
\label{fig:envelopes2}
\end{figure}

\begin{figure}[!ht]
\centering
    \begin{subfigure}[b]{0.49\textwidth}            
            \includegraphics[width=\textwidth]{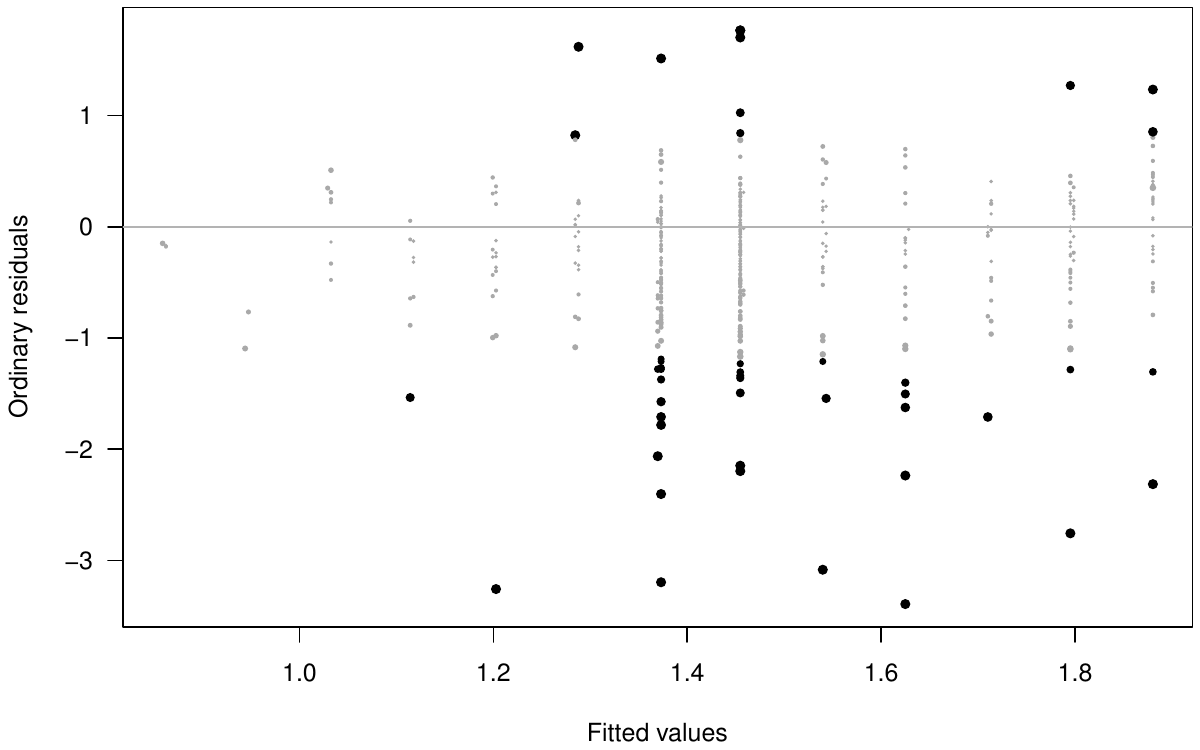}
            \caption{Ordinary residuals vs. fitted values}
            \label{fig:A2residualplot}
    \end{subfigure}%
     %add desired spacing between images, e. g. ~, \quad, \qquad etc.
      %(or a blank line to force the subfigure onto a new line)
    \begin{subfigure}[b]{0.49\textwidth}
            \centering
            \includegraphics[width=\textwidth]{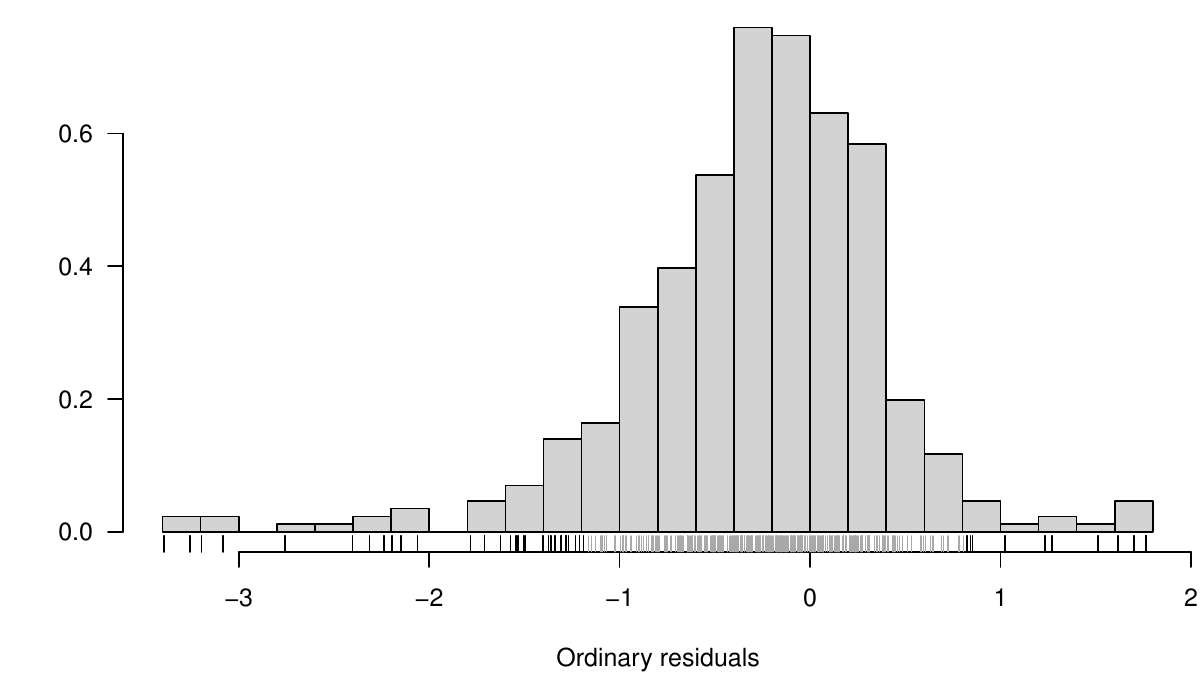}
            \caption{Histogram of ordinary residuals}
            \label{fig:A2residualhist}
    \end{subfigure}
     %add desired spacing between images, e. g. ~, \quad, \qquad etc.
      %(or a blank line to force the subfigure onto a new line)
    % \begin{subfigure}[b]{0.32\textwidth}
    %         \centering
    %         \includegraphics[width=\textwidth]{../A2QRCN.pdf}
    %         \caption{SLcn}
    %         \label{fig:A2_SLcn}
    % \end{subfigure}
    \caption{Mroz data.
  Classical residuals versus fitted values (on the left) and histogram of the classical residuals (on the right).
  Gray and black indicate detected good and bad observations, respectively.
  The size of the bullets in the left plot is proportional to the probability of being an outlier.}
\label{fig:A2coloredresiduals}
\end{figure}

\figurename~\ref{fig:envelopes2} shows the normal probability plot of the quantile residuals for the SLn, SLt, and SLcn models. 
The SLcn and SLt models provide better fits, with the SLcn being slightly better. 
\figurename~\ref{fig:A2coloredresiduals} displays the SLcn model's plot of the (classical) residuals versus the fitted values, along with the histogram of these residuals.
As in Section~\ref{sec:RAND Health Insurance data}, a different color is used based on the outlier detection rule discussed in Section~\ref{sec:Automatic mild outlier detection}. 
These plots allow us to visualize the detected bad observations, which are 45 (10.5\%) out of 428 observed data.

% Heeju Version
% \begin{figure}[!ht]
% 	\begin{center}
% 		\centering \hspace {1cm}\centering \\
% 		\includegraphics[width=5.4cm, height=6cm]{../SLn_envelop_mroz.pdf}
% 		\includegraphics[width=5.4cm, height=6cm]{../SLt_envelop_mroz.pdf}\
%             \includegraphics[width=5.4cm, height=6cm]{../SLcn_envelop_mroz.pdf}\\
% 		\caption{Mroz data. 
%   Normal probability plot for the normalized quantile residuals, along with 95\% envelopes, for the SLn (left), SLt (middle), and SLcn (right) models.  
%   % The normal probability plot for the Normalized Quantile residual $r_{NQ_i}$ with its envelopes for the SLn (left), SLt (middle), and SLcn (right) models for the Mroz data. 
%   % More dots inside the envelope indicate a better fit.
%   } \label{fig:envelopes2}
% 	\end{center}
% \end{figure}

% Heeju Version
% \begin{figure}[!ht]
% 	\begin{center}
% 		\centering \hspace {1cm}\centering \\
% 		\includegraphics[width=7cm, height=6cm]{../SimPlot/Rplot02.pdf}
% 	\includegraphics[width=7cm, height=6cm]{../SimPlot/Histogram2.pdf}
% 		\caption{
%   Mroz data.
%   Classical residuals versus fitted values (on the left) and histogram of the classical residuals (on the right).
%   Orange and black indicate detected good and bad observations, respectively.
%   The size of the bullets in the left plot is proportional to the probability of being a bad point.
%   } 
%   \label{fig:Mroz data_residual_plot1}
% 	\end{center}
% \end{figure}

\newpage

\section{Conclusions}
\label{sec:6}

This paper introduces a novel SL (SLcn) model based on the contaminated normal distribution.  In contrast with the existing literature, where available optimization procedures are used to compute the ML estimates in SL models \cite{marchenko2012heckman, saulo2023bivariate, ogundimu2016sample}, here an ECM algorithm is developed for the proposed SLcn, following the results given in \cite{lachosHeckman}. Our proposed ECM algorithm uses closed-form expressions at the E-step, which rely on formulas for the mean and variance of a truncated multinormal distribution. 
The general formulas for these moments were derived efficiently by \citet{GalarzaCran}, for which we use the \code{MomTrunc} package in \prog{R}. 
Further, the identifiability of the SLcn model and the link of the SLcn model to the family of extended skew contaminated normal distribution is discussed in detail, given the recent results presented by \citet{morales2022moments}.  
The proposed method (including SLn and SLt models) has been coded and implemented in the \prog{R} \code{HechkmanEM} package, which is available for the users on the CRAN repository.

The analysis of two real data sets provides strong evidence of the usefulness and effectiveness of our proposal when compared with the popular SLn and SLt models. Furthermore, rigorous simulation studies demonstrate the robustness and superior performance of the SLcn model regardless of the data distribution, addressing both model misspecification and missing data. 

Future extensions of the work include the extension of the SLcn model to the multivariate SL setting, as discussed in \citet{tauchmann2010consistency}, or the development of some diagnostics measures for the SLcn \citep{Matos11}. Another possible extension includes the generalization of the SLcn model to the class of scale mixtures of skew-normal (SMSN) distribution \citep{Lachos_Ghosh_Arellano_2009}, this rich family of SMSN distributions, includes some well-known multivariate asymmetric heavy-tailed and symmetric distributions, such as the skew-t \citep{AzzaliniC1999} and the family of scale-mixture of normal distributions \citep{Lange1993}.  An in-depth investigation of such extensions is beyond the scope of the present paper, but
certainly is an interesting topic for future research.

%\bigskip
\section*{Acknowledgments}

Victor Lachos acknowledges the partial financial support from UConn - CLAS's Summer Research Funding Initiative 2023 and Research Excellence Program - UConn. José Alejandro Ordoñez acknowledeges the financial support provided by the Agencia Nacional de Investigación y Desarrollo de Chile (ANID) through Grant FONDECYT 3240170.

%\newpage

\section*{Appendix A}

\noindent{\bf Proof of proposition 4.} Note first that:
\begin{eqnarray*}
f\left(Y_{1i} \mid C_i=1,\x_i,\w_i;\btheta\right)&=&\frac{f(Y_{1i}, C_i=1\mid \x_i,\w_i;\btheta)}{f(C_i=1\mid\w_i;\btheta)}=\frac{f(Y_{1i}\mid \x_i;\btheta)f( C_i=1\mid Y_{1i}, \x_i,\w_i;\btheta)}{f(C_i=1\mid\w_i;\btheta)}\\
&=&\frac{f(Y_{1i}\mid \x_i;\btheta)P( Y_{2i}>0\mid Y_{1i}, \x_i,\w_i;\btheta)}{P(Y_{2i}>0\mid\w_i;\btheta)}.
\end{eqnarray*}
From Subsection \ref{Likelihood_tMLC2}, we have that $Y_{1i}\iid \CN(\x^{\top}_i\bbeta,\sigma^2,\nu_1,\nu_2)$ and $Y_{2i}|Y_{1i}=V_{1i}\iid \CN(\mu_{ti},\sigma^2_{t},\omega_{\nu_{2i}},\nu_2)$, with
\begin{equation*}
\mu_{ti}=\w^{\top}_i\bgamma+\frac{\rho}{\sigma}(V_{1i}-\x^{\top}_i\bbeta),\,\,\sigma^2_{t}=(1-\rho^2),\,\, 
\end{equation*}
\begin{equation*}
\omega_{\nu_{2i}}=\frac{\nu_1\phi(V_{1i}\mid\x^{\top}_i\bbeta,\nu_2^{-1}\sigma^2)}{f^{\text{CN}}(V_{1i}\mid\x^{\top}_i\bbeta,\sigma^2,\nu_1,\nu_2)}=\frac{\nu_1\phi(V_{1i}\mid\x^{\top}_i\bbeta,\nu_2^{-1}\sigma^2)}{\nu_1\phi(V_{1i}\mid\x^{\top}_i\bbeta,\nu_2^{-1}\sigma^2)+(1-\nu_1)\phi(V_{1i}\mid\x^{\top}_i\bbeta,\sigma^2)}.
\end{equation*}
So,
\begin{align*}
f\left(Y_{1i} \mid C_i=1,\x_i,\w_i;\btheta\right)&={f}^{\text{CN}}(V_{1i}\mid\x^{\top}_i\bbeta,\sigma^2,\nu_1,\nu_2){F}^{\text{CN}}(-\infty,0\mid-\mu_{ti},\sigma^2_{t},\omega_{\nu_{2i}},\nu_2) / {F}^{\text{CN}}(-\infty,0\mid-\w^{\top}_i\bgamma,1,\nu_1,\nu_2),\\
&=[\nu_1\phi(V_{1i}\mid \x^{\top}_i\bbeta,\nu_2^{-1}\sigma^2)+(1-\nu_1)\phi(V_{1i}\mid \x^{\top}_i\bbeta,\sigma^2)]\frac{\omega_{\nu_{2i}}\Phi(\sqrt{\nu_2}\displaystyle\frac{\mu_{ti}}{\sigma_{t}})+(1-\omega_{\nu_{2i}})\Phi(\frac{\mu_{ti}}{\sigma_{t}})}{\nu_{1}\Phi(\sqrt{\nu_2}\w^{\top}_i\bgamma)+(1-\nu_{1})\Phi(\w^{\top}_i\bgamma)}\\
&=\displaystyle\frac{\nu_{1}\phi(V_{1i}\mid\x^{\top}_i\bbeta,\nu_2^{-1}\sigma^2)\Phi(\sqrt{\nu_2}\Delta_i)+(1-\nu_{1})\phi(V_{1i}\mid\x^{\top}_i\bbeta,\sigma^2)\Phi(\Delta_i)}{\nu_{1}\Phi(\sqrt{\nu_2}\w^{\top}_i\bgamma)+(1-\nu_{1})\Phi(\w^{\top}_i\bgamma)},
\end{align*}
where $\Delta_i=\mu_{ti}/\sigma_{t}$. 
The proof follows by noting that this density corresponds to the pdf of an univariate $\textrm{ESCN}(\mu,\Sigma,\lambda,\nu_1,\nu_2,\tau)$ (see Eq.~\eqref{denESCN}), with $\mu=\x^{\top}_i\bbeta,\Sigma=\sigma^2,\lambda=\rho/\sqrt{1-\rho^2}$ and $\tau=\w^{\top}_i\bgamma/\sqrt{1-\rho^2}$.\\

\noindent{\bf Proof of proposition 5 (Alternative Proof.)}

Note that,
\begin{equation}\label{expectationobs}
    E\left[\varepsilon_{1 i} C_i\right] = E\left[\varepsilon_{1 i} C_i \mid C_i=1\right] P\left(C_{i} = 1\right) 
=E\left[\varepsilon_{1 i} \mid C_i=1\right] P\left(C_i=1\right).
\end{equation}

We start by finding $\EE[\varepsilon_{1i}C_i]$, then, we isolate  $\EE\left(\varepsilon_{1 i} \mid C_i=1\right)$ from Eq.~\eqref{expectationobs}, and finally, we will obtain equation $\EE\left(Y_{1i}|C_i=1,\x_i,\w_i\right)$ by using well-known expectation properties.
  
We have that
\begin{align*}
\EE\left[\varepsilon_{1 i} C_i\right] &= 
\EE\left[\varepsilon_{1 i} \mid C_i=1\right] P\left(C_i=1\right) \\
&= \EE\left[\varepsilon_{1 i} \mid \varepsilon_{2 i}>-\w_i^{T} \gamma\right] P\left(\varepsilon_{2 i}>-\w_i^{\top} r\right) \\
&=\int_{-\w_i^{\top} r}^{\infty}\left[ \int_{-\infty}^{\infty} \varepsilon_{1 i} g\left(\varepsilon_{1 i} \mid \varepsilon_{2 i}\right) d \varepsilon_{1 i}\right] g\left(\varepsilon_{2 i}\right) d \varepsilon_{2 i}, \\
\end{align*}
where $g(\cdot)$ is the pdf of the univariate contaminated normal distribution. 
Now, given that
\begin{equation*}
\varepsilon_{1 i} \mid \varepsilon_{2 i} \iid \CN\left(\rho \sigma \varepsilon_{2 i}, \sigma^2\left(1-\rho_1^2\right), \omega_1, \nu_2\right),
\end{equation*}
 where $\omega_1=\displaystyle\frac{\nu_1 \phi\left(\varepsilon_{2 i} \nu_2\right) }{\nu_1 \phi\left(\varepsilon\nu_2\right)+\left(1-\nu_1\right) \phi\left(\varepsilon_{2 i}\right)}$, then, 
\begin{equation*}
    \EE\left[\varepsilon_{1 i} C_i\right] =
    \int_{-\bw_i^T\bgamma}^{\infty}
    \left[\int_{-\infty}^{\infty} \varepsilon_{1i}f^{\text{CN}}\left(\varepsilon_{1i} \mid \rho\sigma\varepsilon_{2i},\frac{\sigma^2(1-\rho^2)}{\nu_2},\omega_1,\nu_2\right)d\varepsilon_{1i} \right]f^{\text{CN}}(\varepsilon_{2i}|0,1,\nu_1,\nu_2)d\varepsilon_{2i}.
\end{equation*}
If we define $u_i = \varepsilon_{1i} - \rho\sigma\varepsilon_{2i}$, then $\varepsilon_{1i} = u_i +\rho\sigma\varepsilon_{2i}$ and 
\begin{equation*}
    \EE[\varepsilon_{1 i} C_i] =
    \int_{-\bw_i^T\bgamma}^{\infty}\left[\int_{-\infty}^{\infty} \left(u_i + \rho\sigma\varepsilon_{2i}\right) f^{\text{CN}}\left(u_{i} \mid 0,\frac{\sigma^2(1-\rho^2)}{\nu_2},\nu_1,\nu_2\right)d\varepsilon_{1i}\right] f^{\text{CN}}(\varepsilon_{2i}|0,1,\nu_1,\nu_2)d\varepsilon_{2i}.
\end{equation*}

Since 
\begin{align*}
    \int_{-\infty}^{\infty} \left(u_i + \rho\sigma\varepsilon_{2i}\right) f^{\text{CN}}\left(u_{i} \mid 0,\frac{\sigma^2(1-\rho^2)}{\nu_2},\nu_1,\nu_2\right)d\varepsilon_{1i}  = & 
    \int_{-\infty}^{\infty} u_if^{\text{CN}}\left(u_{i} \mid 0,\frac{\sigma^2(1-\rho^2)}{\nu_2},\nu_1,\nu_2\right)d\varepsilon_{1i} + \\
   & \int_{-\infty}^{\infty}  \rho\sigma\varepsilon_{2i} f^{\text{CN}}\left(u_{i} \mid 0,\frac{\sigma^2(1-\rho^2)}{\nu_2},\nu_1,\nu_2\right)d\varepsilon_{1i}\\
    =  & 0 + \rho\sigma\varepsilon_{2i}= \rho\sigma\varepsilon_{2i},
\end{align*}
then
\begin{align*}
    \EE[\varepsilon_{1i}C_i] &= \int_{-\w_i^{\top}\bgamma}^{\infty} \rho\sigma\varepsilon_{2i}f^{\text{CN}}(\varepsilon_{2i}|0,1,\nu_1,\nu_2)d\varepsilon_{2i} \\&= \int_{-\w_i^{\top}\bgamma}^{\infty}\rho\sigma\varepsilon_{2i}\nu_1\phi(\varepsilon_{2i}\mid 0,\frac{1}{\nu_2})d\varepsilon_{2i}  + \int_{-\w_i^{\top}\bgamma}^{\infty}\rho\sigma\varepsilon_{2i}(1-\nu_1)\phi(\varepsilon_{2i})d\varepsilon_{2i}\\
    &= I_1 + I_2.
\end{align*}
Now, by substituting $u = \varepsilon^2_{2i}\nu_2$, $du = 2\varepsilon_{2i}\nu_2d\varepsilon_{2i}$, we have 
\begin{align*}
    \int \varepsilon_{2i}\phi(\varepsilon_{2i}\mid 0,\frac{1}{\nu_2})d\varepsilon_{2i} &= \int \varepsilon_{2i}\exp\left(-\frac{\varepsilon^2_{2i}\nu_2}{2}\right)d\varepsilon_{2i} \\&= 
   \sqrt{\frac{\nu_2}{2\pi}} \frac{1}{2\nu_2}\int \exp\left(-\frac{u}{2}\right)du = -\sqrt{\frac{\nu_2}{2\pi}} \frac{1}{2\nu_2} 2\exp\left(-\frac{u}{2}\right)+ c\\
   & = -\frac{1}{\nu_2}\sqrt{\frac{\nu_2}{2\pi}}\exp\left(-\frac{\varepsilon^2_{2i}\nu_2}{2}\right), 
\end{align*}
then
\begin{align*}
    \int_{-\w_i^{\top}\bgamma}^{\infty}\varepsilon_{2i}\phi(\varepsilon_{2i}\mid 0,\frac{1}{\nu_2})d\varepsilon_{2i} &= -\left.\frac{1}{\nu_2}\sqrt{\frac{\nu_2}{2\pi}}\exp\left(-\frac{\varepsilon^2_{2i}\nu_2}{2}\right)  \right|_{-\w_i^\top\bgamma}^\infty =  \frac{1}{\nu_2}\phi\left(\w_i^{\top}\bgamma\mid 0,\frac{1}{v2}\right) = \frac{1}{\sqrt{\nu_2}}\phi\left(\w_i^{\top}\bgamma\sqrt{\nu_2}\right). 
\end{align*}

As a particular case, when $\nu_2 =1$, we have that
\begin{eqnarray*}
\int_{-\w_i^{\top}\bgamma}^{\infty}\varepsilon_{2i}\phi(\varepsilon_{2i})d\varepsilon_{2i} = \phi\left(\w_i^{\top}\bgamma\right),
\end{eqnarray*}
then
\begin{equation*}
    \EE[\varepsilon_{1i}C_i] = I_1 + I_2 = \nu_1\displaystyle\frac{\rho\sigma}{\sqrt{\nu_2}}\phi\left(\w_i^{\top}\bgamma\sqrt{\nu_2}\right) + (1-\nu_1)\rho\sigma\phi\left(\w_i^{\top}\bgamma\right).
\end{equation*}

Using the symmetry of the contaminated normal distribution, we have that $P(C_i = 1) = {F}^{\text{CN}}(-\infty,0|-\w^{\top}_i\bgamma,1,\nu_1,\nu_2)$, then, isolating from Eq.~ \eqref{expectationobs},
\begin{equation*}
\EE[\varepsilon_{1i}\mid C_i = 1]= \displaystyle\frac{\EE[\varepsilon_{1i}C_i]}{P(C_i = 1)} = \displaystyle\frac{\nu_1\displaystyle\frac{\rho\sigma}{\sqrt{\nu_2}}\phi\left(\w_i^{\top}\bgamma\sqrt{\nu_2}\right) + (1-\nu_1)\rho\sigma\phi\left(\w_i^{\top}\bgamma\right)}{{F}^{\text{CN}}(-\infty,0|-\w^{\top}_i\bgamma,1,\nu_1,\nu_2)}.
\end{equation*}

The expectation operator properties imply that
\begin{equation*}
    \EE\left[Y_{1i}|C_i=1,\x_i,\w_i\right] = \x_i^{\top}\bbeta + \displaystyle\frac{\nu_1\displaystyle\frac{\rho\sigma}{\sqrt{\nu_2}}\phi\left(\w_i^{\top}\bgamma\sqrt{\nu_2}\right) + (1-\nu_1)\rho\sigma\phi\left(\w_i^{\top}\bgamma\right)}{{F}^{\text{CN}}(-\infty,0|-\w^{\top}_i\bgamma,1,\nu_1,\nu_2)}  = \x_i^{\top}\bbeta + \rho\sigma\lambda^{CN}(\w_i^{\top}\bgamma).
\end{equation*}

For the second part of the proposition, we have that
\begin{align*}
    \frac{\partial \lambda^{CN}(x)}{\partial x} &= \frac{\widetilde{F}(\nu_1,\nu_2)\left[-\nu_1 x \sqrt{\nu_2}\phi(\sqrt{\nu_2}x) - (1-\nu_1)x\phi(x)\right] - \widetilde{F}(\nu_1,\nu_2)f^{\text{CN}}(x \mid 0,1,\nu_1,\nu_2)\lambda^{CN}(x)}{\left(\widetilde{F}(\nu_1,\nu_2)\right)^2}\\
    &=- \frac{f^{\text{CN}}(x \mid 0,1,\nu_1,\nu_2)}{\widetilde{F}(\nu_1,\nu_2)}x -\frac{f^{\text{CN}}(x \mid 0,1,\nu_1,\nu_2)}{\widetilde{F}(\nu_1,\nu_2)}\lambda^{CN}(x) \\&= - \frac{f^{\text{CN}}(x \mid 0,1,\nu_1,\nu_2)}{\widetilde{F}(\nu_1,\nu_2)}\left[x+\lambda^{CN}(x)\right].
\end{align*}
where $\widetilde{F}(\nu_1,\nu_2)=F^{\text{CN}}(-\infty,0\mid 0,1,\nu_1,\nu_2)$.

Since $\displaystyle\frac{\partial \x_{i}^{\top}\bbeta}{\partial x_{ik}} = \beta_k$, then
\begin{eqnarray*}
\frac{\partial\EE\left(Y_{1i}|C_i=1,\x_i,\w_i\right)}{\partial x_{ik} } =  \beta_k+\rho\sigma\lambda^{\prime CN}(\w_i^{\top}\bgamma).
\end{eqnarray*}

%$$\hspace{14cm}\square$$  

\noindent{\bf Proof of Theorem 1.}\\

Since the contaminated normal density is a particular case of a mixture of normal densities, we will verify that conditions 1 2, and 3 in theorem 1 hold for the sample mechanism, which also comes from the contaminated normal distribution. 

Note that for {\bf Condition 1}, we have:
\begin{equation}\label{cnorm1}
  \lim _{z \rightarrow-\infty} \displaystyle\frac{F(z)}{e^{\delta z}}  =  \lim _{z \rightarrow-\infty}\frac{\nu_1\Phi(z \mid 0,\nu_2^{-1}) + (1-\nu_1)\Phi(z)}{e^{\delta z}} = \nu_1\lim _{z \rightarrow-\infty}\frac{\Phi(z \mid 0,\nu_2^{-1})}{e^{\delta z}} + (1-\nu_1)\lim _{z \rightarrow-\infty}\frac{\Phi(z)}{e^{\delta z}}.
\end{equation}

On the other hand, {\bf Condition 2} implies that
\begin{align}\label{cnorm2}
   \nonumber &\lim _{z \rightarrow+\infty}
   \frac{F\left(\theta_0+\theta_1 z\right)-F(z)}{e^{-\delta_1 z^2-\delta_2 z}}  
    \\&= \lim _{z \rightarrow+\infty}\frac{\left[\nu_1\Phi(\theta_0 + \theta_1z \mid 0,\nu_2^{-1}) + (1-\nu_1)\Phi(\theta_0 + \theta_1z )\right] - \left[\nu_1\Phi(z \mid 0,\nu_2^{-1}) + (1-\nu_1)\Phi(z )\right] }{e^{-\delta_1 z^2-\delta_2 z}}\nonumber\\
       &= \nu_1 \lim_{z \rightarrow+\infty}
       \frac{\Phi(\theta_0 + \theta_1z \mid 0,\nu_2^{-1}) - \Phi(z \mid 0,\nu_2^{-1})}{e^{-\delta_1 z^2-\delta_2 z}} + (1-\nu_1)\lim_{z \rightarrow+\infty}
       \frac{\Phi(\theta_0 + \theta_1z ) - \Phi(z)}{e^{-\delta_1 z^2-\delta_2 z}}.
\end{align}

Regarding {\bf Condition 3}, the first limit can be written as
\begin{eqnarray}\label{cnorm3}
   && \nu_1 \lim_{z \rightarrow+\infty}
      z^M \left [ \Phi(\theta_0 + \theta_1z \mid 0,\nu_2^{-1}) - \Phi(z \mid 0,\nu_2^{-1})\right ] + (1-\nu_1)\lim_{z \rightarrow+\infty}
       z^M\left [\Phi(\theta_0 + \theta_1z ) - \Phi(z )\right ]. %\iff \\
       %&& \nu_1 \lim_{z \rightarrow+\infty}
      %z^M \left [ \sqrt{\nu_2}\Phi(\sqrt{\nu_2}\theta_0 + \sqrt{\nu_2}\theta_1z ) - \Phi(z \mid 0,\nu_2^{-1})\right ] + (1-\nu_1)\lim_{z \rightarrow+\infty}
       %z^M\left [\frac{1}{\sqrt{\nu_2}}\Phi\left(\frac{\theta_0}{\sqrt{\nu_2}} + \frac{\theta_1}{\sqrt{\nu_2}}z \mid 0, \nu_2^{-1}\right) - \frac{1}{\sqrt{\nu_2}}\Phi\left( \frac{z}{\sqrt{\nu_2}} \mid 0, \nu_2^{-1}\right) \right ] \nonumber \iff \\
       %&& \frac{\nu_1}{\sqrt{\nu_2}^{M-%1}} \lim_{\tilde{z}\to\infty} \tilde{z}\left[\Phi(\tilde{\theta_0} + \theta_1\tilde{z}) - \Phi(\tilde{z})\right] +
       %(1-\nu_1)\sqrt{\nu_2}^{M-1}\lim_{z^{*M}\to \infty} z^{*M}\left[\Phi\left(\theta_0^{*} + \theta_1z^{*}\mid 0, \nu_2^{-1}\right) - \Phi(z^{*}\mid0,\nu_2^{-1})\right] 
\end{eqnarray}

Concerning the second limit, first note that
\begin{eqnarray*}
  \lim_{z \to -\infty} \frac{\Phi(z \mid 0, \nu_2^{-1})}{\Phi(z)} = \lim_{z\to - \infty}  \sqrt{\nu_2} e^{-\frac{z^2}{2}\left(\nu_2-1\right)} = \begin{cases} 
          1, & \nu_2 = 1\\
          +\infty, & 0 < \nu_2< 1 \\
  \end{cases}.
\end{eqnarray*}

This implies that $\Phi(z|0, \nu_2^{-1})>\Phi(z)$ when $z \to -\infty$. 
Then, we have
$$
\Phi(z) \leq \nu_1\Phi(z\mid 0, \nu_2^{-1}) + (1-\nu_1)\Phi(z) \leq \Phi(z\mid 0, \nu_2^{-1}) \iff$$$$ \frac{1}{\Phi(z \mid 0, \nu_2^{-1})} \leq \frac{1}{\nu_1\Phi(z\mid 0,\nu_2^{-1}) + (1-\nu_1)\Phi(z)} \leq \frac{1}{\Phi(z)}.
$$

Multiplying by $\nu_1\Phi(\theta_0 + \theta_1 z \mid 0, \nu_2^{-1}) + (1-\nu_1)\Phi(\theta_0 + \theta_1 z)$, we have
\begin{eqnarray*}
\frac{\nu_1\Phi(\theta_0 + \theta_1 z \mid 0, \nu_2^{-1}) + (1-\nu_1)\Phi(\theta_0 + \theta_1 z)}{\Phi(z \mid 0, \nu_2^{-1})} &\leq& \frac{\nu_1\Phi(\theta_0 + \theta_1 z \mid 0, \nu_2^{-1}) + (1-\nu_1)\Phi(\theta_0 + \theta_1 z)}{\nu_1\Phi(z\mid 0, \nu_2^{-1}) + (1-\nu_1)\Phi(z)} \\ &\leq& \frac{\nu_1\Phi(\theta_0 + \theta_1 z \mid 0,\nu_2^{-1}) + (1-\nu_1)\Phi(\theta_0 + \theta_1 z)}{\Phi(z)}
\end{eqnarray*}
This implies that
\begin{eqnarray*}
&& \frac{\Phi(\theta_0 + \theta_1 z )}{\Phi(z \mid 0, \nu_2^{-1})} \leq \frac{\nu_1\Phi(\theta_0 + \theta_1 z \mid 0,\nu_2^{-1}) + (1-\nu_1)\Phi(\theta_0 + \theta_1 z)}{\nu_1\Phi(z \mid 0, \nu_2^{-1}) + (1-\nu_1)\Phi(z)} \leq \frac{\Phi(\theta_0 + \theta_1 z \mid 0, \nu_2^{-1})}{\Phi(z)}. 
\end{eqnarray*}

Given that $0 \leq\nu_2\leq 1$, this also implies that
\begin{align*}
& \frac{\sqrt{\nu_2}\Phi(\theta_0 + \theta_1 z)}{\Phi(z \mid 0,  \nu_2^{-1})} \leq \frac{\nu_1\Phi(\theta_0 + \theta_1 z \mid 0, \nu_2^{-1}) + (1-\nu_1)\Phi(\theta_0 + \theta_1 z)}{\nu_1\Phi(z \mid 0, \nu_2^{-1}) + (1-\nu_1)\Phi(z)} \leq  \frac{\Phi(\theta_0 + \theta_1 z \mid 0, \nu_2^{-1})}{\sqrt{\nu_2}\Phi(z)} \iff \\
& \frac{\Phi\left(\frac{\theta_0}{\sqrt{\nu_2}} + \frac{\theta_1}{\sqrt{\nu_2}}z \mid 0, \nu_2^{-1}\right)}{\Phi(z \mid 0,  \nu_2^{-1})} \leq \frac{\nu_1\Phi(\theta_0 + \theta_1 z \mid 0, \nu_2^{-1}) + (1-\nu_1)\Phi(\theta_0 + \theta_1 z)}{\nu_1\Phi(z \mid 0, \nu_2^{-1}) + (1-\nu_1)\Phi(z)} \leq  \sqrt{\nu_2}\frac{\Phi\left(\sqrt{\nu_2}\theta_0 + \sqrt{\nu_2}\theta_1 z \right)}{\sqrt{\nu_2}\Phi(z)} \iff \\
& \frac{\Phi\left(\theta^*_0 + \theta^*_1z \mid 0, \nu_2^{-1}\right)}{\Phi(z \mid 0,  \nu_2^{-1})} \leq \frac{\nu_1\Phi(\theta_0 + \theta_1 z \mid 0, \nu_2^{-1}) + (1-\nu_1)\Phi(\theta_0 + \theta_1 z)}{\nu_1\Phi(z \mid 0, \nu_2^{-1}) + (1-\nu_1)\Phi(z)} \leq  \frac{\Phi\left(\tilde{\theta}_0 + \tilde{\theta}_1 z \right)}{\Phi(z)} \iff \\
& 1-\frac{\Phi\left(\theta^*_0 + \theta^*_1z \mid 0, \nu_2^{-1}\right)}{\Phi(z \mid 0,  \nu_2^{-1})} \geq 1-\frac{\nu_1\Phi(\theta_0 + \theta_1 z \mid 0, \nu_2^{-1}) + (1-\nu_1)\Phi(\theta_0 + \theta_1 z)}{\nu_1\Phi(z \mid 0, \nu_2^{-1}) + (1-\nu_1)\Phi(z)} \geq  1-\frac{\Phi\left(\tilde{\theta}_0 + \tilde{\theta}_1 z \right)}{\Phi(z)}.
\end{align*}

Therefore, if $M$ is even, we have
\begin{align}\label{cond321}
 \nonumber \lim_{z\to-\infty}z^{M}\left( 1-\frac{\Phi\left(\theta^*_0 + \theta^*_1z \mid 0, \nu_2^{-1}\right)}{\Phi(z \mid 0,  \nu_2^{-1})} \right)
 &\geq 
 \lim_{z\to-\infty}
 z^{M}\left(1-\frac{\nu_1\Phi(\theta_0 + \theta_1 z \mid 0, \nu_2^{-1}) + (1-\nu_1)\Phi(\theta_0 + \theta_1 z)}{\nu_1\Phi(z \mid 0, \nu_2^{-1}) + (1-\nu_1)\Phi(z)} \right) 
 \\&\geq 
 \lim_{z\to-\infty} z^{M}\left(1-\frac{\Phi\left(\tilde{\theta}_0 + \tilde{\theta}_1 z \right)}{\Phi(z)}\right),  
\end{align}
while if $M$ is odd, we have
\begin{align}\label{cond322}
 \nonumber \lim_{z\to-\infty} z^{M}\left( 1-\frac{\Phi\left(\theta^*_0 + \theta^*_1z \mid 0, \nu_2^{-1}\right)}{\Phi(z \mid 0,  \nu_2^{-1})} \right)
 &\leq 
 \lim_{z\to-\infty}z^{M}\left(1-\frac{\nu_1\Phi(\theta_0 + \theta_1 z \mid 0, \nu_2^{-1}) + (1-\nu_1)\Phi(\theta_0 + \theta_1 z)}{\nu_1\Phi(z \mid 0, \nu_2^{-1}) + (1-\nu_1)\Phi(z)} \right) 
 \\&\leq 
 \lim_{z\to-\infty}z^{M}\left(1-\frac{\Phi\left(\tilde{\theta}_0 + \tilde{\theta}_1 z \right)}{\Phi(z)}\right).  
\end{align}

The sum of limits in equations \eqref{cnorm1} and \eqref{cnorm2} show that conditions 1 and 2 hold for the contaminated normal distribution, if and only if they hold for the normal distribution. Note that Eq.~\eqref{cnorm3} shows that the first limit in Condition 3, can also be written as a linear combination of two limits representing this condition for the normal distribution. 
On the other hand, $\theta_0$ and $\theta_1$ are arbitrary (besides the fact that $\theta_1>0$, which is also true for $\theta_1^*$ and $\tilde{\theta_1}$). 
Therefore, as observed in equations \eqref{cond321} and \eqref{cond322}, the second limit in Condition 3, for the contaminated normal distribution, is limited by two limits that represent the same condition for the normal distribution. 
Consequently, by theorems 4 and 6 in \citet{wang2016iden} (which also hold when including covariates), the SLcn model is identifiable.

%$$\hspace{16cm}\square$$  

\newpage
\section*{Appendix B}

\begin{small}
Additional results of the simulation study discussed in Section~\ref{sec:Simulation study} with $n=250$, $500$, and $1,000$ when varying the missing rate as $10\%$, $25\%$, and $50\%$. 
\begin{figure}[!htb]
	\begin{center}
	\includegraphics[scale=0.25]{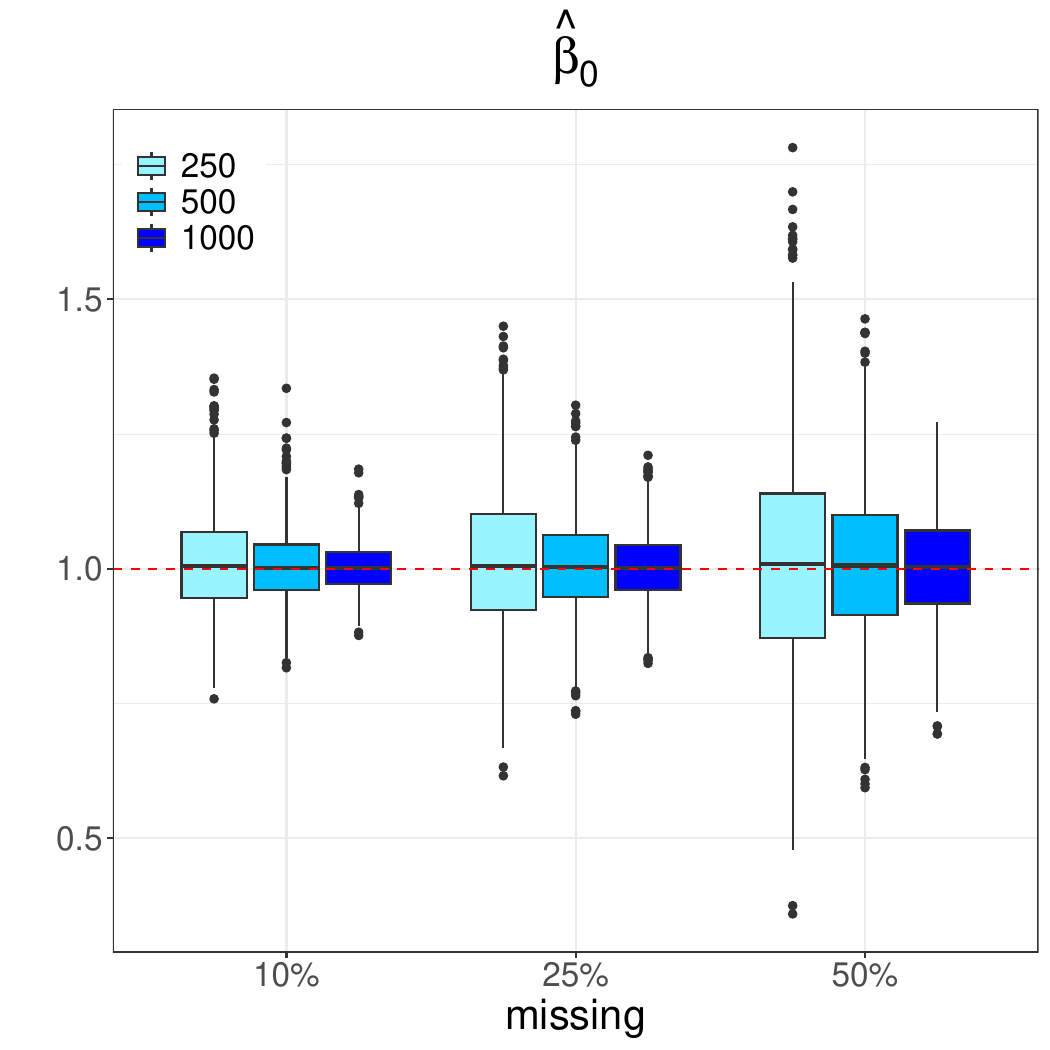}
  	\includegraphics[scale=0.25]{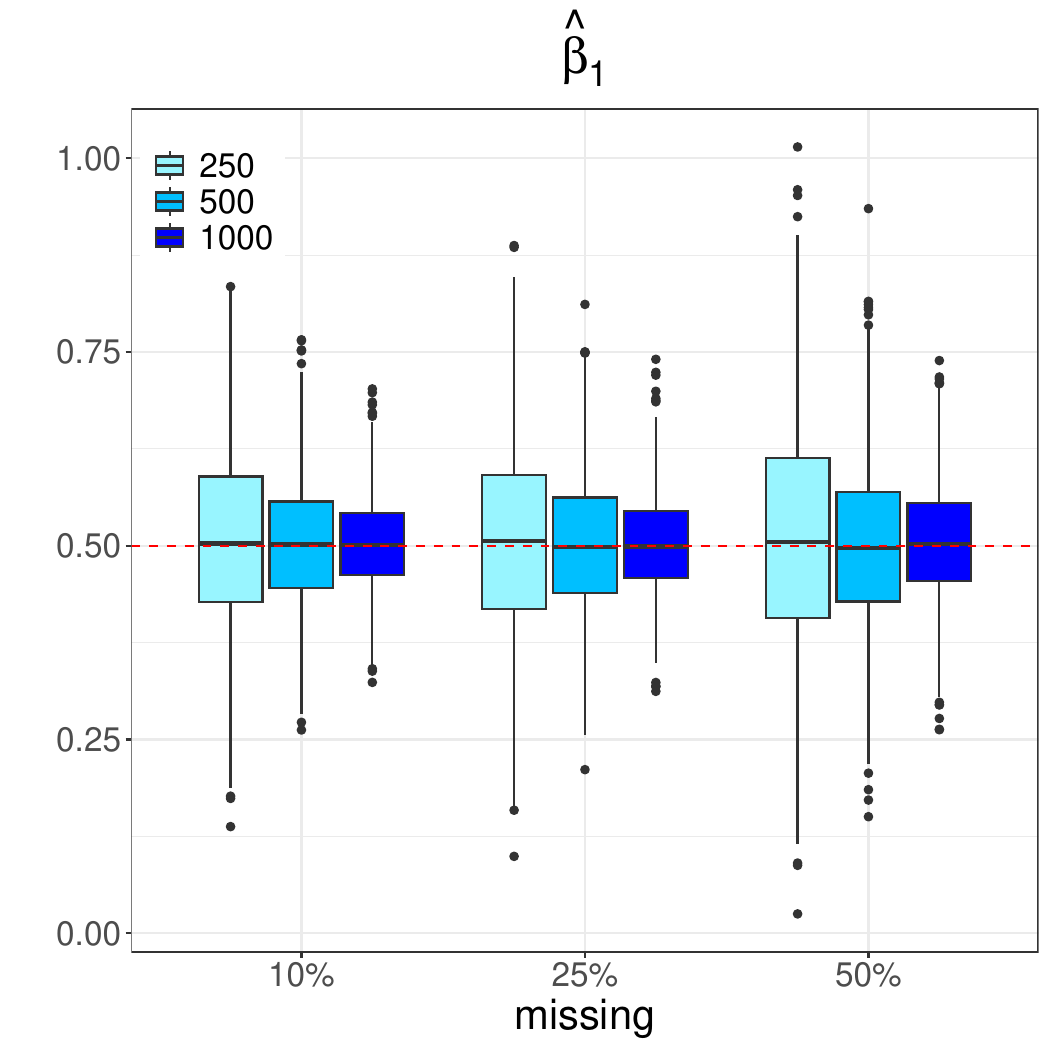}
   	\includegraphics[scale=0.25]{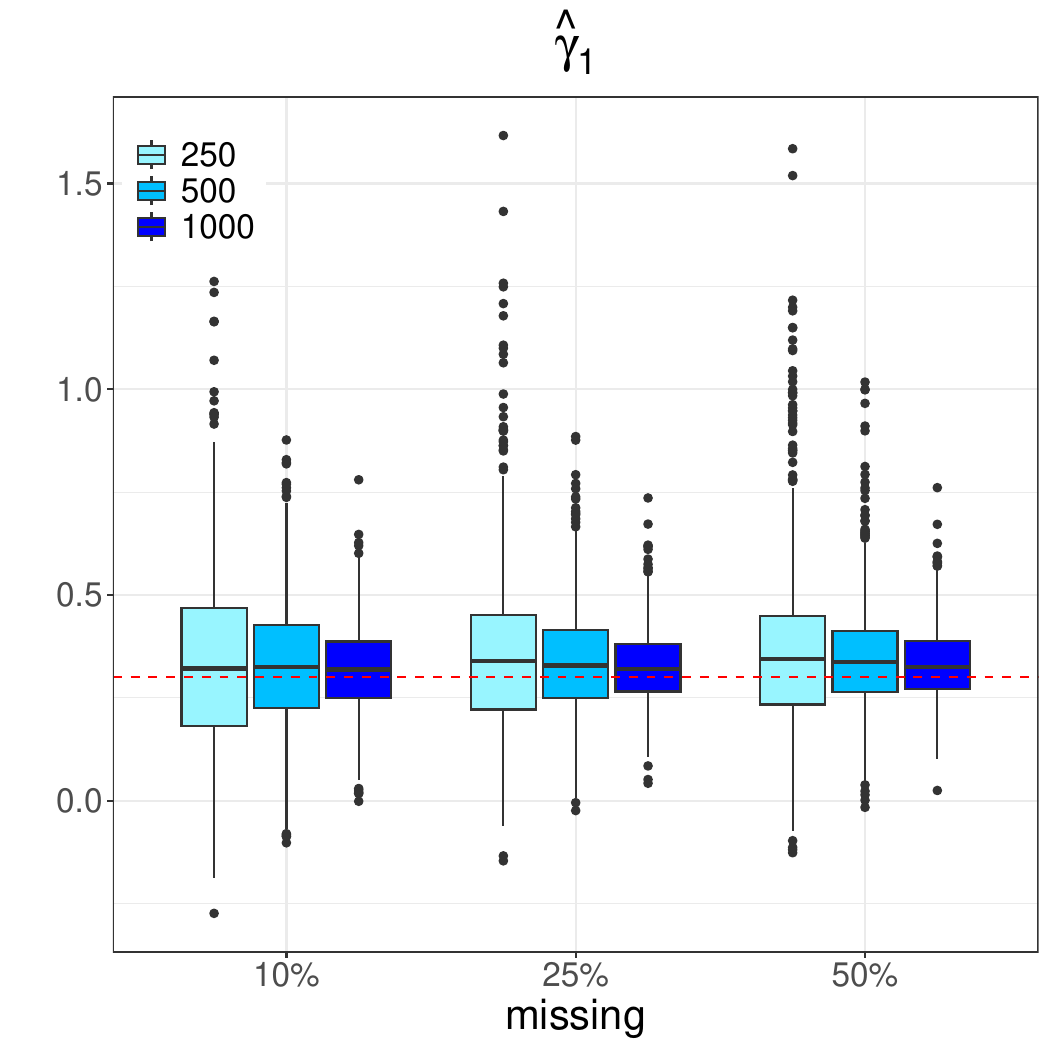}
	\includegraphics[scale=0.25]{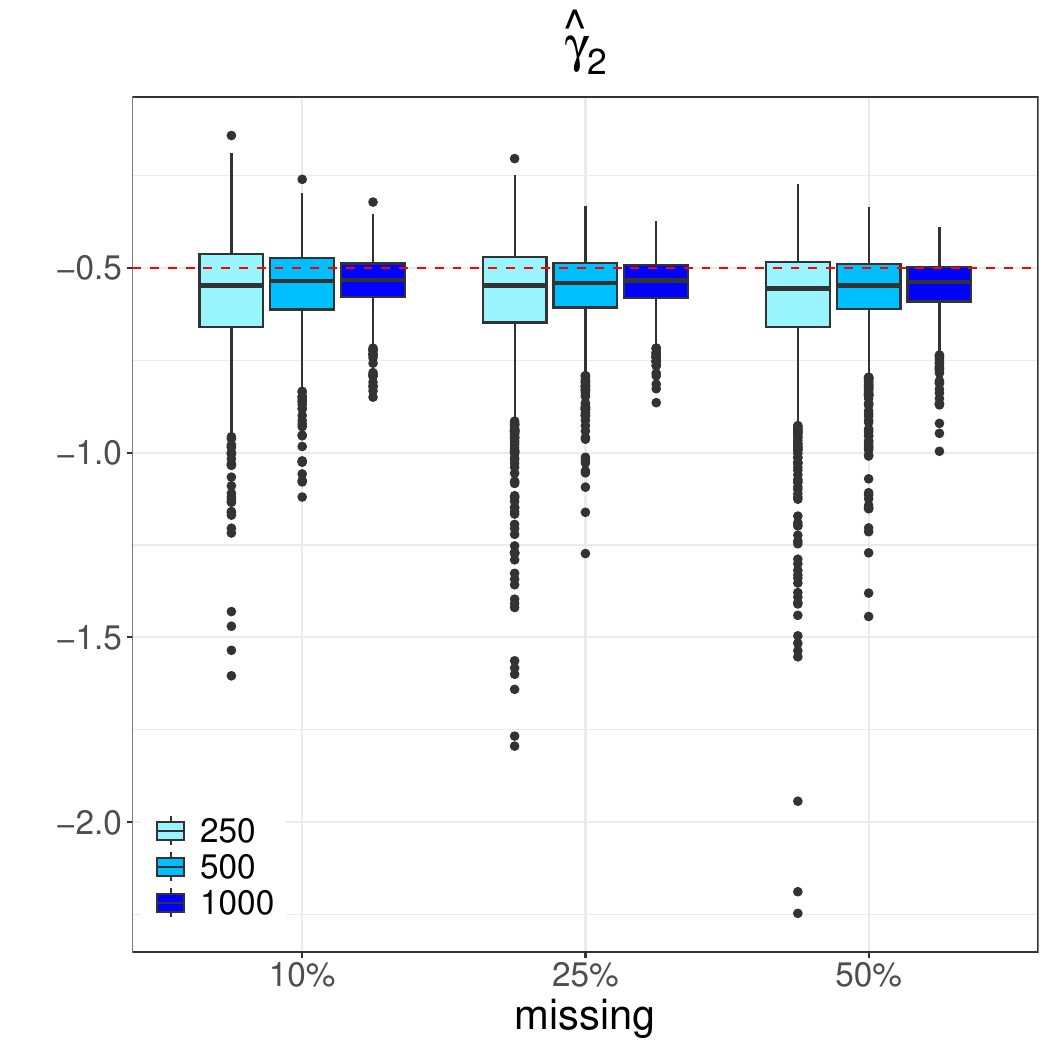}
  	\includegraphics[scale=0.25]{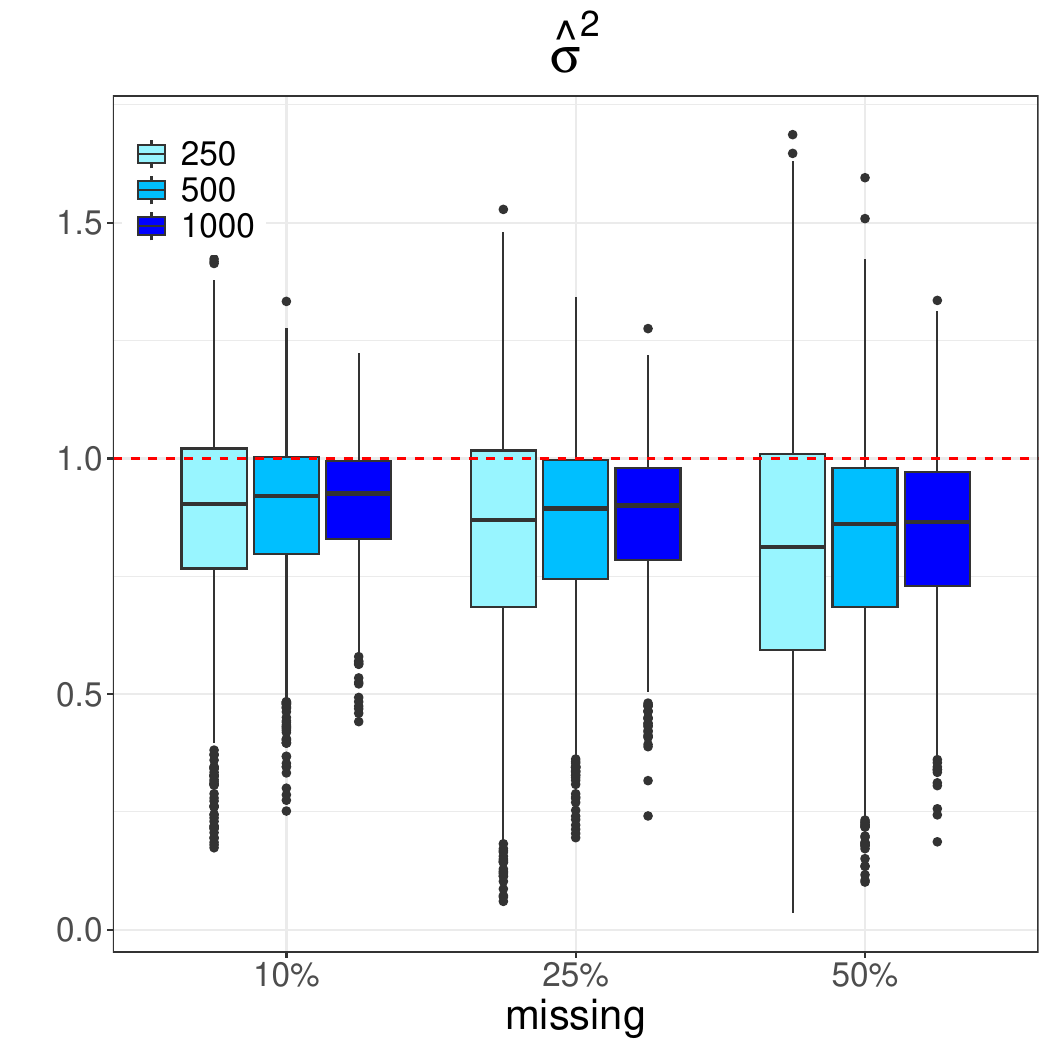}
        \includegraphics[scale=0.25]{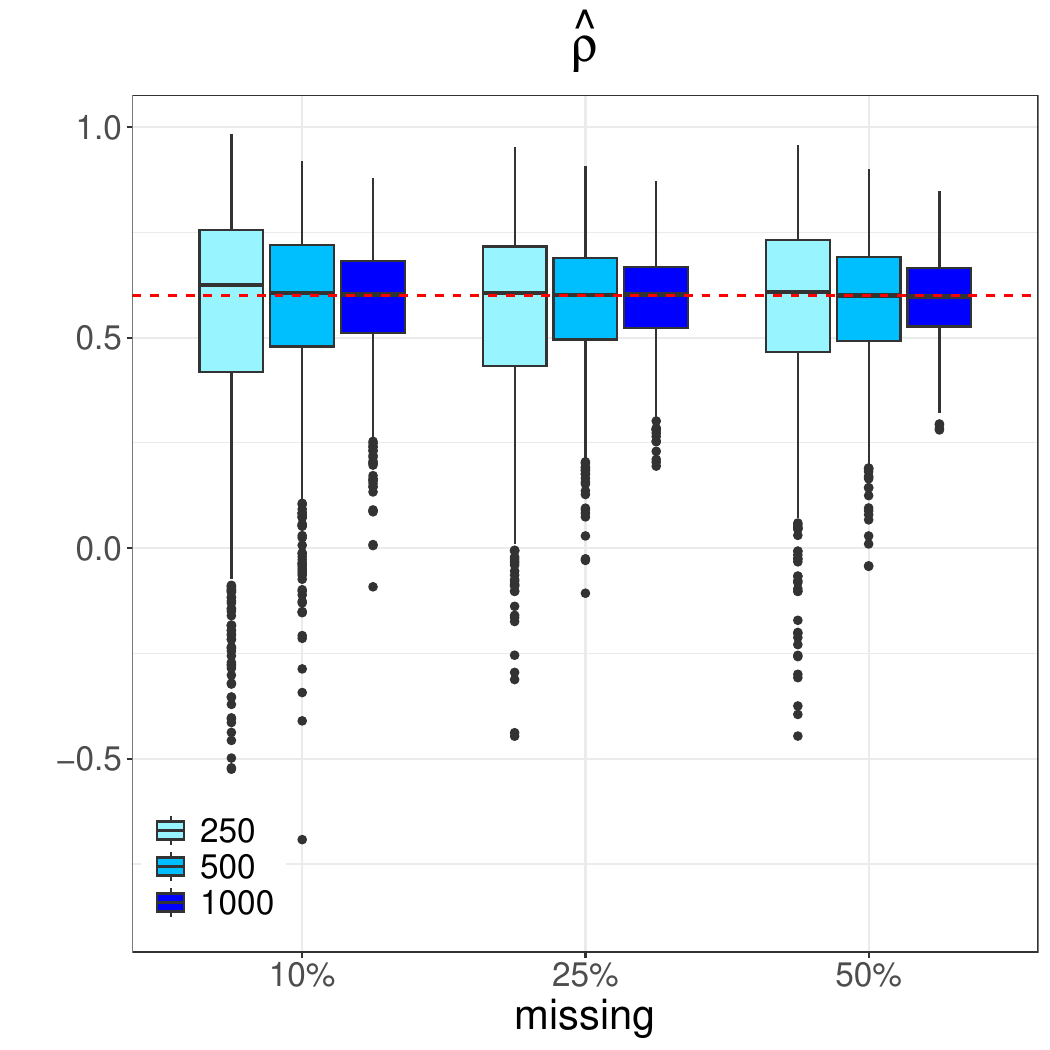}
  \caption{Boxplot of the 1000 Monte Carlo estimates of $\beta_0$, $\beta_1$, $\gamma_1$, $\gamma_2$, $\sigma$ and $\rho$ for the SLcn model, when data are generated from the normal distribution, with $n=250$, $500$, and $1000$ as sample sizes, when varying the missing proportion as $10\%$, $25\%$ and $50\%$.}
    \label{app1}
\end{center}
\end{figure}

\begin{figure}[!htb]
	\begin{center}
	\includegraphics[scale=0.25]{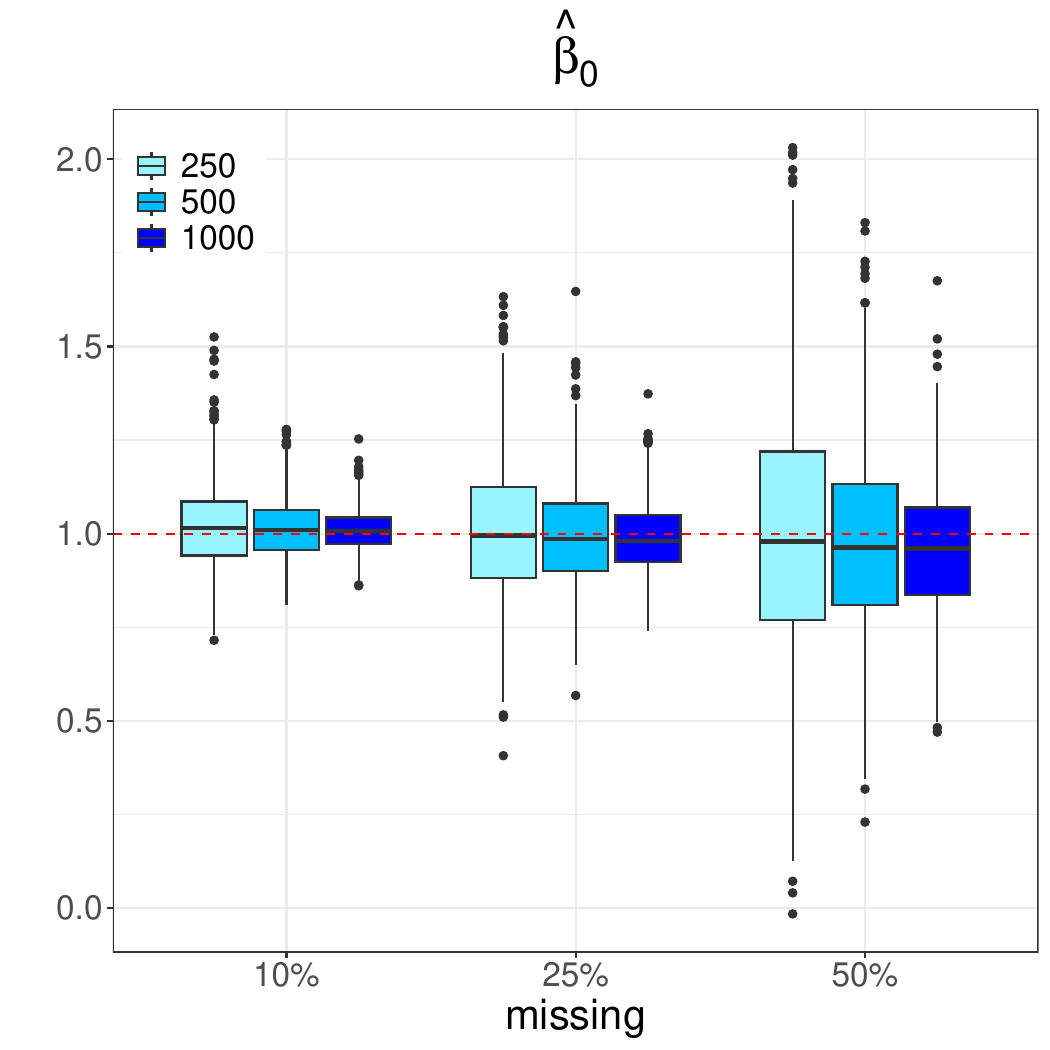}
  	\includegraphics[scale=0.25]{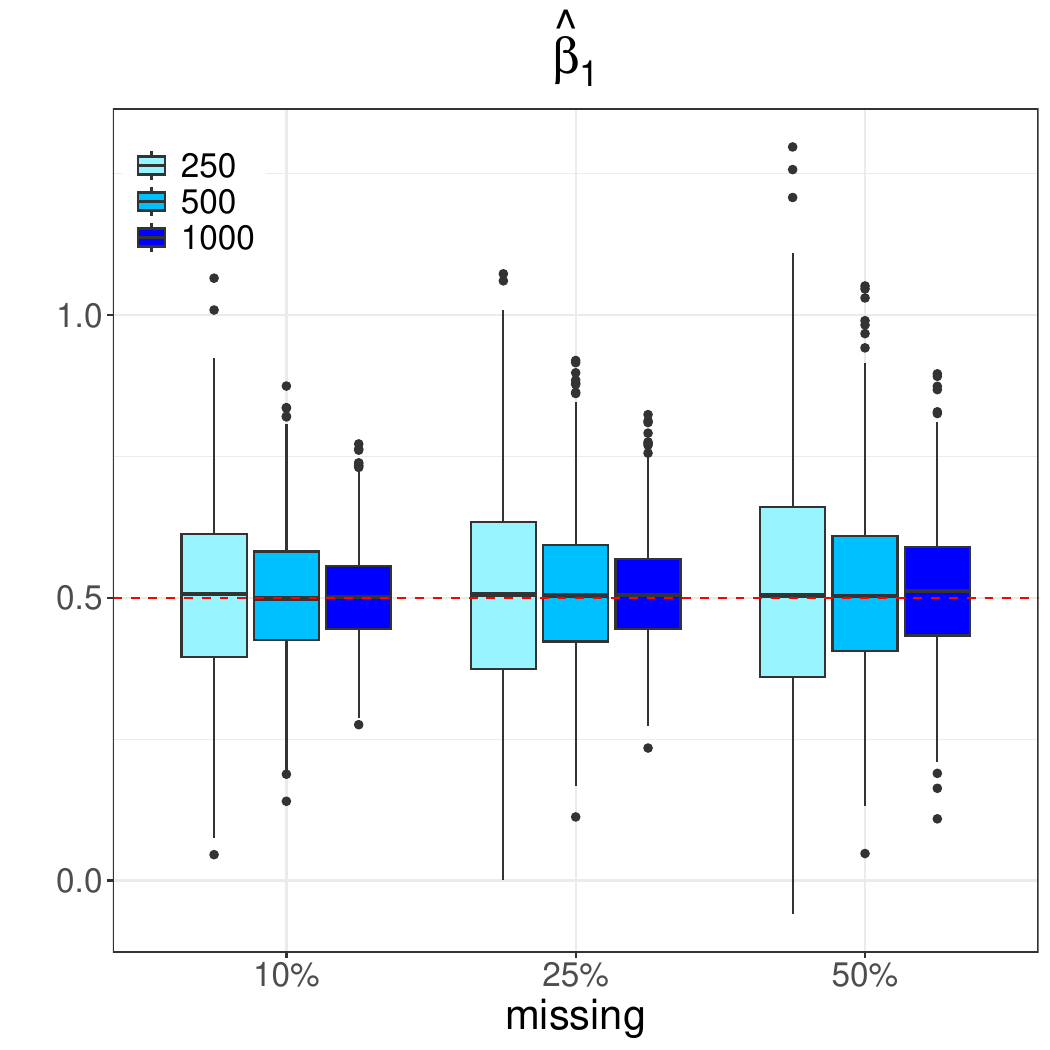}
   	\includegraphics[scale=0.25]{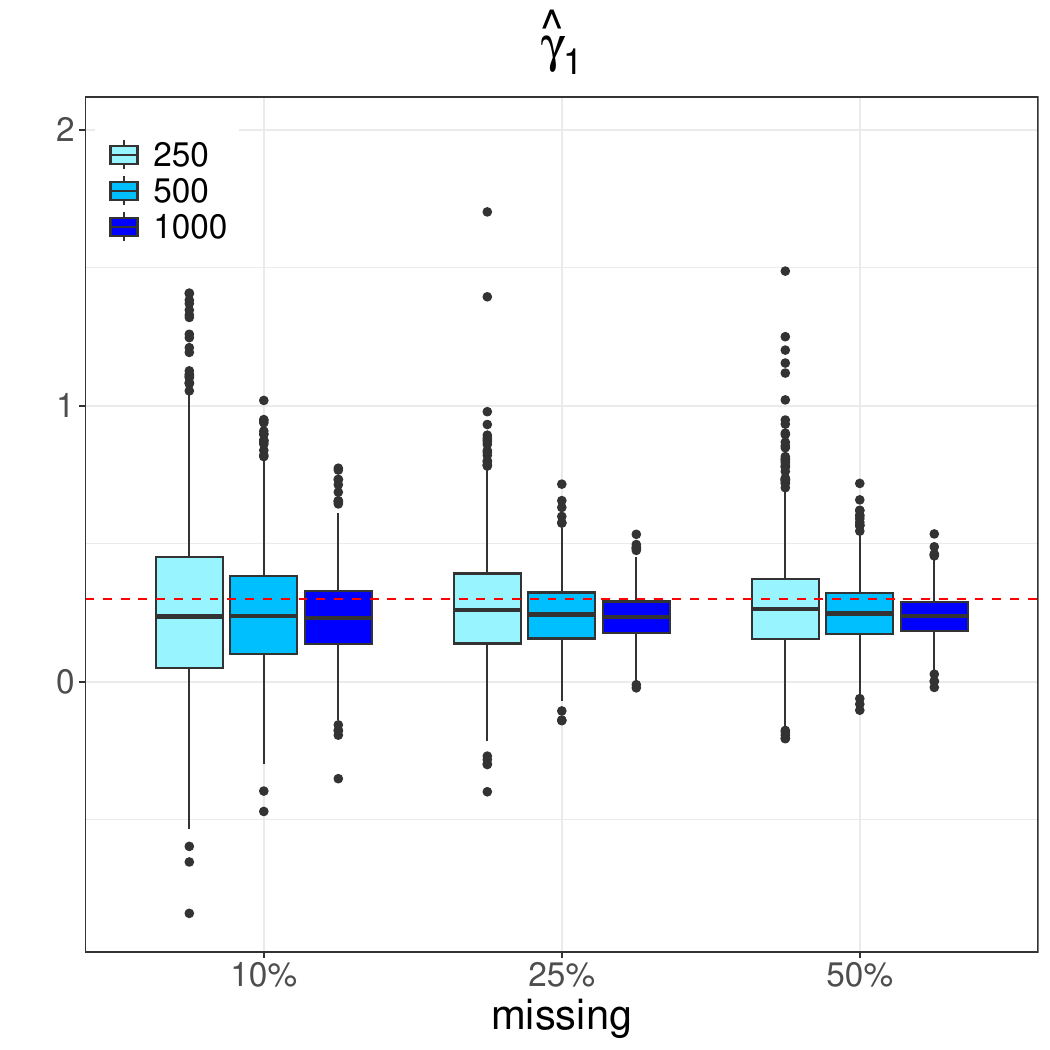}
        \includegraphics[scale=0.25]{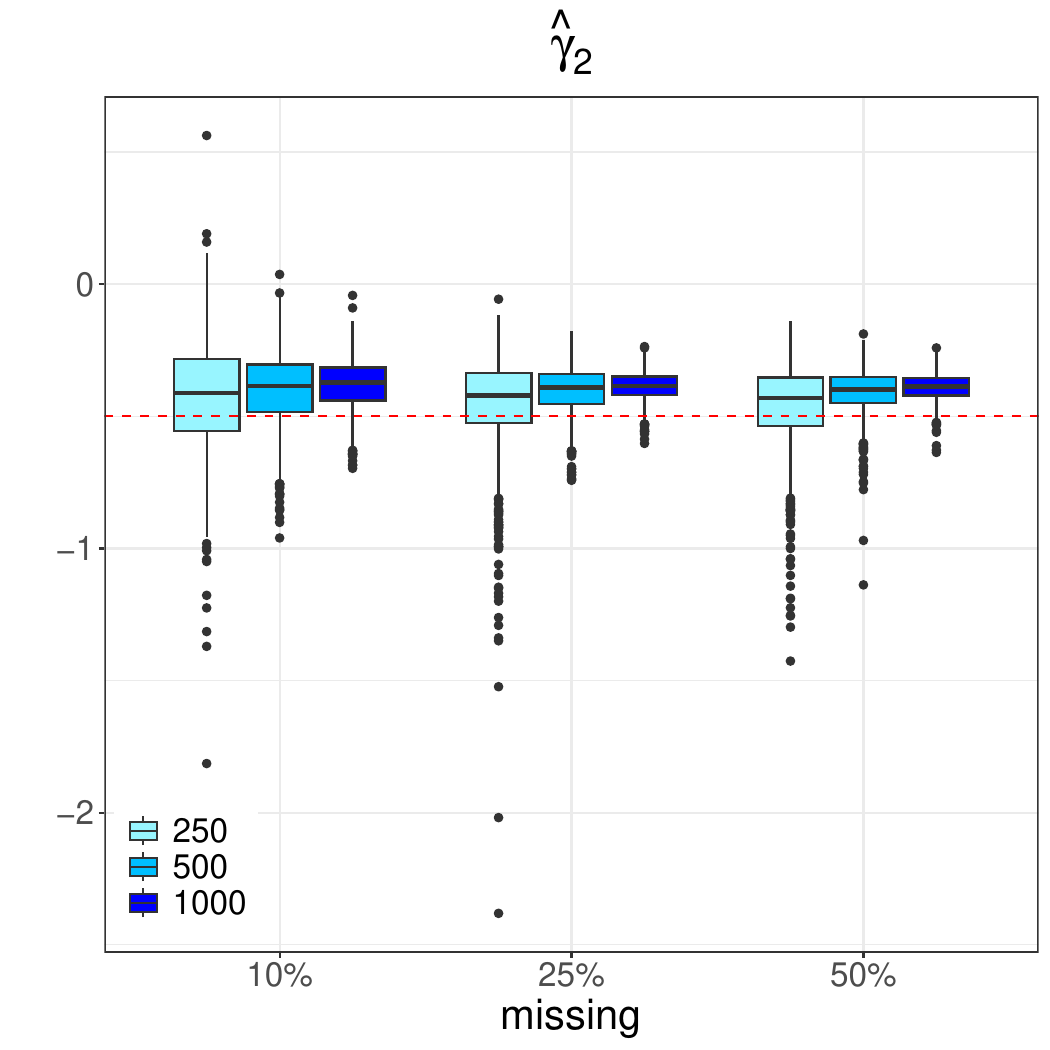}
  	\includegraphics[scale=0.25]{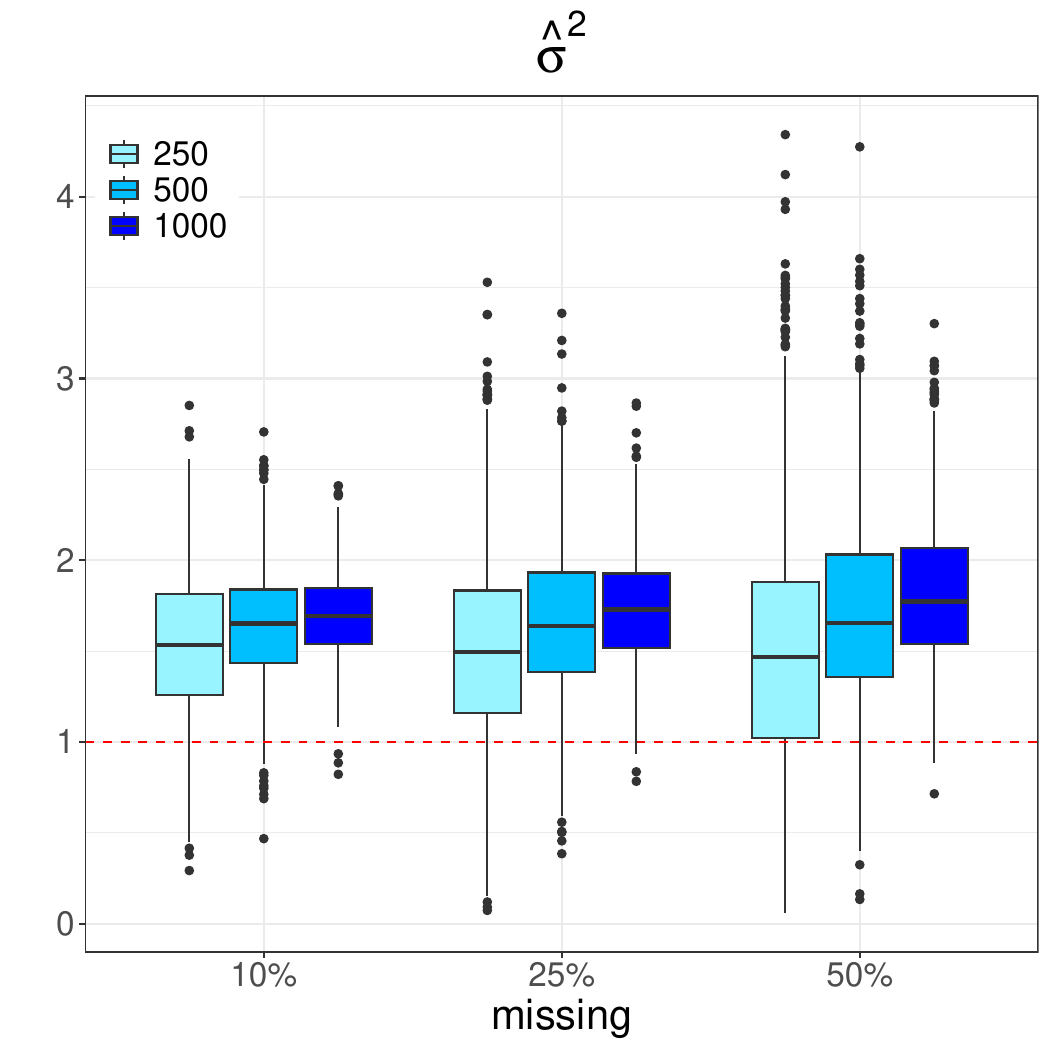}
        \includegraphics[scale=0.25]{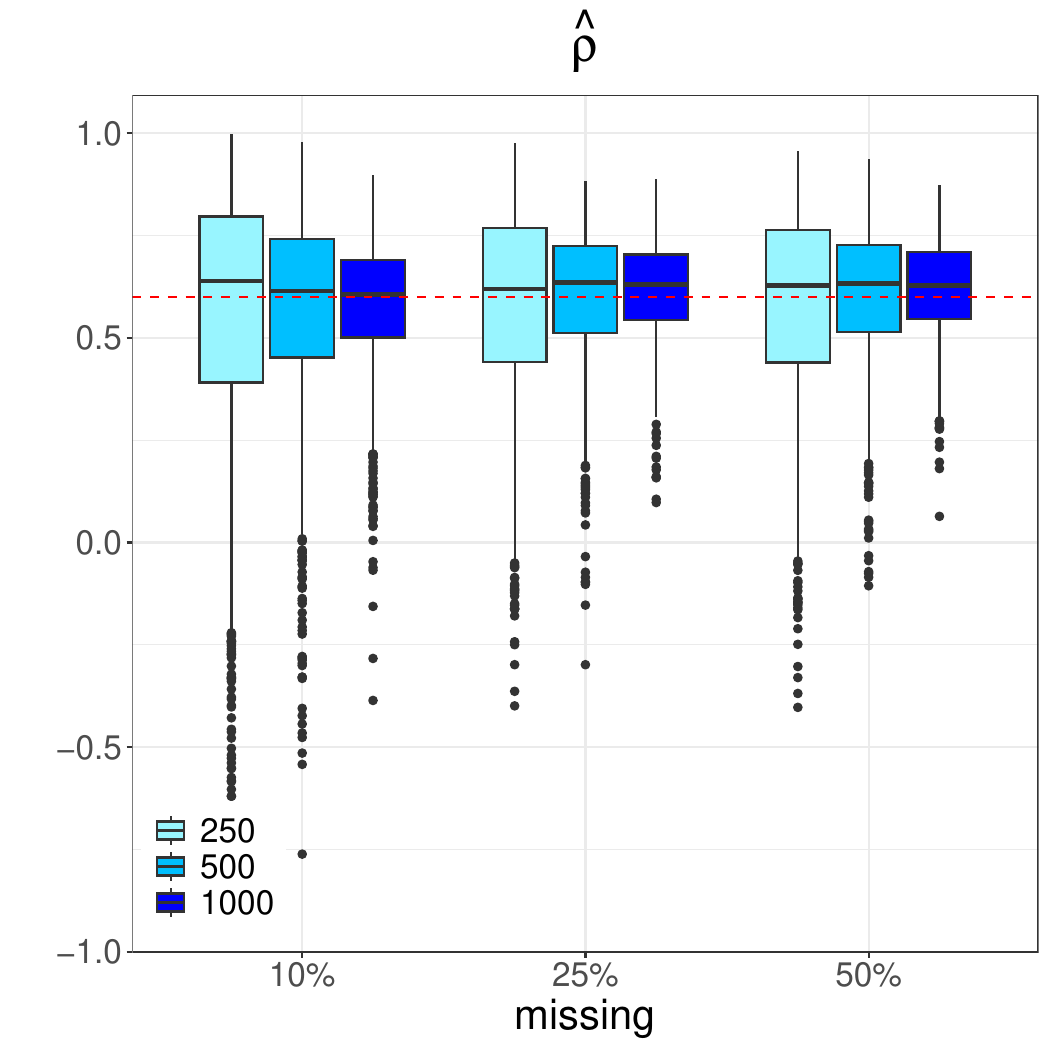}
  \caption{Boxplot of the 1000 Monte Carlo estimates of $\beta_0$, $\beta_1$, $\gamma_1$, $\gamma_2$, $\sigma$ and $\rho$ for the SLcn model, when data are generated from the slash distribution with $1.43$ degrees of freedom, with $n=250$, $500$, and $1000$ as sample sizes, when varying the missing proportion as $10\%$, $25\%$ and $50\%$.}
    \label{app2}
\end{center}
\end{figure}
\end{small}

\begin{small}
Additional results of the simulation study discussed in Section \ref{sec:Simulation study} of the SL models with $n=500$ when varying the missing rate as $10\%$, $25\%$, and $50\%$. 

\begin{figure}[!ht]
	\begin{center}
    \includegraphics[scale=0.25]{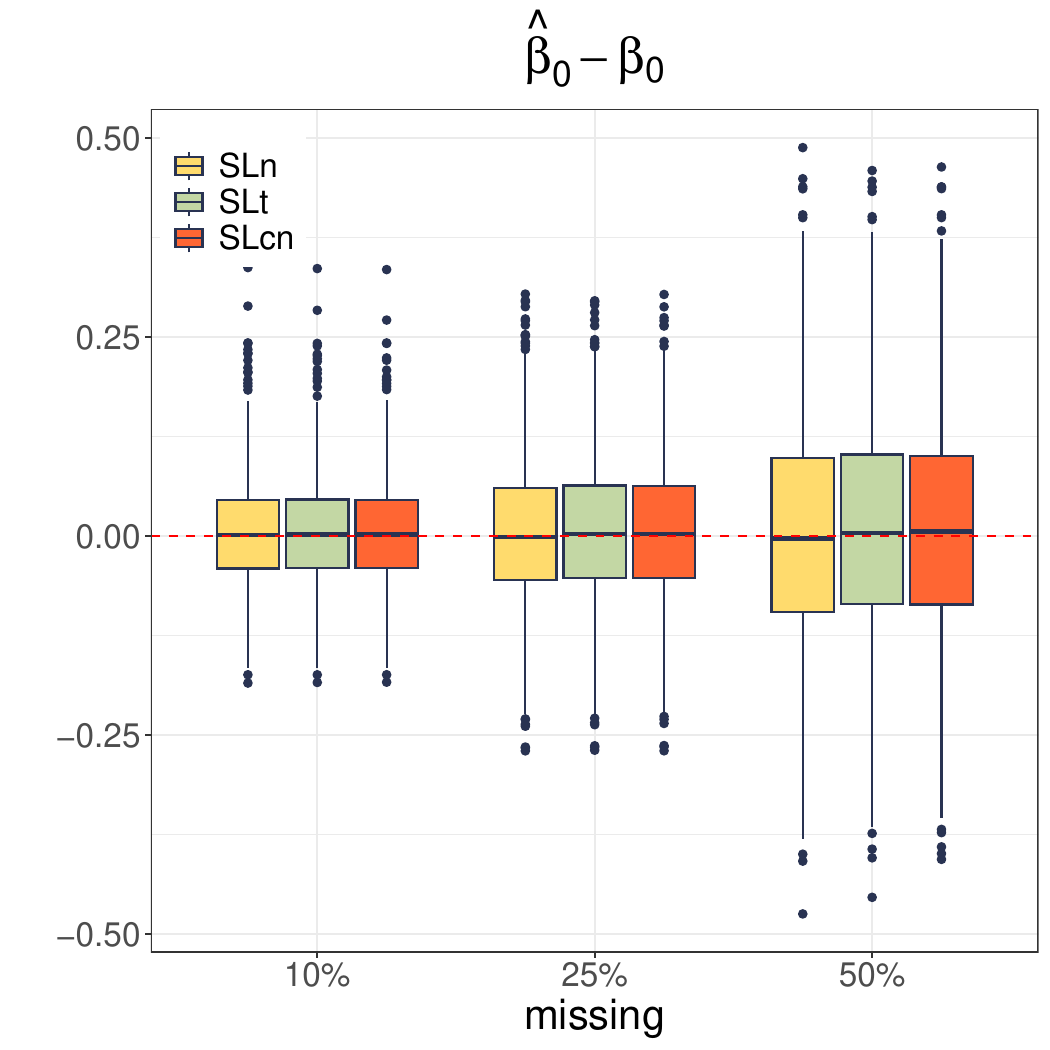}
    \includegraphics[scale=0.25]{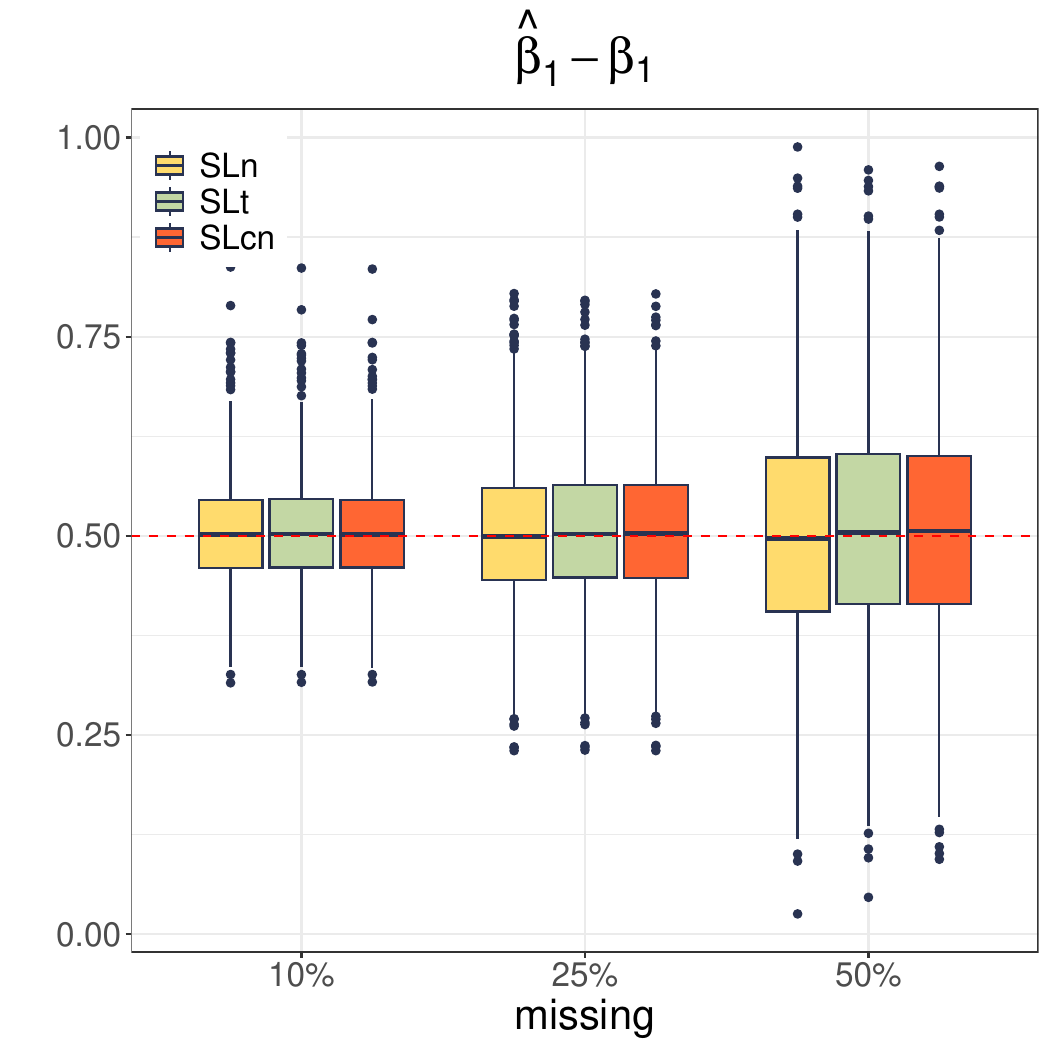}
    \includegraphics[scale=0.25]{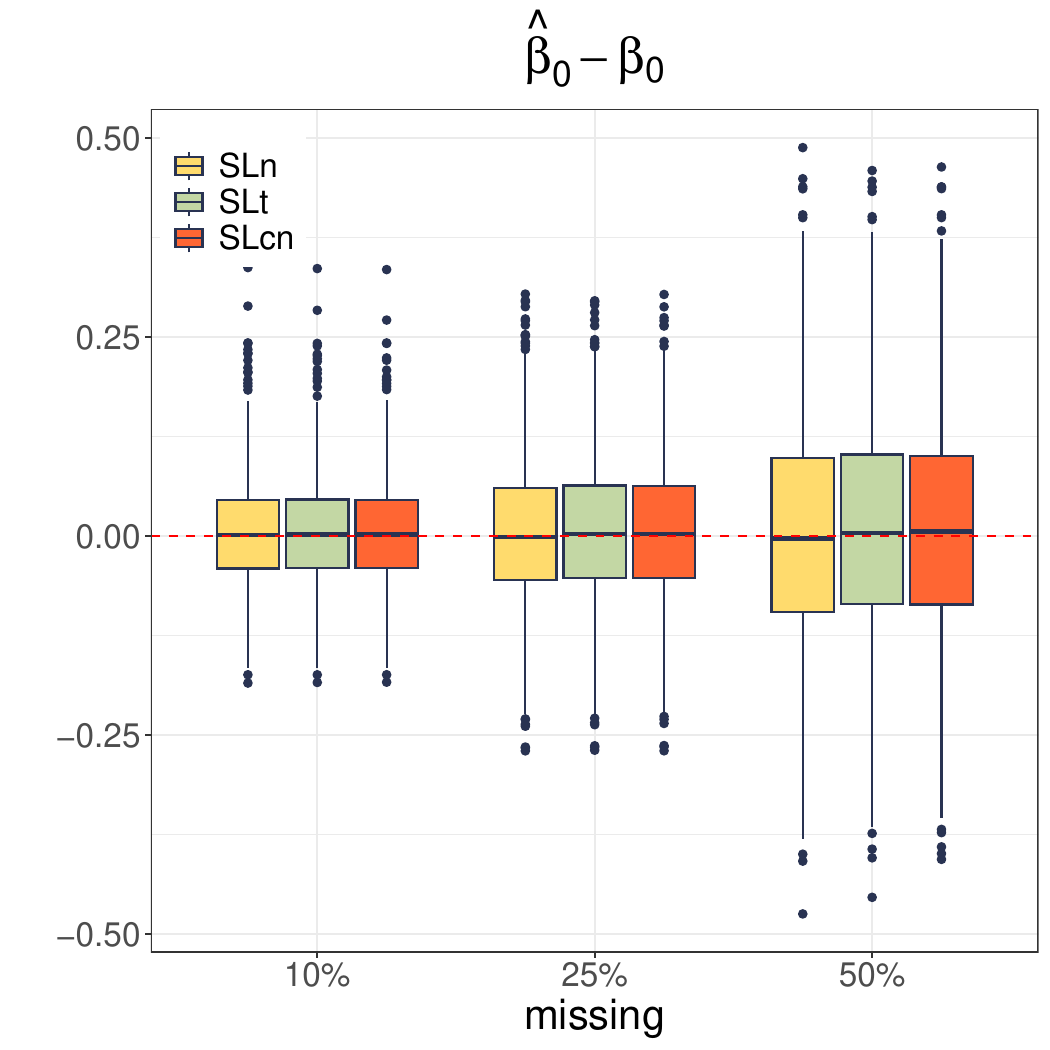}
    \includegraphics[scale=0.25]{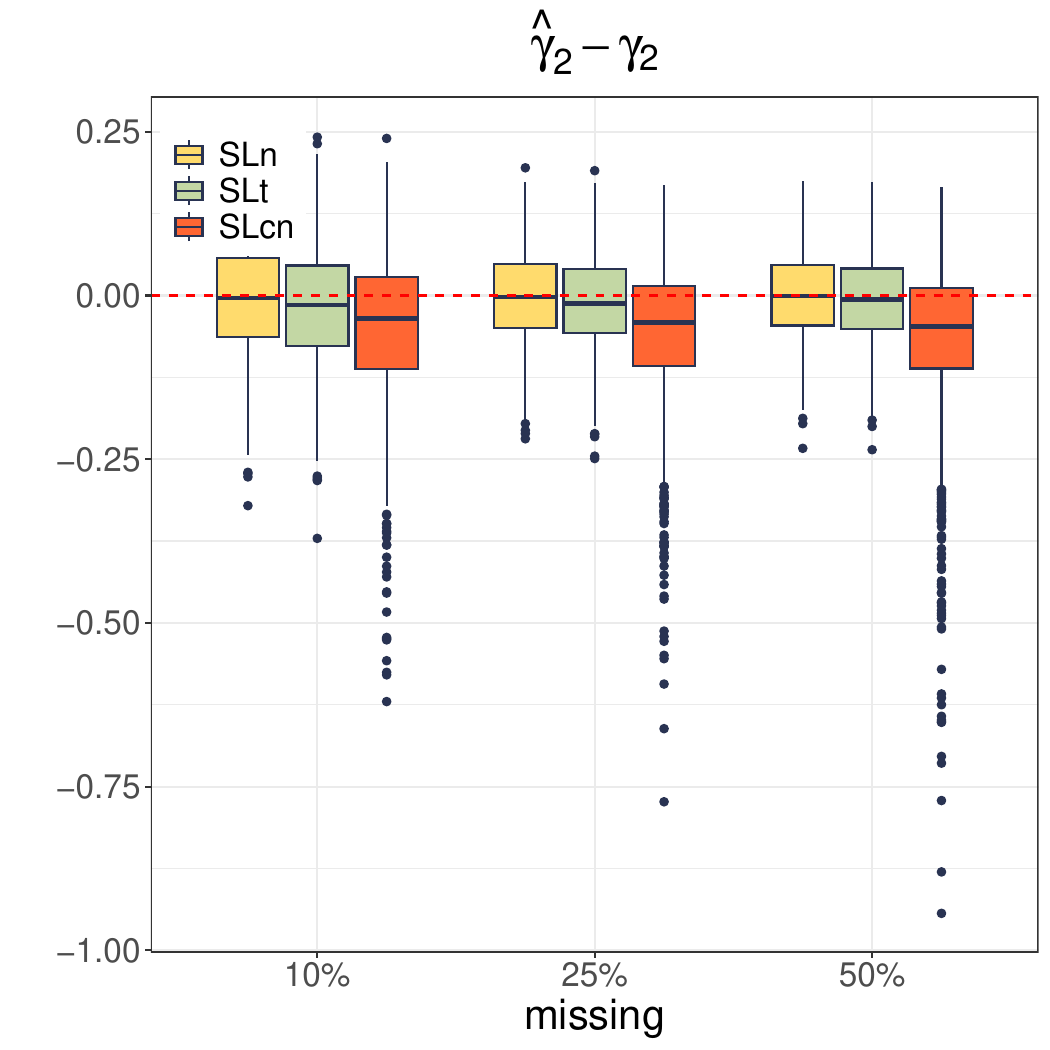}
    \includegraphics[scale=0.25]{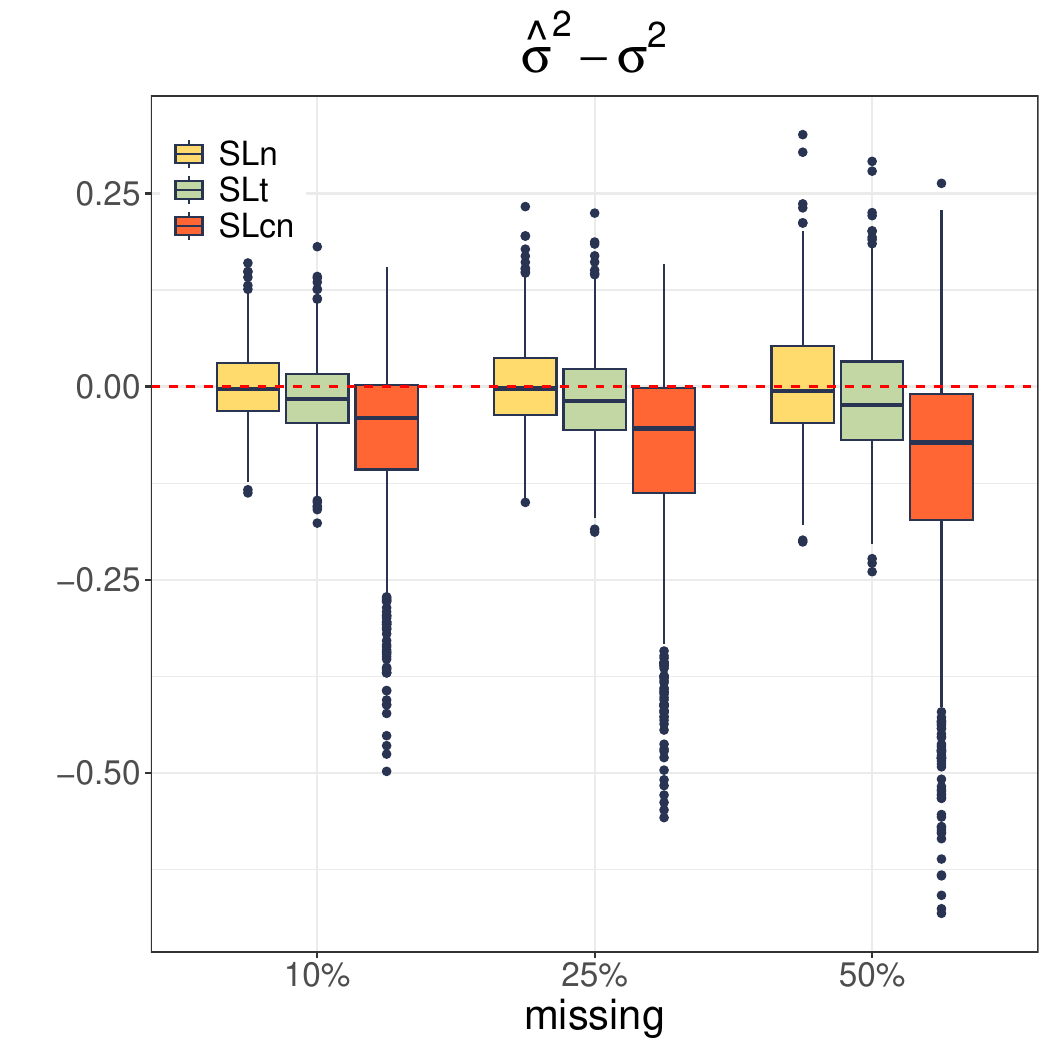}
    \includegraphics[scale=0.25]{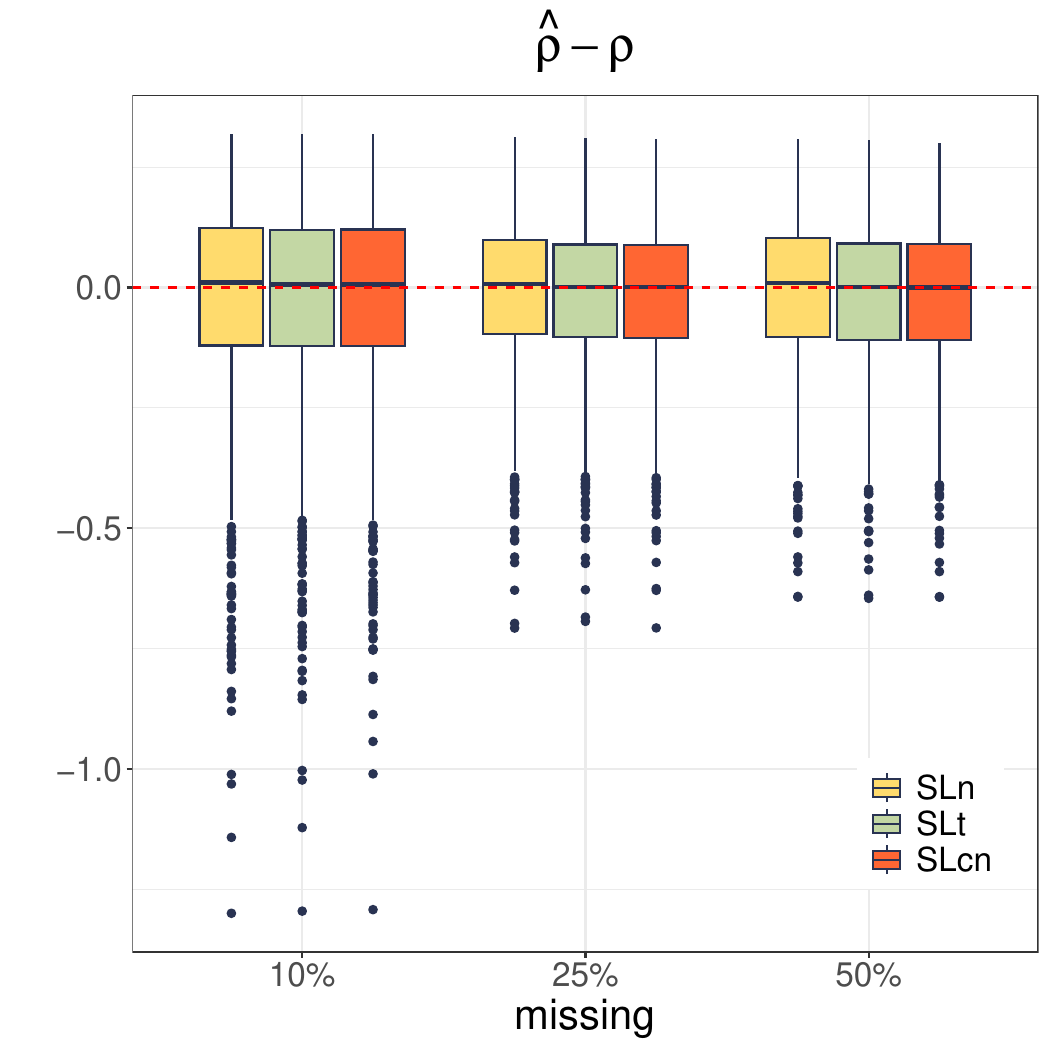}
    \caption{Boxplot of the SL models comparison of the 1000 Monte Carlo estimates for the data generated from the normal distribution. When varying the missing proportion as 10\%, 25\% and 50\%, the estimated parameters are $\beta_0$, $\beta_1$, $\gamma_1$, $\gamma_2$,  $\sigma$ and $\rho$ }
    \label{app3}
\end{center}
\end{figure}

\begin{figure}[!htb]
\begin{center}
    \includegraphics[scale=0.25]{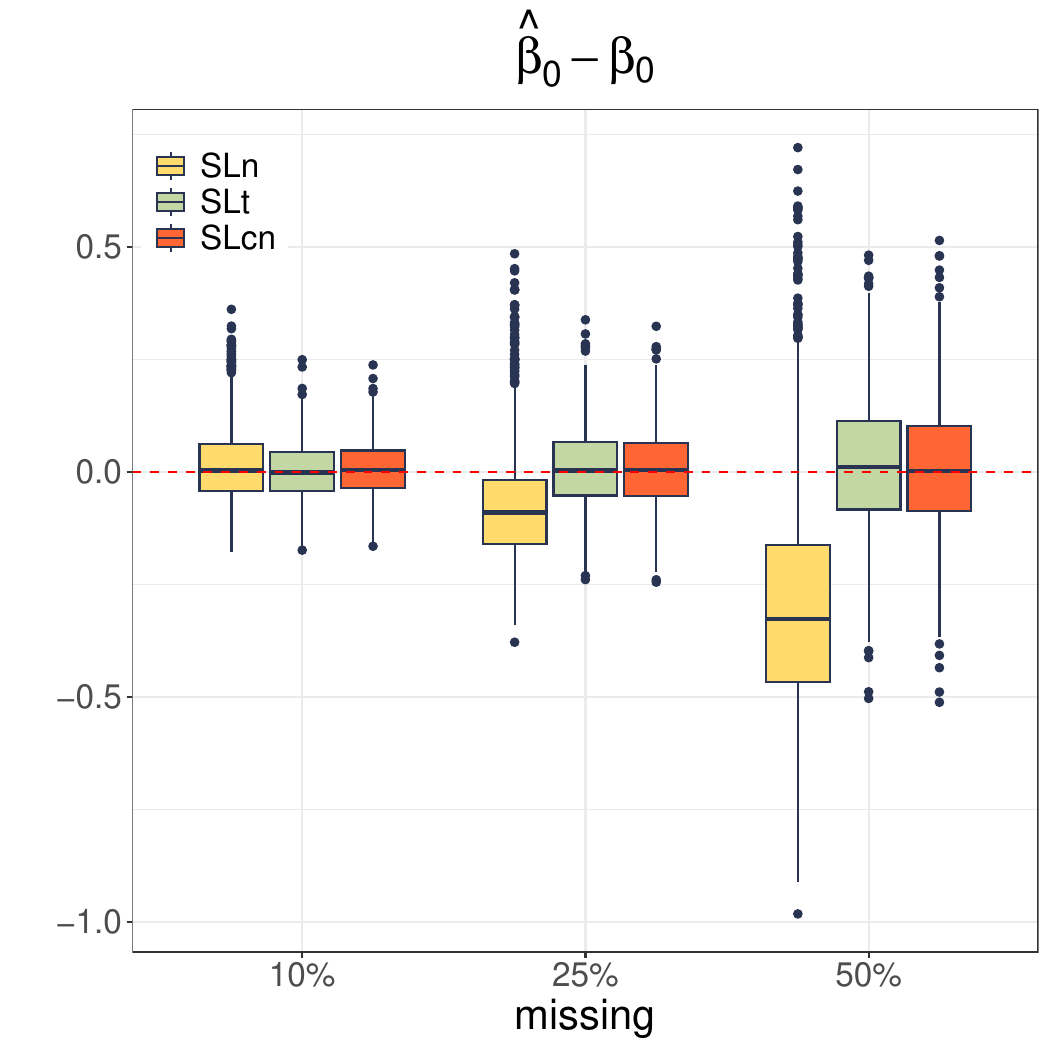}
    \includegraphics[scale=0.25]{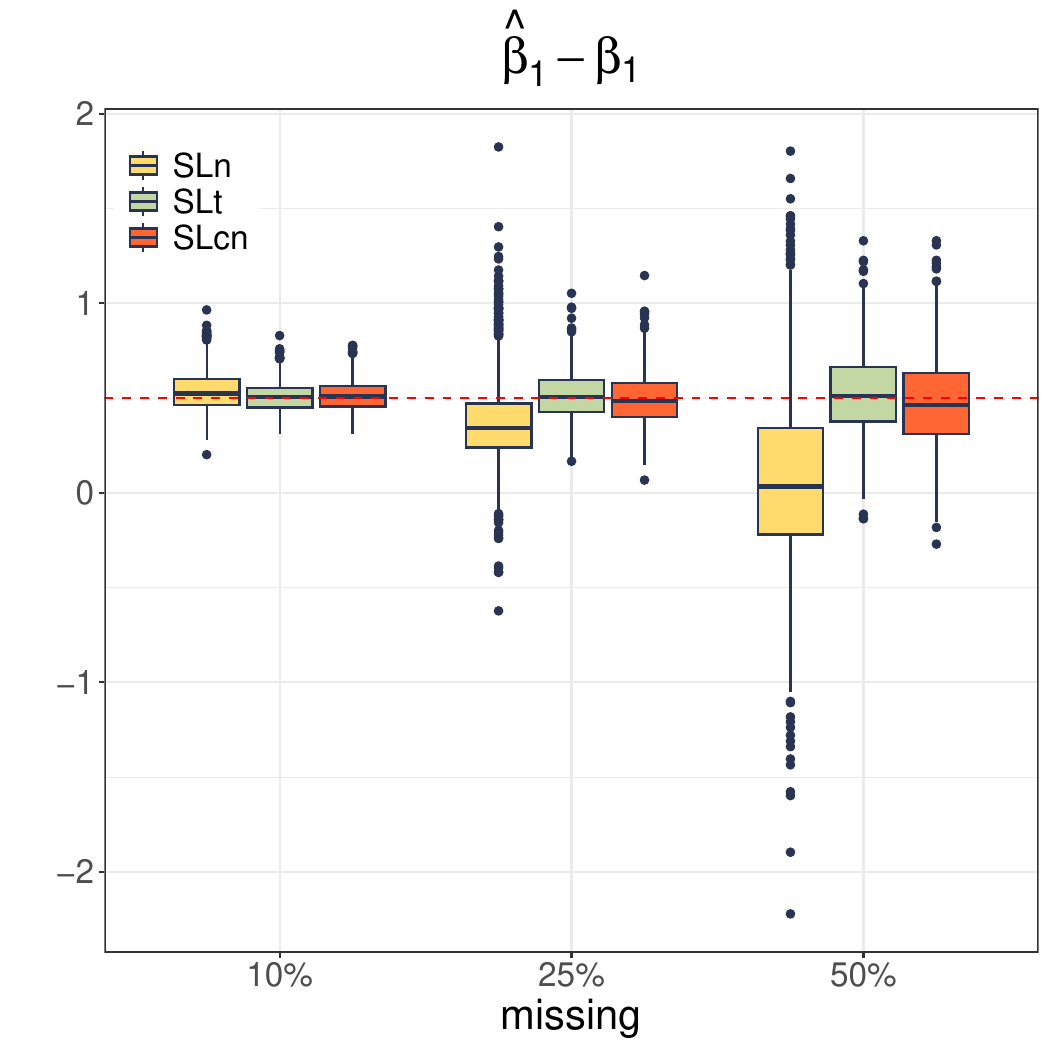}
    \includegraphics[scale=0.25]{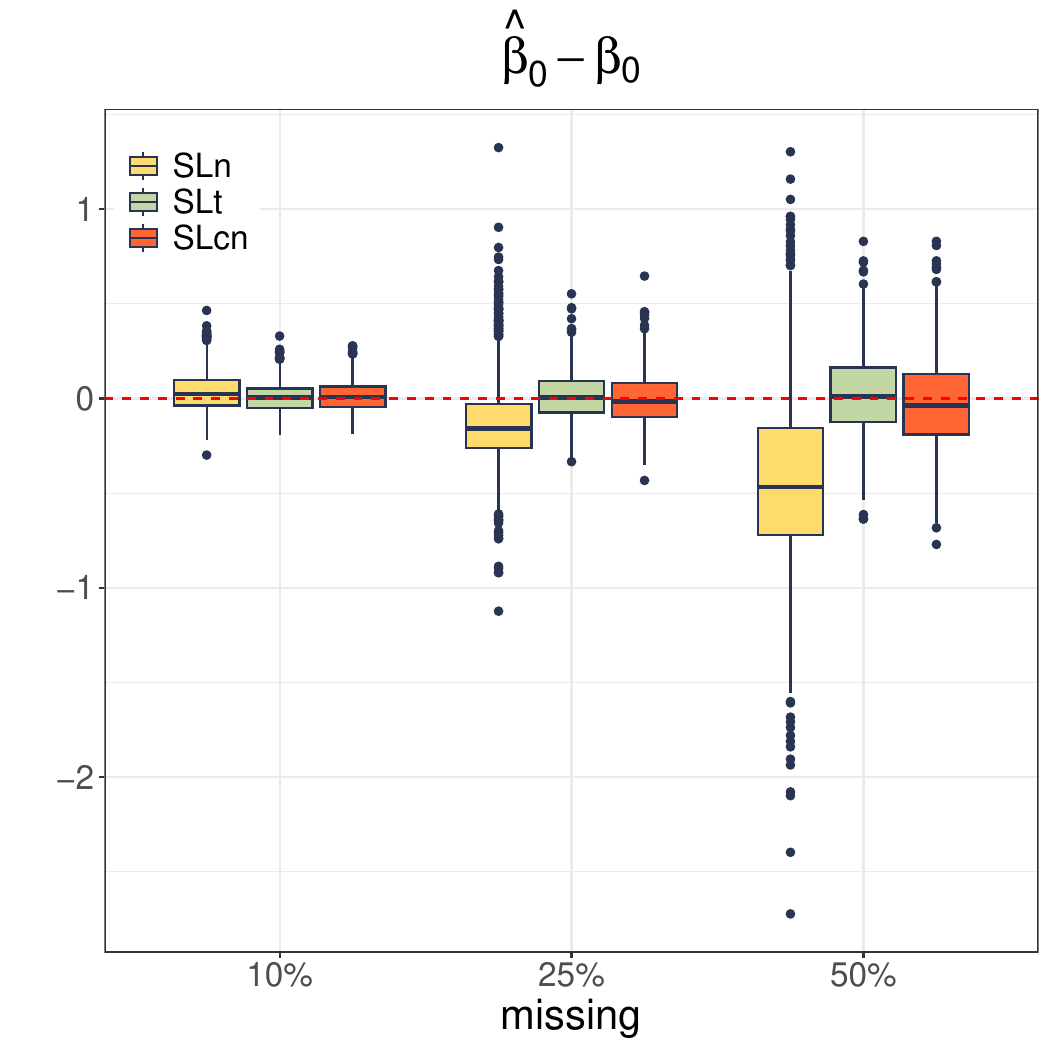}
    \includegraphics[scale=0.25]{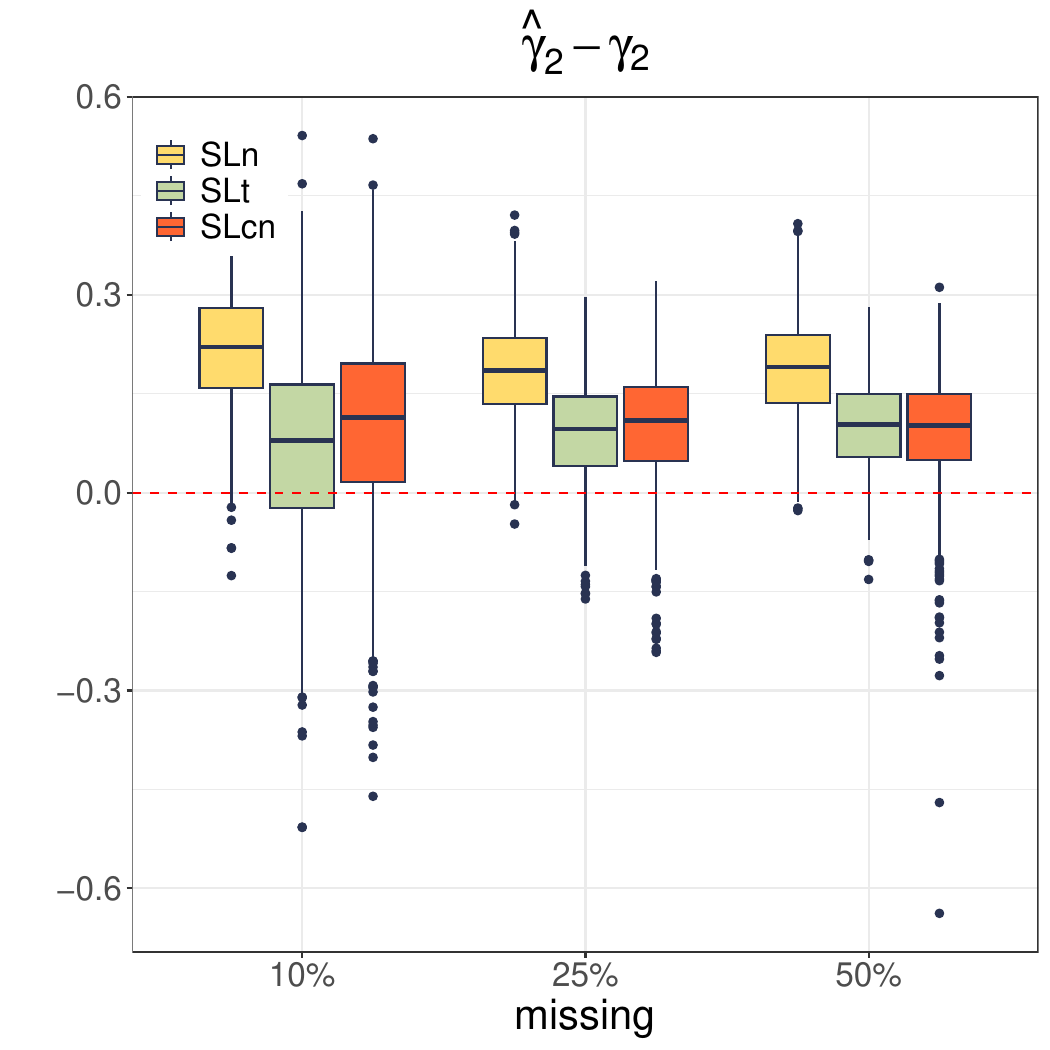}
    \includegraphics[scale=0.25]{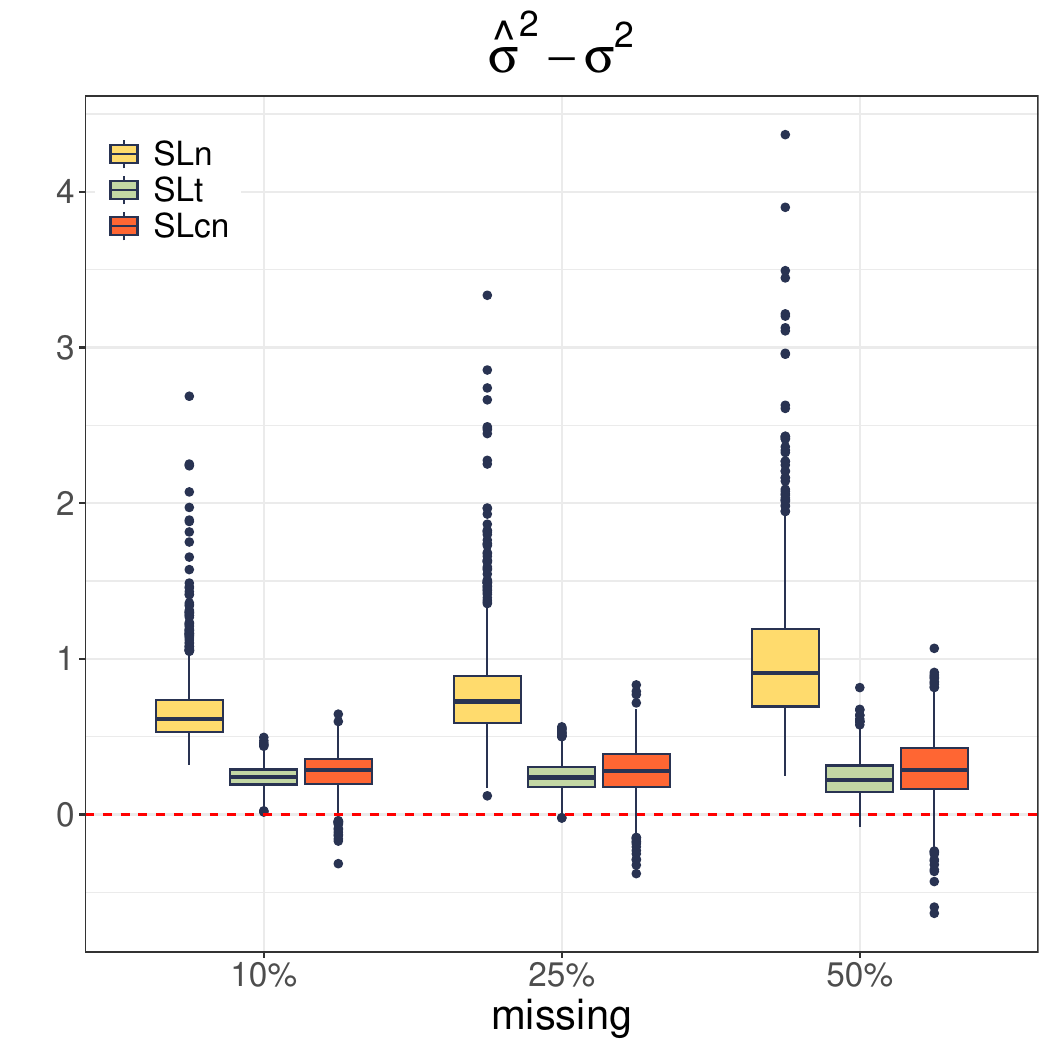}
    \includegraphics[scale=0.25]{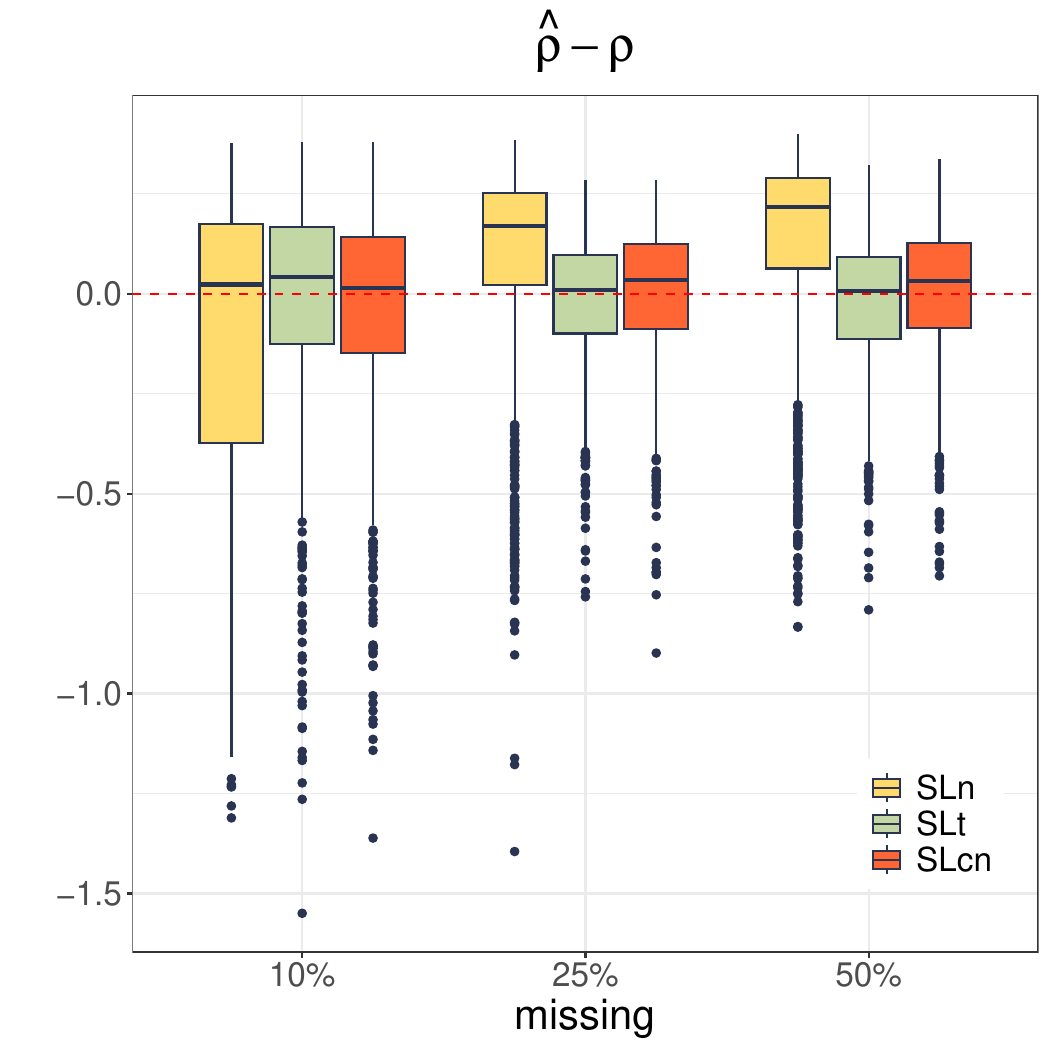}
    \caption{Boxplot of the SL models comparison of the 1000 Monte Carlo estimates for the data generated from the slash distribution. 
    When varying the missing proportion as 10\%, 25\% and 50\%, the estimated parameters are $\beta_0$, $\beta_1$, $\gamma_1$, $\gamma_2$,  $\sigma$ and $\rho$}
    \label{app4}
\end{center}
\end{figure}
\end{small}

\begin{landscape}
\begin{table}[ht]
\setlength{\tabcolsep}{5pt}
\caption{Simulation study 1. Mean estimates (EM), mean standard errors (SE), and Monte Carlo standard error (MC SE) of the $1000$ Monte Carlo replicates for the data generated from the normal distribution with the missing rate(MR) = $10\%$ and $50\%$.}
	\label{tab:sim_app1}
\centering
\small
\begin{tabular}{r|ccc|ccc|ccc||ccc|ccc|ccc}\hline
  \toprule
   & \multicolumn{9}{c}{MR=10\%} & \multicolumn{9}{c}{MR=50\%} \\\hline
 & \multicolumn{3}{c}{SLn} & \multicolumn{3}{c}{SLt} & \multicolumn{3}{c}{SLcn} & \multicolumn{3}{c}{SLn} & \multicolumn{3}{c}{SLt} & \multicolumn{3}{c}{SLcn}\\
 % \cmidrule{3-5} \cmidrule{6-8}    
   TRUE & EM & SE & MC SE & EM & SE & MC SE& EM & SE & MC SE  & EM & SE & MC SE & EM & SE & MC SE& EM & SE & MC SE \\\hline
   & \multicolumn{9}{c}{Sample Size = 250 }  & \multicolumn{9}{c}{Sample Size = 250}\\
 %\midrule
 $\beta_0=1.000$ & 1.014 & 0.096 & 0.101 & 1.014 & 0.094 & 0.100 & 1.011 & 0.093 & 0.097 & 1.001 & 0.221 & 0.221 & 1.012 & 0.216 & 0.216 & 1.015 & 0.212 & 0.213 \\   
 $\beta_1=0.500$ & 0.504 & 0.118 & 0.115 & 0.503 & 0.117 & 0.115 & 0.505 & 0.118 & 0.115 & 0.512 & 0.158 & 0.153 & 0.510 & 0.157 & 0.154 & 0.510 & 0.156 & 0.155 \\
 $\gamma_0=1.281/0$ & 1.307 & 0.126 & 0.131 & 1.333 & 0.132 & 0.141 & 1.436 & 0.209 & 0.279 & 0.001 & 0.084 & 0.080 & 0.001 & 0.085 & 0.081 & 0.002 & 0.101 & 0.100 \\    
 $\gamma_1=0.300$ & 0.298 & 0.177 & 0.187 & 0.307 & 0.182 & 0.192 & 0.331 & 0.204 & 0.217 & 0.305 & 0.141 & 0.141 & 0.309 & 0.142 & 0.143 & 0.361 & 0.181 & 0.203 \\  
 $\gamma_2=-0.500$ & -0.519 & 0.122 & 0.123 & -0.533 & 0.126 & 0.129 & -0.575 & 0.157 & 0.171 & -0.506 & 0.095 & 0.095 & -0.514 & 0.096 & 0.096 & -0.603 & 0.156 & 0.204 \\   
 $\sigma^2=1.000$ & 1.000 & 0.066 & 0.066 & 0.980 & 0.067 & 0.071 & 0.875 & 0.099 & 0.220 & 1.008 & 0.112 & 0.111 & 0.978 & 0.111 & 0.114 & 0.793 & 0.158 & 0.315 \\ 
 $\rho=0.600$ & 0.536 & 0.265 & 0.319 & 0.538 & 0.263 & 0.316 & 0.550 & 0.266 & 0.294 & 0.580 & 0.215 & 0.227 & 0.573 & 0.217 & 0.223 & 0.572 & 0.216 & 0.220 \\  
 $\nu\,\, (\nu_1)$ &  &  &  & 108.626 &  & &   0.461 & 0.032 & 0.123  & & &  & 109.485 &  & &  0.603 & 0.032 & 0.153 \\ 
 $\nu_2$ &  &  &  &  &  &  & 0.782 & 0.319 & 0.269 &  &  & & &  & & 0.721 & 0.313 & 0.309 \\ 
 AIC & \multicolumn{3}{c}{759.647 (23.445)}& \multicolumn{3}{c}{759.349 (23.471)}& \multicolumn{3}{c}{759.057 (23.475)}& \multicolumn{3}{c}{617.434 (27.935)}& \multicolumn{3}{c}{617.225 (27.953)}& \multicolumn{3}{c}{616.806 (27.927)}\\
 BIC & \multicolumn{3}{c}{763.169 (23.445)}& \multicolumn{3}{c}{762.870 (23.471)}& \multicolumn{3}{c}{762.578 (23.475)}& \multicolumn{3}{c}{620.956 (27.935)}& \multicolumn{3}{c}{620.746 (27.953)}& \multicolumn{3}{c}{620.327 (27.927)}\\%\midrule
 
   & \multicolumn{9}{c}{Sample Size = 500 }  & \multicolumn{9}{c}{ Sample Size = 500 }\\
 $\beta_0=1.000$ & 1.007 & 0.066 & 0.069 & 1.008 & 0.065 & 0.069 & 1.007 & 0.065 & 0.068 & 1.002 & 0.150 & 0.148 & 1.009 & 0.148 & 0.145 & 1.011 & 0.148 & 0.145 \\   
 $\beta_1=0.500$ & 0.504 & 0.083 & 0.081 & 0.503 & 0.083 & 0.082 & 0.503 & 0.083 & 0.081 & 0.503 & 0.110 & 0.105 & 0.501 & 0.109 & 0.105 & 0.500 & 0.110 & 0.105 \\  
 $\gamma_0=0.884$ & 1.293 & 0.086 & 0.087 & 1.312 & 0.089 & 0.093 & 1.392 & 0.150 & 0.182 & 0.001 & 0.059 & 0.059 & 0.001 & 0.059 & 0.059 & 0.002 & 0.068 & 0.068 \\  
 $\gamma_1=1.281/0$ & 0.303 & 0.132 & 0.134 & 0.310 & 0.135 & 0.138 & 0.329 & 0.149 & 0.152 & 0.305 & 0.100 & 0.098 & 0.307 & 0.101 & 0.099 & 0.345 & 0.124 & 0.131 \\   
 $\gamma_2=-0.500$ &-0.506 & 0.085 & 0.086 & -0.518 & 0.087 & 0.089 & -0.550 & 0.110 & 0.118 &  -0.502 & 0.065 & 0.065 & -0.507 & 0.066 & 0.066 & -0.571 & 0.105 & 0.132 \\  
 $\sigma^2=1.000$ &1.000 & 0.046 & 0.048 & 0.984 & 0.047 & 0.051 & 0.888 & 0.080 & 0.172 & 1.003 & 0.076 & 0.075 & 0.981 & 0.076 & 0.077 & 0.819 & 0.124 & 0.242 \\   
 $\rho=0.600$ &0.572 & 0.187 & 0.220 & 0.570 & 0.188 & 0.218 & 0.572 & 0.189 & 0.213 & 0.589 & 0.150 & 0.154 & 0.584 & 0.152 & 0.151 & 0.583 & 0.152 & 0.152 \\  
 $\nu\,\, (\nu_1)$ &  &  &  & 110.219 &  & & 0.461 & 0.022 & 0.088&  &  &  & 111.906 &  & &  0.612 & 0.022 & 0.122 \\ 
 $\nu_2$ &  &  &  &  &  &  &  0.806 & 0.312 & 0.222&  &  &  &  &  &  &  0.758 & 0.325 & 0.256 \\ 
 AIC & \multicolumn{3}{c}{1522.364 (33.013)}& \multicolumn{3}{c}{1522.134 (33.061)}& \multicolumn{3}{c}{1521.831 (33.073)}& \multicolumn{3}{c}{1268.664 (38.257)}& \multicolumn{3}{c}{1268.480 (38.274)}& \multicolumn{3}{c}{1268.109 (38.238)}\\ 
 BIC & \multicolumn{3}{c}{1526.579 (33.013)}& \multicolumn{3}{c}{1526.348 (33.061)}& \multicolumn{3}{c}{1526.046 (33.073)}& \multicolumn{3}{c}{1272.879 (38.257)}& \multicolumn{3}{c}{1272.695 (38.274)}& \multicolumn{3}{c}{1272.323 (38.238)}\\%\midrule
  
   & \multicolumn{9}{c}{Sample Size = 1000 }  & \multicolumn{9}{c}{Sample Size = 1000}\\
 $\beta_0=1.000$ & 1.003 & 0.046 & 0.045 & 1.004 & 0.045 & 0.044 & 1.004 & 0.045 & 0.044 & 0.998 & 0.105 & 0.103 & 1.004 & 0.104 & 0.100 & 1.005 & 0.104 & 0.100 \\  
 $\beta_1=0.500$ & 0.501 & 0.060 & 0.059 & 0.501 & 0.060 & 0.059 & 0.501 & 0.060 & 0.059 &  0.502 & 0.079 & 0.077 & 0.501 & 0.078 & 0.077 & 0.501 & 0.079 & 0.077 \\   
 $\gamma_0=1.281/0$ & 1.288 & 0.060 & 0.060 & 1.302 & 0.062 & 0.063 & 1.359 & 0.116 & 0.110 & -0.001 & 0.042 & 0.040 & -0.000 & 0.042 & 0.040 & -0.000 & 0.046 & 0.044 \\ 
 $\gamma_1=0.300$ & 0.302 & 0.095 & 0.095 & 0.307 & 0.096 & 0.097 & 0.320 & 0.105 & 0.104 & 0.302 & 0.073 & 0.076 & 0.305 & 0.073 & 0.076 & 0.332 & 0.091 & 0.090 \\   
 $\gamma_2=-0.500$ &-0.506 & 0.058 & 0.058 & -0.514 & 0.059 & 0.060 & -0.537 & 0.078 & 0.074& -0.500 & 0.045 & 0.047 & -0.504 & 0.046 & 0.047 & -0.550 & 0.081 & 0.078 \\  
 $\sigma^2=1.000$ & 0.999 & 0.033 & 0.034 & 0.987 & 0.033 & 0.036 & 0.906 & 0.070 & 0.124 & 1.002 & 0.054 & 0.053 & 0.984 & 0.054 & 0.055 & 0.841 & 0.115 & 0.186 \\  
 $\rho=0.600$ & 0.588 & 0.128 & 0.139 & 0.586 & 0.129 & 0.138 & 0.586 & 0.130 & 0.138 &  0.597 & 0.105 & 0.103 & 0.592 & 0.106 & 0.101 & 0.592 & 0.107 & 0.101 \\  
 $\nu\,\, (\nu_1)$ &  &  &  & 115.877 &  & &  0.455 & 0.016 & 0.057 &  &  &  & 113.365 &  & &  0.610 & 0.016 & 0.089 \\ 
 $\nu_2$ &  &  &  &  &  &  &  0.831 & 0.321 & 0.177 &  &  &  &  &  & & 0.778 & 0.345 & 0.207  \\ 
 AIC & \multicolumn{3}{c}{3046.684 (45.498)}& \multicolumn{3}{c}{3046.505 (45.555)}& \multicolumn{3}{c}{3046.229 (45.546)}& \multicolumn{3}{c}{2542.714 (53.913)}& \multicolumn{3}{c}{2542.510 (53.959)}& \multicolumn{3}{c}{2542.188 (53.953)}\\ 
 BIC & \multicolumn{3}{c}{3051.592 (45.498)}& \multicolumn{3}{c}{3051.413 (45.555)}& \multicolumn{3}{c}{3051.136 (45.546)}& \multicolumn{3}{c}{2547.622 (53.913)}& \multicolumn{3}{c}{2547.418 (53.959)}& \multicolumn{3}{c}{2547.096 (53.953)}\\
   \bottomrule
\end{tabular}
\end{table}
\end{landscape}

\begin{landscape}
\begin{table}[ht]
\setlength{\tabcolsep}{5pt}
\caption{Simulation study 1. Mean estimates (EM), mean standard errors (SE), and Monte Carlo standard error (MC SE) of the $1000$ Monte Carlo replicates for the data generated from the contaminated normal having $\nu_1$=0.1 and $\nu_2=0.1$ with the missing rate (MR) = $10\%$ and $50\%$.}
	\label{tab:sim_app2}
\centering
\small
\begin{tabular}{r|ccc|ccc|ccc||ccc|ccc|ccc}\hline
  \toprule
   & \multicolumn{9}{c}{MR=10\%} & \multicolumn{9}{c}{MR=50\%} \\\hline
 & \multicolumn{3}{c}{SLn} & \multicolumn{3}{c}{SLt} & \multicolumn{3}{c}{SLcn} & \multicolumn{3}{c}{SLn} & \multicolumn{3}{c}{SLt} & \multicolumn{3}{c}{SLcn}\\
 % \cmidrule{3-5} \cmidrule{6-8}    
   TRUE & EM & SE & MC SE & EM & SE & MC SE& EM & SE & MC SE  & EM & SE & MC SE & EM & SE & MC SE& EM & SE & MC SE \\\hline
   & \multicolumn{9}{c}{Sample Size = 250 }  & \multicolumn{9}{c}{Sample Size = 250}\\
 %\midrule
 $\beta_0=1.000$ &1.035 &0.134 &0.127 &1.009 &0.088 &0.096 & 1.011 & 0.091 & 0.093 & 0.776 & 0.277 & 0.446 & 1.022 & 0.200 & 0.233 & 1.019 & 0.223 & 0.227 \\  
 $\beta_1=0.500$ & 0.496 & 0.159 & 0.148 & 0.494 & 0.125 & 0.125 & 0.494 & 0.126 & 0.123 & 0.544 & 0.235 & 0.224 & 0.498 & 0.173 & 0.172 & 0.495 & 0.172 & 0.168 \\ 
 $\gamma_0=1.493/0$ & 1.350 & 0.127 & 0.136 & 1.701 & 0.203 & 0.242 & 1.670 & 0.221 & 0.289 & -0.004 & 0.084 & 0.083 & 0.005 & 0.092 & 0.093 & 0.005 & 0.095 & 0.098 \\  
 $\gamma_1=0.300$ & 0.236 & 0.184 & 0.186 & 0.335 & 0.254 & 0.266 & 0.324 & 0.248 & 0.261 & 0.269 & 0.139 & 0.138 & 0.309 & 0.153 & 0.153 & 0.321 & 0.161 & 0.165 \\  
 $\gamma_2=-0.500$ & -0.392 & 0.117 & 0.127 & -0.554 & 0.167 & 0.179 & -0.538 & 0.173 & 0.186 & -0.414 & 0.086 & 0.107 & -0.516 & 0.106 & 0.110 & -0.536 & 0.114 & 0.137 \\ 
 $\sigma^2=1.000$ & 1.299 & 0.055 & 0.138 & 0.971 & 0.070 & 0.080 & 0.960 & 0.074 & 0.185 & 1.502 & 0.106 & 0.281 & 0.965 & 0.107 & 0.138 & 0.941 & 0.131 & 0.291 \\
 $\rho=0.600$ & 0.499 & 0.273 & 0.413 & 0.563 & 0.233 & 0.318 & 0.561 & 0.262 & 0.300 & 0.647 & 0.152 & 0.335 & 0.563 & 0.190 & 0.223 & 0.565 & 0.211 & 0.220 \\ 
 $\nu\,\, (\nu_1)$ &  &  &  & 6.147 &  & &  0.143 & 0.029 & 0.106  & & &  & 7.906 &  & &  0.168 & 0.030 & 0.161 \\ 
 $\nu_2$ &  &  &  &  &  &  & 0.131 & 0.069 & 0.104  &  &  && &  & & 0.134 & 0.082 & 0.120 \\ 
 AIC & \multicolumn{3}{c}{874.270 (41.736)}& \multicolumn{3}{c}{843.348 (31.094)}& \multicolumn{3}{c}{840.134 (30.664)}& \multicolumn{3}{c}{698.744 (40.847)}& \multicolumn{3}{c}{678.582 (34.722)}& \multicolumn{3}{c}{675.785 (34.453)}\\
 BIC & \multicolumn{3}{c}{933.566 (53.450)} & \multicolumn{3}{c}{905.324 (31.358)} & \multicolumn{3}{c}{903.392 (31.358)}& \multicolumn{3}{c}{702.266 (40.847)}& \multicolumn{3}{c}{682.103 (34.722)}& \multicolumn{3}{c}{679.306 (34.453)}\\%\midrule
 
   & \multicolumn{9}{c}{Sample Size = 500 }  & \multicolumn{9}{c}{Sample Size = 500}\\
 $\beta_0=1.000$ & 1.019 & 0.089 & 0.090 & 1.001 & 0.060 & 0.062 & 1.006 & 0.063 & 0.061 & 0.710 & 0.170 & 0.277 & 1.016 & 0.133 & 0.147 & 1.006 & 0.150 & 0.145 \\  
 $\beta_1=0.500$ & 0.506 & 0.110 & 0.107 & 0.500 & 0.089 & 0.090 & 0.500 & 0.089 & 0.089 & 0.587 & 0.159 & 0.165 & 0.501 & 0.120 & 0.119 & 0.504 & 0.121 & 0.117 \\  
 $\gamma_0=1.493/0$ & 1.333 & 0.086 & 0.090 & 1.696 & 0.142 & 0.170 & 1.617 & 0.144 & 0.185 & -0.004 & 0.059 & 0.058 & 0.001 & 0.065 & 0.066 & -0.000 & 0.064 & 0.066 \\ 
 $\gamma_1=0.300$ & 0.229 & 0.136 & 0.139 & 0.343 & 0.196 & 0.204 & 0.319 & 0.184 & 0.195 & 0.271 & 0.099 & 0.097 & 0.308 & 0.110 & 0.110 & 0.307 & 0.109 & 0.110 \\  
 $\gamma_2=-0.500$ &-0.383 & 0.081 & 0.089 & -0.563 & 0.119 & 0.128 & -0.526 & 0.119 & 0.129 & -0.408 & 0.059 & 0.076 & -0.520 & 0.074 & 0.075 & -0.515 & 0.072 & 0.074 \\  
 $\sigma^2=1.000$ & 1.308 & 0.035 & 0.099 & 0.961 & 0.049 & 0.052 & 0.975 & 0.050 & 0.116 & 1.519 & 0.067 & 0.201 & 0.950 & 0.073 & 0.084 & 0.983 & 0.084 & 0.184 \\ 
 $\rho=0.600$ & 0.563 & 0.156 & 0.309 & 0.598 & 0.165 & 0.196 & 0.582 & 0.184 & 0.190 & 0.727 & 0.073 & 0.204 & 0.583 & 0.130 & 0.142 & 0.590 & 0.142 & 0.139 \\
 $\nu\,\, (\nu_1)$ &  &  &  & 4.317 &  & & 0.117 & 0.021 & 0.054 &  &  &  & 4.237 &  & &  0.118 & 0.021 & 0.068\\ 
 $\nu_2$ &  &  &  &  &  &  & 0.112 & 0.037 & 0.047 &  &  &  &  &  &  & 0.110 & 0.040 & 0.049  \\ 
 AIC & \multicolumn{3}{c}{1759.885 (58.726)}& \multicolumn{3}{c}{1693.211 (42.269)}& \multicolumn{3}{c}{1687.575 (41.793)}& \multicolumn{3}{c}{1438.771 (56.256)}& \multicolumn{3}{c}{1391.922 (46.615)}& \multicolumn{3}{c}{1387.280 (46.250)}\\ 
 BIC & \multicolumn{3}{c}{1764.099 (58.726)}& \multicolumn{3}{c}{1697.426 (42.269)}& \multicolumn{3}{c}{1691.789 (41.793)}& \multicolumn{3}{c}{1442.986 (56.256)}& \multicolumn{3}{c}{1396.136 (46.615)}& \multicolumn{3}{c}{1391.495 (46.250)}\\%\midrule
  
   & \multicolumn{9}{c}{ Sample Size = 1000 }  & \multicolumn{9}{c}{Sample Size = 1000}\\
 $\beta_0=1.000$ & 1.005 & 0.058 & 0.055 & 0.999 & 0.042 & 0.045 & 1.004 & 0.043 & 0.044 & 0.686 & 0.112 & 0.199 & 1.021 & 0.094 & 0.110 & 1.004 & 0.106 & 0.110 \\ 
 $\beta_1=0.500$ & 0.504 & 0.078 & 0.077 & 0.498 & 0.064 & 0.067 & 0.498 & 0.065 & 0.066 & 0.573 & 0.113 & 0.119 & 0.497 & 0.086 & 0.087 & 0.501 & 0.087 & 0.086 \\  
 $\gamma_0=1.493/0$ & 1.331 & 0.060 & 0.062 & 1.689 & 0.099 & 0.113 & 1.587 & 0.096 & 0.107 & -0.002 & 0.041 & 0.040 & 0.002 & 0.046 & 0.046 & 0.001 & 0.045 & 0.045 \\ 
 $\gamma_1=0.300$ & 0.224 & 0.097 & 0.098 & 0.331 & 0.137 & 0.139 & 0.303 & 0.126 & 0.127 & 0.264 & 0.072 & 0.071 & 0.306 & 0.080 & 0.080 & 0.302 & 0.079 & 0.078 \\  
 $\gamma_2=-0.500$ & -0.382 & 0.054 & 0.060 & -0.563 & 0.080 & 0.085 & -0.515 & 0.078 & 0.083 & -0.401 & 0.040 & 0.055 & -0.512 & 0.052 & 0.051 & -0.504 & 0.050 & 0.051 \\ 
 $\sigma^2=1.000$ & 1.307 & 0.024 & 0.073 & 0.960 & 0.035 & 0.038 & 0.990 & 0.035 & 0.082 &  1.523 & 0.045 & 0.154 & 0.947 & 0.052 & 0.065 & 0.994 & 0.060 & 0.137 \\
 $\rho=0.600$ & 0.618 & 0.078 & 0.194 & 0.609 & 0.113 & 0.140 & 0.590 & 0.126 & 0.136 & 0.752 & 0.041 & 0.141 & 0.580 & 0.094 & 0.106 & 0.592 & 0.100 & 0.104 \\ 
 $\nu\,\, (\nu_1)$ &  &  &  & 4.088 &  & &  0.106 & 0.014 & 0.030 &  &  &  & 3.967 &  & &  0.107 & 0.015 & 0.035 \\ 
 $\nu_2$ &  &  &  &  &  &  & 0.105 & 0.024 & 0.028 &  &  &  &  &  & &   0.106 & 0.026 & 0.030  \\ 
 AIC & \multicolumn{3}{c}{3521.508 (82.493)}& \multicolumn{3}{c}{3386.737 (58.398)}& \multicolumn{3}{c}{3376.411 (57.771)}& \multicolumn{3}{c}{2887.789 (82.413)}& \multicolumn{3}{c}{2792.843 (67.294)}& \multicolumn{3}{c}{2784.650 (66.624)}\\ 
 BIC & \multicolumn{3}{c}{3526.416 (82.493)}& \multicolumn{3}{c}{3391.645 (58.398)}& \multicolumn{3}{c}{3381.319 (57.771)}& \multicolumn{3}{c}{2892.697 (82.413)}& \multicolumn{3}{c}{2797.751 (67.294)}& \multicolumn{3}{c}{2789.558 (66.624)}\\
   \bottomrule
\end{tabular}
\end{table}
\end{landscape}

\begin{landscape}
\begin{table}[ht]
\setlength{\tabcolsep}{5pt}
\caption{Simulation study 1. Mean estimates (EM), mean standard errors (SE), and Monte Carlo standard error (MC SE) of the $1000$ Monte Carlo replicates for the slash with 1.43 degrees of freedom with the missing rate (MR) = $10\%$ and $50\%$.}
	\label{tab:sim_app3}
\centering
\small
\begin{tabular}{r|ccc|ccc|ccc||ccc|ccc|ccc}\hline
  \toprule
   & \multicolumn{9}{c}{MR=10\%} & \multicolumn{9}{c}{MR=50\%} \\\hline
 & \multicolumn{3}{c}{SLn} & \multicolumn{3}{c}{SLt} & \multicolumn{3}{c}{SLcn} & \multicolumn{3}{c}{SLn} & \multicolumn{3}{c}{SLt} & \multicolumn{3}{c}{SLcn}\\
 % \cmidrule{3-5} \cmidrule{6-8}    
   TRUE & EM & SE & MC SE & EM & SE & MC SE& EM & SE & MC SE  & EM & SE & MC SE & EM & SE & MC SE& EM & SE & MC SE \\\hline
   & \multicolumn{9}{c}{Sample Size = 250 }  & \multicolumn{9}{c}{Sample Size = 250}\\
 %\midrule
 $\beta_0=1.000$ &1.060 & 0.198 & 0.144 & 1.017 & 0.111 & 0.114 & 1.019 & 0.117 & 0.114 & 0.695 & 0.411 & 0.679 & 1.019 & 0.289 & 0.330 & 0.995 & 0.315 & 0.339 \\
 $\beta_1=0.500$ & 0.500 & 0.208 & 0.181 & 0.505 & 0.158 & 0.159 & 0.507 & 0.160 & 0.159 & 0.562 & 0.312 & 0.293 & 0.510 & 0.221 & 0.224 & 0.515 & 0.221 & 0.223 \\  
 $\gamma_0=2.266/0$ & 1.512 & 0.138 & 0.149 & 1.926 & 0.231 & 0.284 & 1.967 & 0.299 & 0.410 & -0.011 & 0.083 & 0.079 & 0.001 & 0.089 & 0.084 & 0.001 & 0.103 & 0.099 \\  
 $\gamma_1=0.300$ & 0.170 & 0.207 & 0.212 & 0.257 & 0.303 & 0.323 & 0.264 & 0.313 & 0.334 & 0.211 & 0.138 & 0.135 & 0.241 & 0.148 & 0.147 & 0.278 & 0.176 & 0.194 \\ 
 $\gamma_2=-0.500$ & -0.292 & 0.131 & 0.132 & -0.426 & 0.191 & 0.198 & -0.431 & 0.206 & 0.217 & -0.323 & 0.082 & 0.104 & -0.400 & 0.097 & 0.101 & -0.456 & 0.125 & 0.163 \\ 
 $\sigma^2=1.000$ & 1.653 & 0.072 & 0.257 & 1.250 & 0.086 & 0.107 & 1.527 & 0.102 & 0.416 & 1.961 & 0.147 & 0.536 & 1.262 & 0.150 & 0.203 & 1.490 & 0.193 & 0.707 \\ 
 $\rho=0.600$ & 0.425 & 0.488 & 0.452 & 0.546 & 0.285 & 0.373 & 0.543 & 0.333 & 0.354 & 0.635 & 0.186 & 0.363 & 0.563 & 0.209 & 0.243 & 0.576 & 0.221 & 0.246 \\  
 $\nu\,\, (\nu_1)$ &  &  &  & 7.574 &  & & 0.201 & 0.029 & 0.182 &  &  &  & 11.652 &  & & 0.258 & 0.030 & 0.256 \\ 
 $\nu_2$ &  &  &  &  &  &  & 0.156 & 0.079 & 0.123 &  &  &  &  &  & & 0.146 & 0.088 & 0.116 \\ 
 AIC & \multicolumn{3}{c}{987.223 (58.998)}& \multicolumn{3}{c}{952.314 (30.678)}& \multicolumn{3}{c}{950.631 (30.646)}& \multicolumn{3}{c}{768.858 (50.303)}& \multicolumn{3}{c}{747.628 (36.678)}& \multicolumn{3}{c}{745.656 (36.521)}\\
 BIC & \multicolumn{3}{c}{990.744 (58.998)}& \multicolumn{3}{c}{955.835 (30.678)}& \multicolumn{3}{c}{954.152 (30.646)}& \multicolumn{3}{c}{772.379 (50.303)}& \multicolumn{3}{c}{751.149 (36.678)}& \multicolumn{3}{c}{749.178 (36.521)}\\%\midrule
 
   & \multicolumn{9}{c}{Sample Size = 500 }  & \multicolumn{9}{c}{Sample Size = 500 }\\
 $\beta_0=1.000$ & 1.040 & 0.129 & 0.109 & 1.007 & 0.075 & 0.079 & 1.013 & 0.080 & 0.079 & 0.581 & 0.243 & 0.506 & 1.018 & 0.195 & 0.212 & 0.970 & 0.223 & 0.238 \\   
 $\beta_1=0.500$ & 0.509 & 0.145 & 0.133 & 0.504 & 0.112 & 0.113 & 0.504 & 0.114 & 0.112 & 0.608 & 0.214 & 0.216 & 0.500 & 0.153 & 0.147 & 0.510 & 0.157 & 0.151 \\ 
 $\gamma_0=2.266/0$ & 1.487 & 0.093 & 0.106 & 1.906 & 0.158 & 0.203 & 1.826 & 0.168 & 0.250& -0.009 & 0.058 & 0.059 & 0.001 & 0.063 & 0.063 & 0.001 & 0.064 & 0.065 \\  
 $\gamma_1=0.300$ & 0.170 & 0.150 & 0.149 & 0.268 & 0.228 & 0.235 & 0.250 & 0.214 & 0.224& 0.218 & 0.098 & 0.095 & 0.242 & 0.106 & 0.106 & 0.250 & 0.109 & 0.114 \\   
 $\gamma_2=-0.500$ &-0.281 & 0.089 & 0.093 & -0.429 & 0.134 & 0.143 & -0.398 & 0.131 & 0.145& -0.312 & 0.055 & 0.077 & -0.399 & 0.068 & 0.068 & -0.406 & 0.070 & 0.088 \\   
 $\sigma^2=1.000$ & 1.670 & 0.043 & 0.236 & 1.241 & 0.061 & 0.073 & 1.635 & 0.064 & 0.309 & 2.000 & 0.086 & 0.464 & 1.236 & 0.102 & 0.128 & 1.714 & 0.127 & 0.545 \\ 
 $\rho=0.600$ & 0.488 & 0.290 & 0.370 & 0.582 & 0.198 & 0.263 & 0.561 & 0.232 & 0.256 & 0.718 & 0.091 & 0.260 & 0.581 & 0.145 & 0.159 & 0.605 & 0.153 & 0.171 \\ 
 $\nu\,\, (\nu_1)$ &  &  &  & 4.748 &  & & 0.136 & 0.020 & 0.110 &  &  &  & 5.026 &  & & 0.150 & 0.020 & 0.154 \\ 
 $\nu_2$ &  &  &  &  &  & &  0.130 & 0.040 & 0.081&  &  &  &  &  &  & 0.128 & 0.049 & 0.087 \\  
 AIC & \multicolumn{3}{c}{1987.106 (101.437)}& \multicolumn{3}{c}{1909.860 (43.220)}& \multicolumn{3}{c}{1908.890 (43.920)}& \multicolumn{3}{c}{1578.798 (81.790)}& \multicolumn{3}{c}{1528.395 (51.832)}& \multicolumn{3}{c}{1527.019 (52.186)}\\ 
 BIC & \multicolumn{3}{c}{1991.320 (101.437)}& \multicolumn{3}{c}{1914.074 (43.220)}& \multicolumn{3}{c}{1913.105 (43.920)}& \multicolumn{3}{c}{1583.013 (81.790)}& \multicolumn{3}{c}{1532.610 (51.832)}& \multicolumn{3}{c}{1531.234 (52.186)}\\%\midrule
  
   & \multicolumn{9}{c}{Sample Size = 1000 }  & \multicolumn{9}{c}{Sample Size = 1000 }\\
 $\beta_0=1.000$ & 1.029 & 0.083 & 0.082 & 1.001 & 0.052 & 0.053 & 1.009 & 0.056 & 0.053 &0.505 & 0.149 & 0.375 & 1.019 & 0.136 & 0.143 & 0.954 & 0.158 & 0.172 \\  
 $\beta_1=0.500$ & 0.503 & 0.102 & 0.102 & 0.502 & 0.081 & 0.083 & 0.502 & 0.083 & 0.083 &0.600 & 0.153 & 0.166 & 0.498 & 0.110 & 0.111 & 0.512 & 0.114 & 0.113 \\   
 $\gamma_0=2.266/0$ & 1.478 & 0.065 & 0.077 & 1.897 & 0.110 & 0.139 & 1.759 & 0.107 & 0.166 &-0.009 & 0.041 & 0.041 & 0.002 & 0.044 & 0.043 & 0.002 & 0.044 & 0.043 \\   
 $\gamma_1=0.300$ & 0.167 & 0.106 & 0.105 & 0.262 & 0.160 & 0.164 & 0.232 & 0.143 & 0.147 &0.210 & 0.071 & 0.073 & 0.240 & 0.078 & 0.081 & 0.237 & 0.077 & 0.079 \\  
 $\gamma_2=-0.500$ & -0.276 & 0.059 & 0.063 & -0.429 & 0.091 & 0.097 & -0.380 & 0.085 & 0.096 &-0.305 & 0.037 & 0.058 & -0.399 & 0.047 & 0.049 & -0.389 & 0.046 & 0.052 \\  
 $\sigma^2=1.000$ & 1.677 & 0.026 & 0.172 & 1.235 & 0.044 & 0.051 & 1.696 & 0.044 & 0.225 & 2.034 & 0.054 & 0.350 & 1.222 & 0.072 & 0.090 & 1.815 & 0.088 & 0.397 \\  
 $\rho=0.600$ & 0.533 & 0.155 & 0.301 & 0.604 & 0.136 & 0.164 & 0.577 & 0.158 & 0.169 & 0.767 & 0.043 & 0.178 & 0.585 & 0.103 & 0.104 & 0.620 & 0.106 & 0.120 \\ 
 $\nu\,\, (\nu_1)$ &  &  &  & 4.389 &  & &  0.103 & 0.014 & 0.067&  &  &  & 4.325 &  & & 0.100 & 0.014 & 0.075 \\
 $\nu_2$ &  &  &  &  &  &  & 0.109 & 0.021 & 0.058  &  &  &  &  &  & 0.108 & 0.025 & 0.062 \\ 
 AIC & \multicolumn{3}{c}{3991.221 (150.796)}& \multicolumn{3}{c}{3827.143 (58.769)}& \multicolumn{3}{c}{3828.473 (60.177)}& \multicolumn{3}{c}{3175.487 (120.699)}& \multicolumn{3}{c}{3066.208 (72.290)}& \multicolumn{3}{c}{3066.833 (73.479)}\\ 
 BIC & \multicolumn{3}{c}{3996.129 (150.796)}& \multicolumn{3}{c}{3832.051 (58.769)}& \multicolumn{3}{c}{3833.381 (60.177)}& \multicolumn{3}{c}{3180.394 (120.699)}& \multicolumn{3}{c}{3071.116 (72.290)}& \multicolumn{3}{c}{3071.740 (73.479)}\\
   \bottomrule
\end{tabular}
\end{table}
\end{landscape}

\bibliographystyle{chicago} 
\bibliography{bibliornl}
%\bibliography{chicago}

\end{document}